\documentclass[twocolumn,english,aps,prb,floatfix,amssymb,superscriptaddress,a4paper]{revtex4-1}
\usepackage{lipsum}
\usepackage{amsmath}
\usepackage{amssymb}
\usepackage{graphicx}
\usepackage{physics}
\usepackage{hyperref}
\usepackage[dvipsnames]{xcolor}
\usepackage{float}
\usepackage[caption=false]{subfig}
\hypersetup{
	colorlinks=true,
	hypertexnames=true
}

\newcommand{\Cbar}{\bar{C}}      

\newcommand{\Ras}{\bar{\bf R}}
\newcommand{\Rti}{\tilde{\bf R}}
\newcommand{\ras}{\bar{R}}
\newcommand{\pas}{\bar{\phi}}
\newcommand{\xas}{\bar{x}}
\newcommand{\yas}{\bar{y}}
\newcommand{\uas}{\bar{u}}
\newcommand{\vas}{\bar{v}}
\newcommand{\rti}{\tilde{R}}
\newcommand{\pti}{\tilde{\phi}}
\newcommand{\xti}{\tilde{x}}
\newcommand{\yti}{\tilde{y}}
\newcommand{\uti}{\tilde{u}}
\newcommand{\vti}{\tilde{v}}
\newcommand{\epin}{e_{\mathrm{pin}}}

\newcommand{\fpin}{f_{\mathrm{pin}}}
\newcommand{\Fpin}{F_{\mathrm{pin}}}

\newcommand{\Uti}{\mathcal{U}_{\Rti}}
\newcommand{\Bas}{\mathcal{B}_{\Ras}}
\newcommand{\Vas}{\mathcal{V}_{\Ras}}
\newcommand{\Jti}{\mathcal{J}_\mathrm{\Rti}}
\newcommand{\Lti}{\mathcal{L}_\mathrm{\Rti}}
\newcommand{\Rjp}{\Rti_{\mathrm{jp}}}
\newcommand{\Rlp}{\Rti_{\mathrm{lp}}}
\newcommand{\Rsjp}{\Rti_{s,\mathrm{jp}}}
\newcommand{\Rslp}{\Rti_{s,\mathrm{lp}}}
\newcommand{\ujp}{\uti_{\mathrm{jp}}}
\newcommand{\vjp}{\vti_{\mathrm{jp}}}
\newcommand{\ulp}{\uti_{\mathrm{lp}}}
\newcommand{\vlp}{\vti_{\mathrm{lp}}}
\newcommand{\usjp}{\uti_{s,\mathrm{jp}}}
\newcommand{\vsjp}{\vti_{s,\mathrm{jp}}}

\newcommand{\xjp}{\xti_{\mathrm{jp}}}

\newcommand{\xlp}{\xti_{\mathrm{lp}}}

\begin{document}

\title{Strong pinning transition with arbitrary defect potentials}

\author{Filippo Gaggioli}
\affiliation
{Institut f\"ur Theoretische Physik, ETH Z\"urich,
CH-8093 Z\"urich, Switzerland}

\author{Gianni Blatter}
\affiliation
{Institut f\"ur Theoretische Physik, ETH Z\"urich,
CH-8093 Z\"urich, Switzerland}

\author{Martin Buchacek}
\affiliation
{Institut f\"ur Theoretische Physik, ETH Z\"urich,
CH-8093 Z\"urich, Switzerland}

\author{Vadim B. Geshkenbein}
\affiliation
{Institut f\"ur Theoretische Physik, ETH Z\"urich,
CH-8093 Z\"urich, Switzerland}

\date{\today}

 \begin{abstract}
Dissipation-free current transport in type II superconductors requires
vortices, the topological defects of the superfluid, to be pinned by defects
in the underlying material. The pinning capacity of a defect is quantified by
the Labusch parameter $\kappa \sim f_p/\xi\Cbar$, measuring the pinning force
$f_p$ relative to the elasticity $\Cbar$ of the vortex lattice, with $\xi$
denoting the coherence length (or vortex core size) of the superconductor.
The critical value $\kappa = 1$ separates weak from strong pinning, with a
strong defect at $\kappa > 1$ able to pin a vortex on its own.  So far, this
weak-to-strong pinning transition has been studied for isotropic defect
potentials, resulting in a critical exponent $\mu = 2$ for the onset of the
strong pinning force density $F_\mathrm{pin} \sim n_p f_p
(\xi/a_0)^2(\kappa-1)^\mu$, with $n_ p$ denoting the density of defects and
$a_0$ the intervortex distance. This result is owed to the special rotational
symmetry of the defect producing a {\it finite} trapping area $S_\mathrm{trap}
\sim \xi^2$ at the strong-pinning onset.  The behavior changes dramatically
when studying anisotropic defects with no special symmetries: the strong
pinning then originates out of isolated points with length scales growing as
$\xi (\kappa - 1)^{1/2}$, resulting in a different force exponent $\mu = 5/2$.
Our analysis of the strong pinning onset for arbitrary defect potentials
$e_p(\mathbf{R})$, with $\mathbf{R}$ a planar coordinate, makes heavy use of
the Hessian matrix describing its curvature and leads us to interesting
geometrical structures: the strong pinning onset is characterized by the
appearance of {\it unstable} areas of elliptical shape whose boundaries mark
the locations where vortices jump. The associated locations of asymptotic
vortex positions define areas of {\it bistable} vortex states; these bistable
regions assume the shape of a crescent with boundaries that correspond to the
spinodal lines in a thermodynamic first-order transition and cusps
corresponding to critical end-points.  Both, unstable and bistable areas grow
with $\kappa > 1$ and join up into larger domains; for a uniaxially
anisotropic defect, two face to face crescents merge into the ring-shaped area
previously encountered for the isotropic defect.  Both, onset and merger
points are defined by local differential properties of the Hessian's
determinant $D(\mathbf{R})$, specifically, its minima and saddle points.
Extending our analysis to the case of a random two-dimensional pinning
landscape, we discuss the topological properties of unstable and bistable
regions as expressed through the Euler characteristic, with the latter related
to the local differential properties of $D(\mathbf{R})$ through Morse theory.
\end{abstract}

\maketitle


\section{Introduction}

Vortex pinning by material defects \cite{CampbellEvetts_2006} determines the
phenomenological properties of all technically relevant (type II)
superconducting materials, e.g., their dissipation-free transport or magnetic
response. Similar applies to the pinning of dislocations in metals
\cite{Kassner_2015} or domain walls in magnets \cite{Gorchon_2014}, with the
commonalities found in the topological defects of the ordered phase being
pinned by defects in the host material: these topological defects are the
vortices \cite{Abrikosov_1957}, dislocations \cite{Burgers_1940}, or domain
walls \cite{Bloch_1932, LandauLifshitz_1935} appearing within the respective
ordered phases---superconducting, crystalline, or magnetic.  The theory
describing the pinning of topological defects has been furthest developed in
superconductors, with the strong pinning paradigm
\cite{Labusch_1969,LarkinOvch_1979} having been strongly pushed during the
last decade \cite{Koopmann_2004, Thomann_2012, Willa_2015_PRL, Buchacek_2019}.
In its simplest form, it boils down to the setup involving a single vortex
subject to one defect and the cage potential \cite{ErtasNelson_1996,
Vinokur_1998} of other vortices.  While still exhibiting a remarkable
complexity, it produces quantitative results which benefit the comparison
between theoretical predictions and experimental findings
\cite{Buchacek_2019_exp}. So far, strong pinning has focused on isotropic
defects, with the implicit expectation that more general potential shapes
would produce small changes.  This is not the case, as first demonstrated by
Buchacek et al.\ \cite{Buchacek_2020} in their study of correlation effects
between defects that can be mapped to the problem of a string pinned to an
anisotropic pinning potential.  In the present work, we generalize strong
pinning theory to defect potentials of arbitrary shape.  We find that this
simple generalization has pronounced (geometric) effects near the onset of
strong pinning that even change the growth of the pinning force density
$F_\mathrm{pin} \propto (\kappa - 1)^\mu$ with increasing pinning strength
$\kappa > 1$ in a qualitative manner, changing the exponent $\mu$ from $\mu =
2$ for isotropic defects \cite{Labusch_1969,Koopmann_2004} to $\mu = 5/2$ for
general anisotropic pinning potentials.

The pinning of topological defects poses a rather complex problem that has been
attacked within two paradigms, weak-collective- and strong pinning. These have
been developed in several stages: originating in the sixties of the last
century, weak pinning and creep \cite{LarkinOvch_1979} has been further
developed with the discovery of high temperature superconductors as a subfield
of vortex matter physics \cite{Blatter_1994}.  Strong pinning was originally
introduced by Labusch \cite{Labusch_1969} and by Larkin and Ovchinnikov
\cite{LarkinOvch_1979} and has been further developed recently with several
works studying critical currents \cite{Koopmann_2004}, current--voltage
characteristics \cite{Thomann_2012, Thomann_2017}, magnetic field penetration
\cite{Willa_2015_PRL, Willa_2016, Gaggioli_2022}, and creep \cite{Buchacek_2018,
Buchacek_2019, Gaggioli_2022}; results on numerical simulations involving strong pins have
been reported in Refs.\ \onlinecite{Kwok_2016, Willa_2018a, Willa_2018b}.  The
two theories come together at the onset of strong pinning: an individual
defect is qualified as weak if it is unable to pin a vortex, i.e., a vortex
traverses the pin smoothly.  Crossing a strong pin, however, the vortex
undergoes jumps that mathematically originate in bistable distinct vortex
configurations, `free' and `pinned'.  Quantitatively, the onset of strong
pinning is given by the Labusch criterion $\kappa = 1$, with the Labusch
parameter $\kappa \equiv \max[-e_p^{\prime \prime}]/\Cbar \sim f_p/\xi\Cbar$,
the dimensionless ratio of the negative curvature $e_p^{\prime\prime}$ of the
isotropic pinning potential and the effective elasticity $\Cbar$ of the
vortex lattice.  Strong pinning appears for $\kappa > 1$, i.e., when the
lattice is soft compared to the curvatures in the pinning landscape.

So far, the strong pinning transition at $\kappa = 1$ has been described for
defects with isotropic pinning potentials; it can be mapped
\cite{Koopmann_2004} to the magnetic transition in the $h$-$T$
(field--temperature) space, with the strong-pinning phenomenology at $\kappa >
1$ corresponding to the first-order Ising magnetic transition at $T < T_c$ and
the critical point at $T = T_c$ corresponding to the strong pinning transition
at $\kappa = 1$.  The role of the reduced temperature $T/T_c$ is then assumed
by the Labusch parameter $\kappa$ and the bistabilities associated with the
ferromagnetic phases at $T/T_c < 1$ translate to the bistable pinned and free
vortex states at $\kappa > 1$, with the bistability disappearing on
approaching the critical point, $T/T_c =1$ and $\kappa = 1$, respectively.

\begin{figure}[t]
        \includegraphics[width = 1.\columnwidth]{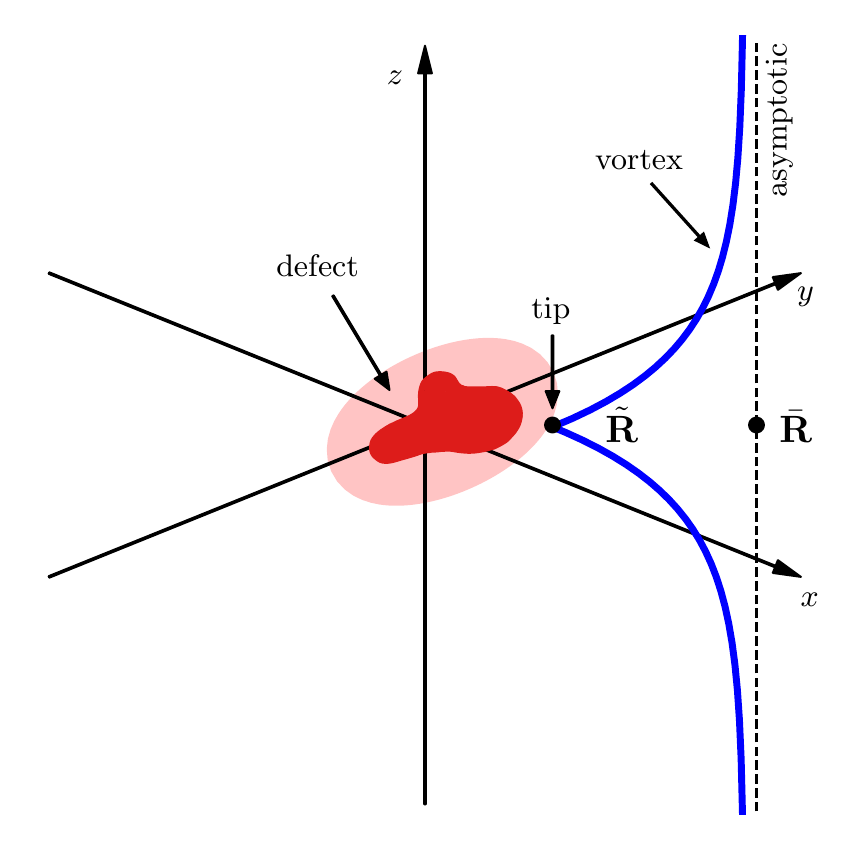}
	\caption{Sketch of a vortex interacting with a defect located at the origin.
	The vortex approaches the asymptotic position $\Ras$ at $z \to \pm
	\infty$ and is attracted to the defect residing at the origin; the cusp
	at $z=0$ defines the tip position $\Rti$ and its angle quantifies the
	pinning strength.}
    \label{fig:3D_setup}
\end{figure}

A first attempt to account for correlations between defects has been done in
Ref.\ \onlinecite{Buchacek_2020}. The latter analysis takes into account the
enhanced pinning force excerted by pairs of isotropic defects that can be cast
in the form of {\it anisotropic effective} pinning centers. Besides shifting
the onset of strong pinning to $\kappa = 1/2$ (with $\kappa$ defined for one
individual defect), the analysis unravelled quite astonishing (geometric)
features that appeared as a consequence of the symmetry reduction in the
pinning potential.  In the present paper, we take a step back and study the
transition to strong pinning for anisotropic defect potentials $e_p({\bf R})$,
with $\mathbf{R}$ a planar coordinate, see Fig.\ \ref{fig:3D_setup}. Note that
collective effects of many weak defects can add up to effectively strong pins
that smoothen the transition at $\kappa = 1$, thereby turning the strong
pinning transition into a weak-to-strong pinning crossover.

We find that the onset of strong pinning proceeds quite differently when going
from the isotropic defect to the anisotropic potential of a generic defect
without special symmetries and further on to a general random pinning
landscape.  The simplest comparison is between an isotropic and a uniaxially
anisotropic defect, acting on a vortex lattice that is directed along the
magnetic field ${\bf B} \parallel {\bf e}_z$ chosen parallel to the $z$-axis;
for convenience, we place the defect at the origin of our coordinate system
${\bf r} = ({\bf R}, z)$ and have it act only in the $z = 0$-plane.  In this
setup, see Fig.\ \ref{fig:3D_setup}, the pinning potential $e_p({\bf R})$ acts
on the {\it nearest} vortex with a force ${\bf f}_p({\bf R}) = -\nabla_{\bf R}
e_p|_{z=0}$ attracting the vortex to the defect; the presence of the {\it
other vortices} constituting the lattice renormalizes the vortex elasticity
$\Cbar$.  With the pinning potential acting in the $z = 0$ plane, the vortex
is deformed with a pronounced cusp at $z=0$, see Fig.\ \ref{fig:3D_setup}; we
denote the tip position of the vortex where the cusp appears by $\Rti$, while
the asymptotic position of the vortex at $z \to \pm \infty$ is fixed at
$\Ras$.  With this setup the problem can be reduced to a planar one, with the
tip coordinate $\Rti$ and the asymptotic coordinate $\Ras$ determining the
location and full shape (and hence the pinning force) of the vortex line.

\begin{figure}[t]
        \includegraphics[width = 1.\columnwidth]{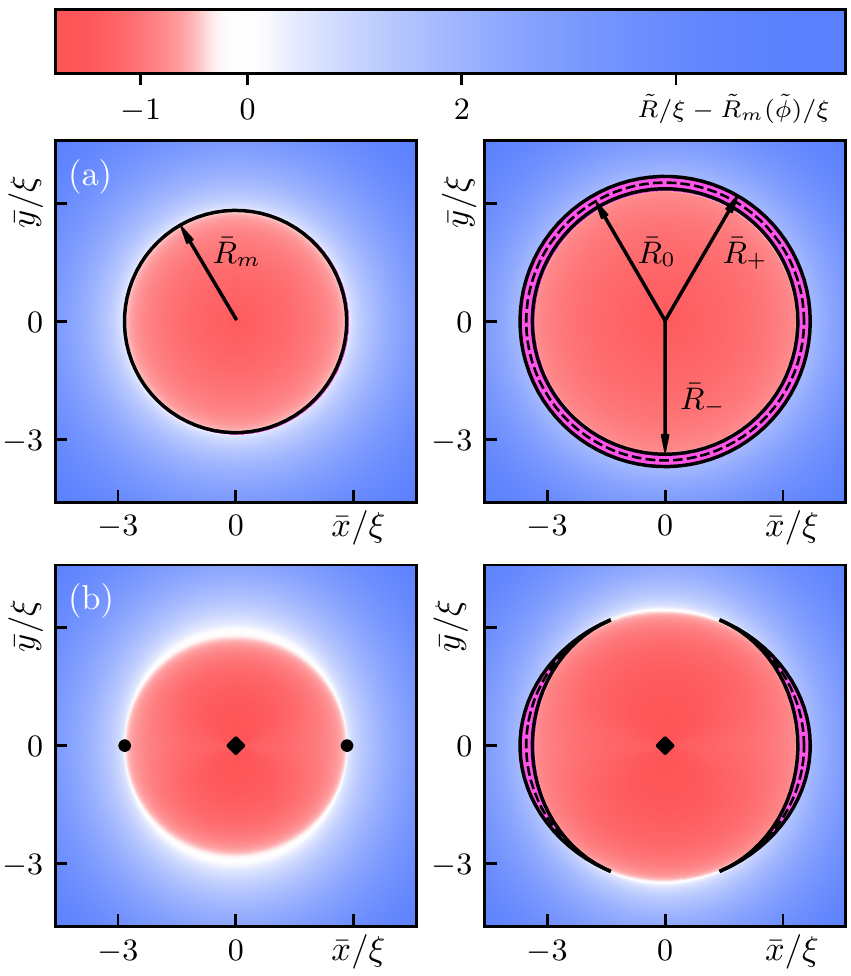}
	\caption{Illustration of bistable regions in asymptotic $\Ras$-space
	for a vortex pinned to a defect located at the origin. (a) For an
	isotropic defect (Lorentzian shape with $\kappa = 1,~1.5$), pinning
	appears at $\kappa = 1$ along a ring with radius $\ras_m$, with the
	red area corresponding to pinned states and free states colored in
	blue.  With increasing pinning strength $\kappa$, see right panel at
	$\kappa = 1.5$, a bistable region (in magenta) appears in a ring
	geometry, with vortices residing inside, $\ras < \ras_-$, being pinned
	and vortices outside, $\ras > \ras_+$, remaining free.  Vortices with
	asymptotic positions inside the ring ($\ras_- < \ras < \ras_+$)
	exhibit bistable states, pinned and free. The dashed circle $\ras_0$
	marks the crossing of pinned and free branches, see Fig.\
	\ref{fig:e_pin}.  (b) For a uniaxially anisotropic defect, see Eq.\
	\eqref{eq:uniax_potential_formal} with $\epsilon = 0.3$ and largest
	(negative) curvature along $x$, pinning appears in two points $(\pm
	\xas_m,0)$ along the $x$-axis. As the pinning strength increases
	beyond unity, see right panel, bistable regions (magenta) develop in a
	crescent-shape geometry. Pinned- and free-like states are smoothly
	connected as indicated by the crossover of colors (see Sec.\
	\ref{sec:Bas} for the precise description of coloring in terms of an
	`order parameter').  As $\kappa_m$ further increases, the cusps of the
	two crescents merge on the $y$-axis, changing the topology of the
	$\Ras$-plane through separation into inner and outer regions (not
	shown).  A ring-like bistable region appears as in $(\mathrm{a})$,
	with the inner (outer) region corresponding to unique vortex states
	that are pinned (free), while vortices residing inside the ring-shaped
	domain exhibit bistable states, pinned and free.}
    \label{fig:Ras-plane}
\end{figure}

In the case of an {\it isotropic} pin, e.g., produced by a point-like defect
\cite{Thomann_2012}, strong pinning first appears on a circle of finite radius
$R_m \sim \xi$, typically of order of the vortex core radius $\xi$, see left
panel of Fig.\ \ref{fig:Ras-plane}(a).  This is owed to the fact that, given
the radial symmetry, the Labusch criterion $\kappa =
\max_R[-e_p^{\prime\prime}(R)]/\Cbar = 1$ is satisfied on a circle $R = R_m$
where the (negative) curvature $-e_p^{\prime\prime} >0$ is maximal.
Associated with the radius $R_m$ where the tip is located at $\kappa = 1$,
$\rti(\kappa = 1) \equiv \rti_m = R_m$, there is an asymptotic vortex position
$\ras(\kappa = 1) = \ras_m > \rti_m$.  Increasing the Labusch parameter beyond
$\kappa = 1$, the circle of radius $\ras_m$ transforms into a ring $\ras_- <
\ras < \ras_+$ of finite width. Vortices placed inside the ring at small
distances $\ras < \ras_-$ near the defect are qualified as `pinned', while
vortices at large distances $\ras > \ras_+$ away from the pin are described as
`free', see right panel in Fig.\ \ref{fig:Ras-plane}(a); physically, we denote
a vortex configuration as `free' when it is smoothly connected to the
asymptotic undeformed state, while a `pinned' vortex is localized to a finite
region around the defect.  Vortices placed inside the bistable ring at $\ras_-
< \ras < \ras_+$ acquire two possible states, pinned and free (colored
magenta in Fig.\ \ref{fig:Ras-plane}, the superposition of red (pinned state)
and blue (free state) colors).

The onset of strong pinning for the {\it uniaxially anisotropic} defect
proceeds in several stages.  Let us consider an illustrative example and
assume a defect with an anisotropy aligned with the axes and a steeper
potential along $x$. In this situation, strong pinning as defined by the
criterion $\kappa_m = 1$, with a properly generalized Labusch parameter
$\kappa_m$, appears out of two points $(\pm \xas_m,0)$ where the Labusch
criterion $\kappa_m = 1$ is met first, see Fig.\ \ref{fig:Ras-plane}(b) left.
Increasing $\kappa_m > 1$ beyond unity, two bistable domains spread around
these points and develop two crescent-shaped areas (with their large extent
along $\yas$) in asymptotic $\Ras$-space, see Fig.\ \ref{fig:Ras-plane}(b)
right.  Vortices with asymptotic positions within these crescent-shaped
regions experience bistability, while outside these regions the vortex state
is unique.  Classifying the bistable solutions as `free' and `pinned' is not
possible, with the situation resembling the one around the gas--liquid
critical point with a smooth crossover (from blue to white to red) between
phases.  With $\kappa_m$ increasing further, the cusps of the crescents
approach one another.  As the arms of the two crescents touch and merge at
a sufficiently large value of $\kappa_m$, the topology of the bistable area
changes: the two merged crescents now define a ring-like geometry and separate
$\Ras$-space into an inside region where vortices are pinned, an outside
region where vortices are free and the bistable region with pinned and
free states inside the ring-like region.  As a result, the pinning geometry
of the isotropic defect is recovered, though with the perfect ring replaced by
a deformed ring with varying width.  Using the language describing a
thermodynamic first-order transition, the cusps of the crescents correspond to
critical points while its boundaries map to spinodal lines; the merging of
critical points changing the topology of the bistable regions of the pinning
landscape goes beyond the standard thermodynamic analogue of phase diagrams.

The bistable area is defining the trapping area where vortices get pinned to
the defect; this trapping area is one of the relevant quantities determining
the pinning force density $F_\mathrm{pin}$, the other being the jumps in
energy associated with the difference between the bistable states
\cite{Labusch_1969, Koopmann_2004}, see the discussion in Secs.\
\ref{sec:F_pin_gen}, \ref{sec:F_pin_iso}, and \ref{sec:F_pin_anis} below.  It is
the change in the bistable- and hence trapping geometry that modifies the
exponent $\mu$ in $F_\mathrm{pin} \propto (\kappa - 1)^\mu$, replacing the
exponent $\mu = 2$ for isotropic defects by the new exponent $\mu = 5/2$ for
general anisotropic pinning potentials.

While the existence of bistable regions $\Bas$ in the space of asymptotic
vortex positions $\Ras$ is an established element of strong pinning theory by
now, in the present paper, we introduce the new concept of unstable domains
$\Uti$ in tip-space.  The two coordinates $\Rti$ and $\Ras$ represent dual
variables in the sense of the thermodynamic analog, with the asymptotic
coordinate $\Ras$ corresponding to the driving field $h$ in the Ising model
and the tip position $\Rti$ replacing the magnetic response $m$; from a
thermodynamic perspective it is then quite natural to change view by going
back and forth between intensive ($h$) and extensive ($m$) variables.  In tip
space $\Rti$, the onset of pinning appears at isolated points $\Rti_m$ that
grow into ellipses as $\kappa$ is increased beyond unity.  These ellipses
describe {\it unstable areas} $\Uti$ in the $\Rti$-plane across which vortex
tips jump when flipping between bistable states; they relate to the {\it
bistable crescent-shaped} areas $\Bas$ in asymptotic space through the force
balance equation; the latter determines the vortex shape with elastic and
pinning forces compensating one another.  The unstable regions $\Uti$ in tip
space are actually more directly accessible than the bistable regions $\Bas$
in asymptotic space and play an equally central role in the discussion of the
strong pinning landscape.

The simplification introduced by the concept of unstable domains $\Uti$ in tip
space $\Rti$ is particularly evident when going from individual defects as
described above to a generic pinning landscape.  Here, we focus on a model
pinning potential landscape (or short pinscape) confined to the
two-dimensional (2D) $\mathbf{R}$ plane at $z=0$; such a pinscape can be
produced, e.g., by defects that reside in the $z = 0$ plane.  The pinned
vortex tip $\Rti$ then still resides in the $z=0$ plane as well and the strong
pinning problem remains two-dimensional. For a 2D random pinscape, unstable
ellipses appear sequentially out of different (isolated) points and at
different pinning strength $\kappa_m$; their assembly defines the unstable
area $\Uti$, with each newly appearing ellipse changing the topology of
$\Uti$, specifically, its number of components. Increasing $\kappa_m$, the
ellipses first grow in size, then deform away from their original elliptical
shapes, and finally touch and merge in a hyperbolic geometry.  Such mergers
change, or more precisely reduce, the number of components in $\Uti$ and hence
correspond again to topological transitions as described by a change in the
Euler characteristic $\chi$ associated with the shape of $\Uti$.  Furthermore,
these mergers tend to produce $\Uti$ shapes that are non-simply connected,
again implying a topological transition in $\Uti$ with a change in $\chi$.
Such non-simply connected parts of $\Uti$ separate the tip space into `inner'
and `outer' regions that allows to define proper `pinned' states (localized
near a potential minimum) in the `inner' of $\Uti$, while `free' states
(smoothly connected to asymptotically undeformed vortices) occupy the regions
outside of $\Uti$.

The discussion below is dominated by three mathematical tools: for one, it is
the Hessian matrix $\mathrm{H}({\bf R})$ of the pinning potential
\cite{Buchacek_2020,Willa_2022} $e_p({\bf R})$, its eigenvalues
$\lambda_\pm({\bf R})$ and eigenvectors ${\bf v}_\pm({\bf R})$, its
determinant $\det[\mathrm{H}]({\bf R})$ and trace $\tr[\mathrm{H}]({\bf R})$.
The Hessian matrix involves the curvatures $\mathrm{H}_{ij} =
\partial_i\partial_j e_p({\bf R})$, $i, j \in  \{x, y\}$, of the pinning
potential, that in turn are the quantities determining strong pinning, as can
be easily conjectured from the form of the Labusch parameter $\kappa \propto
-e_p^{\prime\prime}$ for the isotropic defect.  The second tool is the
Landau-type expansion of the total pinning energy near the strong-pinning
onset around $\Rti_m$ at $\kappa_m = 1$ (appearance of a critical point) as
well as near merging around $\Rti_s$ at $\kappa(\Rti_s) \equiv \kappa_s = 1$
(disappearance of a pair of critical points); the standard manipulations as
they are known from the description of a thermodynamic first-order phase
transition produce most of the new results. Third, the topological structure
of the unstable domain $\Uti$ associated with a generic 2D pinning landscape,
i.e., its components and their connectedness, is conveniently described
through its Euler characteristic $\chi$ with the help of Morse theory.

The structure of the paper is as follows: In Section \ref{sec:intro}, we
briefly introduce the concepts of strong pinning theory with a focus on the
isotropic defect.  The onset of strong pinning by a defect of arbitrary shape
is presented in Sec.\ \ref{sec:arb_shape}; we start with a translation and
extension of the strong pinning ideas from the isotropic situation to a
general anisotropic one, that leads us to the Hessian analysis of the pinning
potential as our basic mathematical tool. Close to onset, we find (using a
Landau-type expansion, see Sec.\ \ref{sec:ell_expansion}) that the unstable
(Sec.\ \ref{sec:Uti}) and bistable (Sec.\ \ref{sec:Bas}) domains are
associated with minima of the determinant of the Hessian curvature matrix and
assume the shape of an ellipse and a crescent, respectively. Due to the
anisotropy, the geometry of the trapping region depends non-trivially on the
Labusch parameter and the critical exponent for the pinning force is changed
from $\mu=2$ to $\mu=5/2$, see Sec.\ \ref{sec:F_pin_anis}.  The analytic
solution of the strong pinning onset for a weakly uniaxial defect presented in
Sec.\ \ref{sec:uniax_defect} leads us to define new hyperbolic points
associated with saddle points of the determinant of the Hessian curvature
matrix. These hyperbolic points describe the merging of unstable and bistable
domains, see Sec.\ \ref{sec:hyp_expansion}, and allow us to relate the new
results for the anisotropic defect to our established understanding of
isotropic defects.  In a final step, we extend the local perspective on the
pinscape, as acquired through the analysis of minima and saddles of the
determinant of the Hessian curvature matrix, to a global description in terms
of the topological characteristics of the unstable domain $\Uti$: in  Sec.\
\ref{sec:2D_landscape}, we discuss strong pinning in a two-dimensional pinning
potential of arbitrary shape, e.g., as it appears when multiple pinning defects
overlap (though all located in one plane). We follow the evolution of the
unstable domain $\Uti$ with increasing pinning strength $\kappa_m$ and express
its topological properties through the Euler characteristic $\chi$; the latter
is related to the local differential properties of the pinscape's curvature,
its minima, saddles, and maxima, through Morse theory.  Finally, in Appendix
\ref{sec:eff_1D}, we map the two-dimensional Landau-type theories (involving
two order parameters) describing onset and merging, to effective
one-dimensional Landau theories and rederive previous results following
standard statistical mechanics calculations as they are performed in the
analysis of the critical point in the van der Waals gas.

\section{Strong pinning theory}\label{sec:intro}

We start with a brief introduction to strong pinning theory, keeping a focus
on the transition region at moderate values of $\kappa > 1$. We consider an
isotropic defect (Sec.\ \ref{sec:iso_def}) and determine the unstable and
bistable ring domains for this situation in Sec.\ \ref{sec:U-B-domains}. We
derive the general expression for the pinning force density $\Fpin$ in Sec.\
\ref{sec:F_pin_gen}, determine the relevant scales of the strong pinning
characteristic near the crossover in Sec.\ \ref{sec:sp_char}, and apply the
results to derive the scaling $\Fpin \propto (\kappa - 1)^2$ for the isotropic
defect (Sec.\ \ref{sec:F_pin_iso}). In Sec.\ \ref{sec:Landau}, we relate the
strong pinning theory for the isotropic defect to the Landau mean-field
description for the Ising model in a magnetic field.

\subsection{Isotropic defect}\label{sec:iso_def}

The standard strong-pinning setup involves a vortex lattice directed along $z$
with a lattice constant $a_0$ determined by the induction $B = \phi_0/a_0^2$
that is interacting with a dilute set of randomly arranged defects of density
$n_p$. This many-body problem can be reduced \cite{Koopmann_2004, Willa_2016,
Buchacek_2019} to a much simpler effective problem involving an elastic string
with effective elasticity $\Cbar$ that is pinned by a defect potential
$e_p({\bf R})$ acting in the origin, as described by the energy function
\begin{equation}\label{eq:en_pin_tot}
   \epin(\Rti;\Ras) = \frac{\Cbar}{2}(\Rti-\Ras)^2 + e_p(\Rti)
\end{equation}
depending on the tip- and asymptotic coordinates $\Rti$ and $\Ras$ of the
vortex, see Fig.\ \ref{fig:3D_setup}.  The energy (or Hamiltonian)
$\epin(\Rti;\Ras)$ of this setup involves an elastic term and the pinning
energy $e_p({\bf R})$ evaluated at the location $\Rti$ of the vortex tip.  We
denote the depth of the pinning potential by $e_p$. A specific example is the
point-like defect that produces an isotropic pinning potential which is
determined by the form of the vortex \cite{Thomann_2012} and assumes a
Lorentzian shape $e_p(R) = -e_p/(1+ R^2/2\xi^2)$ with $R = \abs{{\bf R}}$; in
Sec.\ \ref{sec:arb_shape} below, we will consider pinning potentials of
arbitrary shape $e_p({\bf R})$ but assume a small (compared to the coherence
length $\xi$) extension along $z$.  `Integrating out' the vortex lattice, the
remaining string or vortex is described by the effective elasticity $\bar{C}
\approx \nu \varepsilon (a_0^2/\lambda_{\rm\scriptscriptstyle L})
\sqrt{c_{66}c_{44}(0)} \sim \varepsilon \varepsilon_0/a_0$.  Here,
$\varepsilon_0 = (\phi_0/4\pi \lambda_{\rm\scriptscriptstyle L})^2$ is the
vortex line energy, $\lambda_{\rm\scriptscriptstyle L}$ denotes the London
penetration depth, $\varepsilon < 1$ is the anisotropy parameter for a
uniaxial material \cite{Blatter_1994}, and $\nu$ is a numerical, see Refs.\
\onlinecite{Kwok_2016, Willa_2018b}.
\begin{figure}
        \includegraphics[width = 1.\columnwidth]{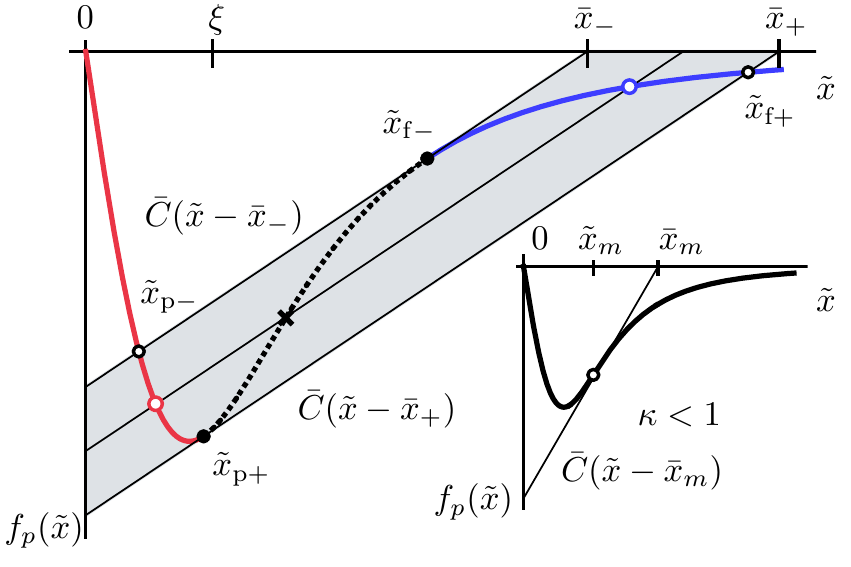}
	\caption{ Graphical illustration\cite{Buchacek_2019} of the
	self-consistent solution of the microscopic force-balance equation
	Eq.\ \eqref{eq:force_balance} for a Lorentzian potential with $\kappa
	= 2.5$.  The vortex coordinates $\xti$ and $\xas$ are expressed in
	units of $\xi$. When moving the asymptotic vortex position $\xas$
	across the bistable interval $[\xas_-,\xas_+]$, we obtain three
	solutions describing pinned $\xti_\mathrm{p} \lesssim \xi$, free
	$\xti_\mathrm{f}$ close to $\xas$, and unstable $\xti_\mathrm{us}$
	states; they define the corresponding pinned (red), free (blue), and
	unstable (black dotted) branches. The tip-positions at the edges of
	the bistable interval denoted by $\xti_\mathrm{p+}$ and
	$\xti_\mathrm{f-}$ denote jump points where the vortex tip turns
	unstable, see Eq.\ \eqref{eq:der_force_balance}; they are defined by
	the condition $f^\prime_p (\xti_\mathrm{p+}) = f^\prime_p
	(\xti_\mathrm{f-}) = \Cbar$ (black solid dots). The associated
	positions $\xti_\mathrm{f+}$ and $\xti_\mathrm{p-}$ denote the tip
	landing points after the jump (open circles); they are given by the
	second solution of Eq.\ \eqref{eq:force_balance} at the same
	asymptotic position $\xas$. The open red/blue circles and the cross
	mark the positions of metastable minima and the unstable maximum in
	Fig.\ \ref{fig:e_pin}.  The lower right inset shows the weak-pinning
	situation at $\kappa < 1$, here implemented with a larger $\Cbar$,
	where the tip solution $\xti$ is unique for all $\xas$.}
    \label{fig:self-cons-sol}
\end{figure}

The most simple pinning geometry is for a vortex that traverses the defect
through its center. Given the rotational symmetry of the isotropic defect, we
choose a vortex that impacts the defect in a head-on collision from the left
with asymptotic coordinate $\Ras = (\xas,0)$ and increase $\xas$ along the
$x$-axis; finite impact parameters $\yas \neq 0$ will be discussed later.  The
geometry then simplifies considerably and involves the asymptotic vortex
position $\xas$ and the tip position $\xti$ of the vortex, reducing the
problem to a one-dimensional one; the full geometry of the deformed string can
be determined straightforwardly \cite{Willa_2016} once the tip position $\xti$
has been found.  The latter follows from minimizing \eqref{eq:en_pin_tot} with
respect to $\xti$ at fixed asymptotic position $\xas$ and leads to the
non-linear equation
\begin{equation}\label{eq:force_balance}
   \Cbar(\xti-\xas)=-\partial_x e_p|_{x=\xti} = f_p(\xti).
\end{equation}
This can be solved graphically, see Fig.\ \ref{fig:self-cons-sol}, and
produces either a single solution or multiple solutions---the appearance of
multiple tip solutions is the signature of strong pinning. The relevant
parameter that distinguishes the two cases is found by taking the derivative
of \eqref{eq:force_balance} with respect to $\xas$ that leads to
\begin{equation}\label{eq:der_force_balance}
   \partial_{\xas} \xti = \frac{1}{1-f_p'(\xti)/\Cbar},
\end{equation}
where prime denotes the derivative, $f'_p(x) =\partial_x f_p(x) = -
\partial_x^2 e_p(x)$. Strong pinning involves vortex instabilities, i.e.,
jumps in the tip coordinate $\xti$, that appear when the denominator in
\eqref{eq:der_force_balance} vanishes; this leads us to the strong pinning
parameter $\kappa$ first introduced by Labusch \cite{Labusch_1969},
\begin{equation}\label{eq:Lab_par}
   \kappa = \max_{\xti} \frac{f'_p(\xti)}{\Cbar} = \frac{f'_p(\xti_m)}{\Cbar},
\end{equation}
with $\xti_m$ defined as the position of maximal force derivative $f_p'$, i.e.,
$f_p''(\xti_m) = 0$, or maximal negative curvature $-e_p''$  of the defect
potential.  Defining the force scale $f_p \equiv e_p/\xi$ and estimating the
force derivative or curvature $f_p^\prime  = -e_p^{\prime\prime} \sim f_p/\xi$
produces a Labusch parameter $\kappa \sim e_p/\Cbar\xi^2$; for the Lorentzian
potential, we find that $f_p^\prime(\xti_m) = e_p /4 \xi^2$ at $\xti_m =
\sqrt{2}\,\xi$ and hence $\kappa = e_p/4\Cbar\xi^2$. We see  that strong
pinning is realized for either large pinning energy $e_p$ or small effective
elasticity $\Cbar$.

As follows from Fig.\ \ref{fig:self-cons-sol} (inset), for $\kappa < 1$ (large
$\Cbar$) the solution to Eq.\ \eqref{eq:force_balance} is unique for all
values of $\xas$ and pinning is weak, while for $\kappa > 1$ (small $\Cbar$),
multiple solutions appear in the vicinity of $\xti_m$ and pinning is strong.
These multiple solutions appear in a finite interval $\xas \in
[\xas_-,\xas_+]$ and we denote them by $\xti = \xti_\mathrm{f},
\xti_\mathrm{p}, \xti_\mathrm{us}$, see Fig.\ \ref{fig:self-cons-sol}; they
are associated with free (weakly deformed vortex with $\xti_\mathrm{f}$ close
to $\xas$), pinned (strongly deformed vortex with $\xti_\mathrm{p} < \xi$),
and unstable vortex states.

\begin{figure}
        \includegraphics[width = 1.\columnwidth]{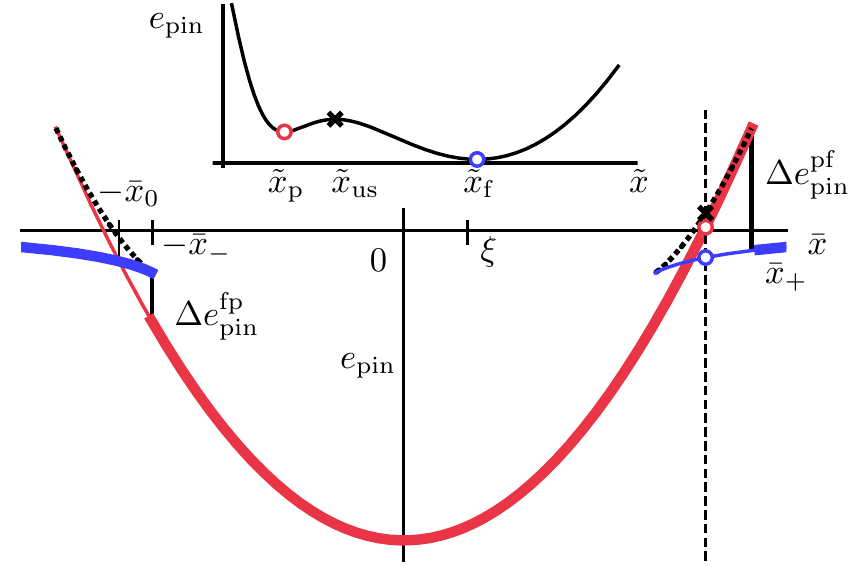}
	\caption{ Multi-valued pinning energy landscape
	$\epin^\mathrm{i}(\xas)$ for a defect producing a Lorentzian-shaped
	potential with $\kappa = 2.5$; the branches
	$\mathrm{i}=\mathrm{p,f,us}$ correspond to the pinned (red), free
	(blue), and unstable (black dotted) vortex states.  The bistability
	extends over the intervals $|\xas| \in \left[\xas_-, \xas_+\right]$
	where the different branches coexist; pinned and free vortex branches
	cut at the branch crossing point $\xas=\xas_0$. A vortex traversing
	the defect from left to right assumes the free and pinned states
	marked with thick colored lines and undergoes jumps $\Delta
	e_\mathrm{pin}^\mathrm{fp}$ and $\Delta e_\mathrm{pin}^\mathrm{pf}$ in
	energy (vertical black solid lines) at the boundaries $-\xas_-$ and
	$\xas_+$. The asymmetric occupation of states produces a finite
	pinning force density $\Fpin$.  Inset: Total energy $\epin(\xti;\xas)$
	versus vortex tip position $\xti$ for a fixed vortex position $\xas$
	(vertical dashed line in the main figure).  The points
	$\xti_\mathrm{f}$, $\xti_\mathrm{p}$, and $\xti_\mathrm{us}$ mark the
	free, pinned, and unstable solutions of the force-balance equation
	\eqref{eq:force_balance}; they correspond to local minima and the
	maximum in $\epin(\xti;\xas)$ and are marked with corresponding
	symbols in Fig.\ \ref{fig:self-cons-sol}.}
    \label{fig:e_pin}
\end{figure}

Inserting the solutions $\xti(\xas) = \xti_\mathrm{f}(\xas),
\xti_\mathrm{p}(\xas), \xti_\mathrm{us}(\xas)$ of Eq.\
\eqref{eq:force_balance} at a given vortex position $\xas$ back into the pinning
energy $\epin(\xti;\xas)$, we find the energies of the corresponding branches,
\begin{equation}\label{eq:e_pin^i}
   e^\mathrm{i}_\mathrm{pin} (\xas) \equiv e_\mathrm{pin}[\xti_\mathrm{i}(\xas);\xas],
   \quad \mathrm{i} = \mathrm{f,p,us}.
\end{equation}
The pair $e_p(\xti)$ and $e^\mathrm{i}_\mathrm{pin}(\xas)$ of energies in
tip- and asymptotic spaces then has its correspondence in the force: associated
with $f_p(\xti)$ in tip space are the force branches
$f^\mathrm{i}_\mathrm{pin}(\xas)$ in asymptotic $\xas$-space defined as
\begin{equation}\label{eq:f_pin^i}
   f^\mathrm{i}_\mathrm{pin}(\xas) = f_p[\xti_\mathrm{i}(\xas)],
   \quad \textrm{i} = \mathrm{f,p,us}.
\end{equation}
Using Eq.\ \eqref{eq:force_balance}, it turns out that the force $\fpin$
can be written as the total derivative of $\epin$,
\begin{equation}\label{eq:f_pin}
   f_\mathrm{pin}(\xas) = - \frac{d e_\mathrm{pin}[\xti(\xas);\xas]}{d\xas}.
\end{equation}
The multiple branches $e^\mathrm{i}_\mathrm{pin}$ and
$f^\mathrm{i}_\mathrm{pin}$ associated with a strong pinning situation at
$\kappa > 1$ are shown in Figs.\ \ref{fig:e_pin} and \ref{fig:f_pin}$(\mathrm{b})$.

\subsection{Unstable and bistable domains $\mathcal{U}_{\Rti}$ and
$\mathcal{B}_{\Ras}$}\label{sec:U-B-domains}

\begin{figure}
        \includegraphics[width = 1.\columnwidth]{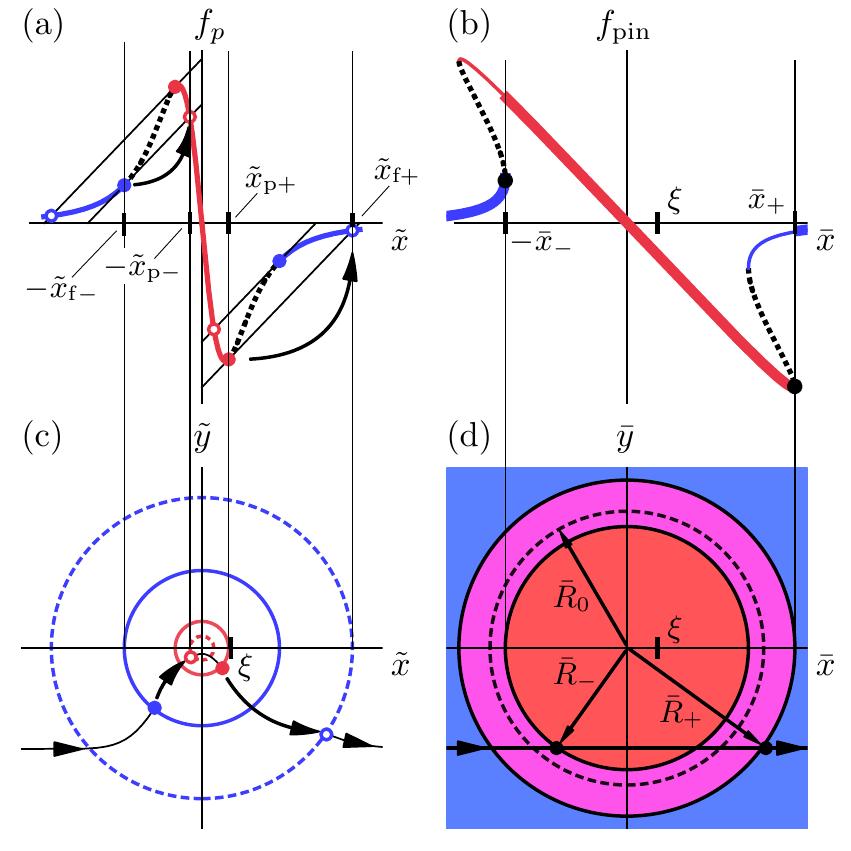}
	\caption{(a) and (b): Force profiles $f_p(\xti)$ and $f_\mathrm{pin}
	(\xas)$ in tip- and asymptotic coordinates for a Lorentzian-shaped
	potential with $\kappa = 2.5$. The tip of a vortex moving from left to
	right along the $x$-axis approaches the defect on the free branch
	(thick blue line) undergoes a jump (arrow) from $-\xti_\mathrm{f-}$ to
	$-\xti_\mathrm{p-}$, follows the pinned branch (red) until
	$\xti_\mathrm{p+}$ and then jumps back (arrow) to the free (blue)
	state at $\xti_\mathrm{f+}$. Extending these jump positions to the
	$(\xti,\yti)$-plane, see (c), defines jump (solid) and landing
	(dashed) circles, with the jump circles enclosing an unstable domain
	$\mathcal{U}_{\Rti}$ characteristic of strong pinning.  The force
	profile $f_\mathrm{pin} (\xas)$ in $(\mathrm{b})$ includes free
	(blue), pinned (red), and unstable branches (black dotted). (d)
	Extending the bistable intervals $[-\xas_+,-\xas_-]$ and
	$[\xas_-,\xas_+]$ to the $[\xas,\yas]$-plane defines a bistable ring
	$\mathcal{B}_{\Ras}$ (magenta), again a strong pinning characteristic.
	The dashed circle of radius $\ras_0$ in (d) marks the branch crossing
	point.  Vortices passing the defect with a finite impact parameter
	$\yas \neq 0$ move on a straight line in asymptotic space, see (d);
	the associated trajectory in tip space is nontrivial, see
	$(\mathrm{c})$ and undergoes jumps at pinning (circle $\rti_\mathrm{f-}$) and
	depinning (circle $\rti_\mathrm{p+}$).
}
    \label{fig:f_pin}
\end{figure}

Next, we identify the unstable (in $\xti$) and bistable (in $\xas$) domains of
the pinning landscape that appear as signatures of strong pinning when
$\kappa$ increases beyond unity. Figure \ref{fig:f_pin}(a) shows the force
profile $f_p(\xti)$ as experienced by the tip coordinate $\xti$. A vortex
passing the defect on a head-on trajectory from left to right undergoes a
forward jump in the tip from $-\xti_\mathrm{f-}$ to $-\xti_\mathrm{p-}$;
subsequently, the tip follows the pinned branch until $\xti_\mathrm{p+}$ and
then returns back to the free state with a forward jump from
$\xti_\mathrm{p+}$ to $\xti_\mathrm{f+}$. The {\it jump positions}
(later indexed by a subscript `$\mathrm{jp}$') are determined by the two
solutions of the equation
\begin{equation}\label{eq:uti_iso_j}
   f_p'(x)\Big|_{-\xti_\mathrm{f-}, \xti_\mathrm{p+}} = \Cbar
\end{equation}
that involves the curvature of the pinning potential $e_p(x)$; the {\it
landing positions} $-\xti_\mathrm{p-}$ and $\xti_\mathrm{f+}$ (later indexed
by a subscript `$\mathrm{lp}$'), on the other hand, are given by the second
solution of the force-balance equation \eqref{eq:force_balance} that involves
the driving term $\Cbar(\xti-\xas)$ and hence depends on the asymptotic
position $\xas$.  Finally, the positions in asymptotic space $\xas$ where the
vortex tip jumps are obtained again from the force balance equation
\eqref{eq:force_balance},
\begin{eqnarray}\label{eq:xas_pm}
   \xas_- &=& \xti_\mathrm{f-} - f_p(\xti_\mathrm{f-})/\Cbar,  \\ \nonumber
   \xas_+ &=& \xti_\mathrm{p+} - f_p(\xti_\mathrm{p+})/\Cbar.
\end{eqnarray}
Note that the two pairs of tip jump and landing positions,
$\xti_\mathrm{p+},~\xti_\mathrm{f+}$ and $\xti_\mathrm{f-},~\xti_\mathrm{p-}$
are associated with only two asymptotic positions $\xas_+$ and $\xas_-$.

Let us generalize the geometry and consider a vortex moving parallel to
$\xas$, impacting the defect at a finite distance $\yas$. We then have to
extend the above discussion to the entire $z=0$ plane, see Fig.\
\ref{fig:f_pin}. For an isotropic defect, the jump- and landing points now
define jump circles with radii $\rti_\mathrm{jp}$ given by $\rti_\mathrm{f-} =
\xti_\mathrm{f-}$ and $\rti_\mathrm{p+} = \xti_\mathrm{p+}$ (solid circles in
Fig.\ \ref{fig:f_pin}$(\mathrm{c})$) and landing circles with radii
$\rti_\mathrm{lp}$ given by $\rti_\mathrm{f+} = \xti_\mathrm{f+}$,
$\rti_\mathrm{p-} = \xti_\mathrm{p-}$ (dashed circles in Fig.\
\ref{fig:f_pin}$(\mathrm{c})$).  Their combination defines an unstable ring
$\rti_\mathrm{p+} < \rti < \rti_\mathrm{f-}$ in tip space where tips cannot
reside. The existence of unstable domains $\mathcal{U}_{\Rti}$ in tip space is
a signature of strong pinning.

Figures \ref{fig:f_pin}$(\mathrm{b})$ and $(\mathrm{d})$ show the
corresponding results in asymptotic coordinates $\xas$ and $\Ras$,
respectively. The pinning force $f_\mathrm{pin}(\xas) = f_p[\xti(\xas)]$ shown
in $(\mathrm{b})$ is simply an `outward tilted' version of $f_p(\xti)$, with
$S$-shaped overhangs that generate bistable intervals $[-\xas_+, -\xas_-]$ and
$[\xas_-, \xas_+]$.  Extending them to the asymptotic $\Ras$-plane with radii
$\ras_- \equiv \xas_-$ and $\ras_+ \equiv \xas_+$, see Fig.\
\ref{fig:f_pin}$(\mathrm{d})$, we obtain a ring $\ras_- < \ras < \ras_+$ that
marks the location of bistability. Again, the appearance of bistable domains
$\mathcal{B}_{\Ras}$ in asymptotic space is a signature of strong pinning.
Both, the size of the unstable- and bistable rings depend on the Labusch
parameter $\kappa$; they appear out of circles with radii $\rti = \xti_m$ and
$\ras = \xas_m = \xti_m - f_p(\xti_m)/\Cbar$ at $\kappa = 1$ and grow in
radius and width when $\kappa$ increases. The unstable and bistable domains
$\mathcal{U}_{\Rti}$ and $\mathcal{B}_{\Ras}$ (see Ref.\
\onlinecite{Buchacek_PhD}) will exhibit interesting non-trivial behavior as a
function of $\kappa$ when generalizing the analysis to defect potentials of
arbitrary shape.

\subsubsection{Alternative strong pinning formulation}\label{sec:alt_sp}

An alternative formulation of strong pinning physics is centered on the local
differential properties of the pinning energy $\epin(\xti; \xas)$,
i.e., its extremal points in $\xti$ at different values of the asymptotic
coordinate $\xas$.  We start from equation \eqref{eq:en_pin_tot} restricted to
one dimension and rearrange terms to arrive at the expression
\begin{equation}\label{eq:e_pin_eff}
   \epin(\xti;\xas) = e_\mathrm{eff}(\xti) - \Cbar \xas\>\xti +\Cbar \xas^2/2
\end{equation}
with the effective pinning energy
\begin{equation}\label{eq:e_eff}
   e_\mathrm{eff}(\xti)  = e_p(\xti) + \Cbar \xti^2/2
\end{equation}
involving both pinning and elastic terms. Equation \eqref{eq:e_pin_eff}
describes a particle at position $\xti$ subject to the potential
$e_\mathrm{eff}(\xti)$ and the force term $f \> \xti = -\Cbar\xas \> \xti$,
see also Ref.\ \onlinecite{Willa_2022}.  The potential $e_\mathrm{eff}(\xti)$
can trap two particle states if there is a protecting maximum with negative
curvature $\partial_{\xti}^2 e_\mathrm{eff} = \partial_{\xti}^2 \epin < 0$,
preventing its escape from the metastable state at forces $f = \pm \Cbar \xas$
with $\xas \in [\xas_+,\xas_-]$; the maximum in $\epin$ at $\xti_\mathrm{us}$
then separates two minima in $\epin$ defining distinct branches with different
tip coordinates $\xti_\mathrm{p}$ and $\xti_\mathrm{f}$, see the inset of Fig.\
\ref{fig:e_pin}.

As the asymptotic position $\xas$ approaches the boundaries $\xas_\pm$, one of
the minima joins up with the maximum to define an inflection point with
\begin{equation}\label{eq:e_eff_jp}
   [\partial_{\xti}^2 e_\mathrm{eff}]_{\xjp}  = [\partial_{\xti}^2 \epin]_{\xjp} = 0,
\end{equation}
that corresponds to the instability condition \eqref{eq:uti_iso_j} where the
vortex tip jumps; the persistent second minimum in $\epin(\xti;\xas)$ defines
the landing position $\xlp$ and the condition for a flat inflection point
$[\partial_{\xti} \epin]_{\xjp} = 0$ defines the associated asymptotic
coordinate $\pm \xas_\pm$.

Finally, strong pinning vanishes at the Labusch point $\kappa = 1$, with the
inflection point in $e_\mathrm{eff}(\xti)$ coalescing with the second minimum
at $\xti_m$, hence
\begin{eqnarray}\label{eq:e_eff_m}
   [\partial_{\xti}^2 e_\mathrm{eff}]_{\xti_m} &=& 0 \quad \textrm{and}\\
   \nonumber
   [\partial_{\xti}^3 e_\mathrm{eff}]_{\xti_m} &=& [\partial_{\xti}^3 e_p]_{\xti_m} = 0.
\end{eqnarray}
Note the subtle use of $\epin$ versus $e_\mathrm{eff}$ versus $e_p$ in the
above discussion; as we go to higher derivatives, first the asymptotic
coordinate $\xas$ turns irrelevant in the second derivative $\partial_{\xti}^2
\epin = \partial_{\xti}^2 e_\mathrm{eff}$ and then all of the elastic
response, i.e., $\Cbar$, drops out in the third derivative $[\partial_{\xti}^3
\epin] = [\partial_{\xti}^3 e_p]$.

The above alternative formulation of strong pinning turns out helpful in
several discussions below, e.g., the derivation of strong pinning
characteristics near the transition in Secs.\ \ref{sec:sp_char} and
\ref{sec:ell_expansion} and in the generalization of the instability condition
to an anisotropic defect in Sec.\ \ref{sec:arb_shape} and furthermore provides
an inspiring link to the Landau theory of phase transitions discussed below in
Sec.\ \ref{sec:Landau}.

\subsection{Pinning force density $\Fpin$}\label{sec:F_pin_gen}

Next, we determine the pinning force density $\Fpin$ at strong pinning,
assuming a random homogeneous distribution of pins with a small density $n_p$,
$n_p a_0\xi^2 \ll 1$, see Refs.\ \onlinecite{Willa_2016, Buchacek_2019}. The
derivation of $\Fpin$ is conveniently done in asymptotic $\Ras$ coordinates
where vortex trajectories follow simple straight lines. Vortices approach
the pin by following the free branch until its termination, jump to the pinned
branch to again follow this to its termination, and finally jump back to the
free branch. This produces an asymmetric pinned-branch occupation
$p_{c}(\Ras)$ that leads to the pinning force density (we assume vortices
approaching the defect along $\xas$ from the left; following convention, we
include a minus sign)
\begin{align}\label{eq:F_pin_vec}
  \mathbf{F}_c &= - n_p \int \frac{d^2\Ras}{a_0^2}\bigl[
   p_{c}(\Ras)\mathbf{f}^\mathrm{p}_\mathrm{pin}(\Ras) + (1-p_{c}(\Ras))
   \mathbf{f}^\mathrm{f}_\mathrm{pin}(\Ras)\bigr]\nonumber\\
   &= - n_p \int \frac{d^2\Ras}{a_0^2} p_{c}(\Ras)
   [\partial_x\Delta e^\mathrm{fp}_\mathrm{pin}(\Ras)]\,\mathbf{e}_{\xas},
\end{align}
with the energy difference $\Delta e^\mathrm{fp}_\mathrm{pin}(\Ras) =
e^\mathrm{f}_\mathrm{pin}(\Ras) - e^\mathrm{p}_\mathrm{pin}(\Ras)$ and
$\mathbf{e}_{\xas}$ the unit vector along $\xas$; the $\yas$-component of the
pinning force density vanishes due to the antisymmetry in
$f_{\mathrm{pin},\yas}$. For the isotropic defect, the jumps $\Delta
e^\mathrm{fp}_\mathrm{pin} (\Ras)$ in energy appearing upon changing branches
are independent of angle and the average in \eqref{eq:F_pin_vec} separates in
$\xas$ and $\yas$ coordinates; note that the energy jumps are no longer
constant for an anisotropic defect and hence such a separation does not occur.
Furthermore, i) all vortices approaching the defect within the transverse
length $|\yas| < \ras_-$ get pinned, see Fig.\ \ref{fig:f_pin}(d), while those
passing further away follow a smooth (weak pinning) trajectory that does not
undergo jumps and hence do not contribute to the pinning force, and ii) all
vortices that get pinned contribute the same force that is most easily
evaluated for a head-on vortex--defect collision on the $\xas$-axis with
$p_c(\xas) = \Theta(\xas + \xas_-) - \Theta(\xas - \xas_+)$ and
\begin{align}\label{eq:f_pin_av_def}
  \langle f_\mathrm{pin} \rangle &= - \!\! \int_{-a_0/2}^{a_0/2} \frac{d\xas}{a_0} \>
   \bigl[ p_{c}(\xas) f^\mathrm{p}_\mathrm{pin}(\xas) + (1-p_{c}(\xas))
   f^\mathrm{f}_\mathrm{pin}(\xas)\bigr]\nonumber\\
   &= \frac{\Delta e^\mathrm{fp}_\mathrm{pin}(-\xas_-) +
  \Delta e^\mathrm{pf}_\mathrm{pin}(\xas_+)}{a_0},
\end{align}
where we have replaced $-\Delta e^\mathrm{fp}_\mathrm{pin}(\xas_+)$ by $\Delta
e^\mathrm{pf}_\mathrm{pin}(\xas_+) > 0$.  Hence, the average pinning force
$\langle f_\mathrm{pin} \rangle$ is given by the jumps in the pinning energy
$e_\mathrm{pin}^\mathrm{i}(\xas)$ associated with different branches
$\mathrm{i} = \mathrm{p,f}$, see Fig.\ \ref{fig:e_pin}.

Finally, accounting for trajectories with finite impact parameter 
$|\yas| < \ras_-$, we arrive at the result for the pinning force density 
$\Fpin$ acting on the vortex system,
\begin{equation}\label{eq:F_pin}
  F_\mathrm{pin} = n_p \frac{2\ras_-}{a_0} \langle f_\mathrm{pin} \rangle
  = n_p \frac{2 \ras_-}{a_0} \frac{\Delta e^\mathrm{fp}_\mathrm{pin} +
  \Delta e^\mathrm{pf}_\mathrm{pin}}{a_0},
\end{equation}
where the factor $2 \Ras_\mathrm{-}/{a_0}$ accounts for the averaging of the
pinning force along the $y$-axis.  As strong pins act independently, a
consequence of the small defect density $n_p$, the pinning force density is
linear in the defect density, $F_\mathrm{pin} \propto n_p$.  If pinning is
weak, i.e., $\kappa < 1$, we have no jumps, $\langle f_\mathrm{pin} \rangle =
0$, and $F_\mathrm{pin}|_\mathrm{strong} = 0$.  A finite pinning force then
only arises from correlations between pinning defects and scales in density as
\cite{LarkinOvch_1979,Koopmann_2004} $F_\mathrm{pin}|_\mathrm{weak} \propto n_p^{2}$. This
contribution to the pinning force density $\Fpin$ continues beyond $\kappa =
1$, hence, while the strong pinning onset at $\kappa = 1$ can be formulated in
terms of a transition, weak pinning goes to strong pinning in a smooth
crossover.

Knowing the pinning force density $F_\mathrm{pin}$, the motion of the vortex
lattice follows from the bulk dynamical equation
\begin{equation}\label{eq:macroscopic_force_balance}
  \eta \mathbf{v} = \mathbf{F}_{\rm \scriptscriptstyle
  L}(\mathbf{j})-\mathbf{F}_\mathrm{pin}.
\end{equation}
Here, $\eta = B H_{c2}/\rho_n c^2$ is the Bardeen-Stephen viscosity
\cite{Bardeen_1965} (per unit volume; $\rho_n$ is the normal state
resistivity) and $\mathbf{F}_{\rm \scriptscriptstyle L} = \mathbf{j} \times
\mathbf{B}/c$ is the Lorentz force density driving the vortex system.  The
pinning force density $\mathbf{F}_\mathrm{pin}$ is directed along
$\mathbf{v}$, in our case along $x$.

Next, we determine the strong pinning characteristics $\xas_-$, $\xas_+$,
$\xti_{\mathrm{f}\pm}$, $\xti_{\mathrm{p}\pm}$, $\Delta
e^\mathrm{fp}_\mathrm{pin}$ and $\Delta e^\mathrm{pf}_\mathrm{pin}$ as a
function of the Labusch parameter $\kappa$ close to the strong pinning
transition, i.e., $\kappa \gtrsim 1$.

\subsection{Strong pinning characteristics near the transition} \label{sec:sp_char}

Near the strong pinning transition at $\kappa \gtrsim 1$, we can derive
quantitative results for the strong pinning characteristics by expanding the
pinning energy $\epin(\xti;\xas)$ in $\xti$ at fixed $\xas$; this reminds
about the Landau expansion of the free energy $f(\phi,h)$ in the order
parameter $\phi$ at a fixed field $h$ in a thermodynamic transition, see Sec.\
\ref{sec:Landau} below for a detailed discussion. 

We expand $\epin(\xti;\xas)$ in $\xti$ around the point of first instability
$\xti_m$ by introducing the relative tip and asymptotic positions $\uti = \xti
- \xti_m$ and $\uas = \xas - \xas_m$ and make use of our alternative strong
pinning formulation summarized in Sec.\ \ref{sec:alt_sp}.  At $\xti_m$ and
close to $\kappa = 1$, we have $[\partial_{\xti}^2 \epin]_{\xti_m} =
[\partial_{\xti}^2 e_p]_{\xti_m} + \Cbar = \Cbar (1-\kappa)$ and
$[\partial_{\xti}^3 \epin]_{\xti_m} = 0$, hence,
\begin{eqnarray}\label{eq:e_pin_expans}
   e_\mathrm{pin}(\xti;\xas) \approx \frac{\Cbar}{2}
   (1-\kappa)\> \uti^{2} + \frac{\gamma}{24} \>
   \uti^{4} - \Cbar \uas \uti,
\end{eqnarray}
where we have introduced the shape parameter $\gamma = [\partial^4_{x}
e_p]_{\xti_m}$ describing the quartic term in the expansion and we have made
use of the force balance equation \eqref{eq:force_balance} to rewrite
$f_p(\xti_m) = \Cbar (\xti_m - \xas_m)$; furthermore, we have dropped all
irrelevant terms that do not depend on $\uti$.

We find the jump and landing positions $\xjp$ and $\xlp$ exploiting the
differential properties of $\epin(\xti)$ at a fixed $\xas$: As discussed
above, the vortex tip jumps at the boundaries $\xas_\pm$ of the bistable
regime, where $\epin$ develops a flat inflection point at $\xjp$ with one
minimum joining up with the unstable maximum and the second minimum at the
landing position $\xlp$ staying isolated. Within our fourth-order expansion
the jump positions at (de)pinning are placed symmetrically with respect to the
onset at $\xti_m$,
\begin{equation}\label{eq:jp_pos}
   \xti_\mathrm{p+} = \xti_m + \uti_\mathrm{jp}, ~~~
   \xti_\mathrm{f-} = \xti_m - \uti_\mathrm{jp}
\end{equation}
and imposing the condition $[\partial_{\uti}^2\epin]_{\xjp} = 0$ (that is
equivalent to the jump condition $f_p'[\xti_\mathrm{f-}] =
f_p'[\xti_\mathrm{p+}] = \Cbar$ of Eq.\ \eqref{eq:uti_iso_j}, see also Fig.\
\ref{fig:self-cons-sol}), we find that
\begin{equation}\label{eq:ujp}
   \uti_\mathrm{jp} \approx 
   - \sqrt{\frac{2\Cbar}{\gamma}} (\kappa-1)^{1/2}.
\end{equation}

In order to find the (symmetric) landing positions, it is convenient to shift
the origin of the expansion to the jump position, $\uti \to \uti - \ujp \equiv
\uti'$, and define the jump distance $\Delta \uti$,
\begin{eqnarray}\label{eq:lp_pos}
   \xti_\mathrm{f+} = \xti_\mathrm{p+} + \Delta\uti, ~~~ \xti_\mathrm{p-} =
   \xti_\mathrm{f-} - \Delta\uti.
\end{eqnarray}
At the jump position, the linear and quadratic terms in $\uti'$ vanish,
resulting in the expansion (up to an irrelevant constant)
\begin{equation}\label{eq:e_pin_expans_jp}
   e_\mathrm{pin}(\xti_\mathrm{p+} + \uti';\xas_+) \approx 
   \frac{\gamma}{6} \ujp \uti^{\prime\, 3}
   + \frac{\gamma}{24} \uti^{\prime\, 4}
\end{equation}
and similar at $\xti_\mathrm{f-}$ and $\xas_-$ for a left moving vortex. This
expression is minimal at the landing position $\xti_\mathrm{f+}$, i.e., at
$\uti' = \Delta \uti$, $[\partial_{\uti'} e_\mathrm{pin}]_{\Delta \uti} = 0$,
and we find the jump distance
\begin{equation}\label{eq:j_dist}
   \Delta\uti = - 3 \uti_\mathrm{jp}. 
\end{equation}
Inserting this result back into \eqref{eq:e_pin_expans_jp}, we obtain the jump
in energy $\Delta \epin^\mathrm{pf} = e_\mathrm{pin}(\xti_\mathrm{p+};\xas_+) -
e_\mathrm{pin}(\xti_\mathrm{f+};\xas_+)$,
\begin{equation}\label{eq:d_epin^pf}
   \Delta \epin^\mathrm{pf} (\xas_+) \approx \frac{\gamma}{72}(\Delta\uti)^4
   \approx \frac{9\Cbar^2}{2\gamma}(\kappa - 1)^2,
\end{equation}
and similar at $\xas_-$.  Note that all these results have been obtained
without explicit knowledge of the asymptotic coordinates $\xas_\pm$ where
these tip jumps are triggered. The latter follow from the force equation
\eqref{eq:force_balance} that corresponds to the condition
$[\partial_{\xti}\epin]_{\xjp} = 0$ for a flat inflection point. Using the
expansion \eqref{eq:e_pin_expans} of the pinning energy, we find that
\begin{equation}\label{eq:bs_pos}
   \xas_{\pm} - \xas_m = \mp \frac{2}{3} \uti_\mathrm{jp}(\kappa - 1)
   = \pm \frac{2}{3} \sqrt{\frac{2\Cbar}{\gamma}} (\kappa - 1)^{3/2}.
\end{equation}

The pair $\xas_m$ and $\xti_m$ of asymptotic and tip positions depends on the
details of the potential; while $\xti_m$ derives solely from the shape
$e_p(\xti)$, $\xas_m$ as given by \eqref{eq:force_balance} involves $\Cbar$
and shifts $\propto (\kappa - 1)$.  For a Lorentzian potential, we find that
\begin{equation}\label{eq:xti_xas_m}
   \xti_m = \sqrt{2}\xi, \quad \xas_m = 2\sqrt{2} \xi + \sqrt{2}\xi (\kappa - 1).
\end{equation}
The shape coefficient is $\gamma = 3e_p/4\xi^4$ and the Labusch parameter is
given by $\kappa = e_p/4\Cbar\xi^2$ (hence $\Cbar^2/\gamma = e_p/12
\kappa^2$), providing us with the results
\begin{equation}\label{eq:ujp_depin}
   \uti_\mathrm{jp} \approx -\xi \, [2(\kappa-1)/3]^{1/2} \mathrm{~~and~~}
   \Delta \epin^\mathrm{pf} \approx \frac{3}{8} e_p (\kappa - 1)^2.
\end{equation}

\subsection{Pinning force density for the isotropic defect}\label{sec:F_pin_iso}

Using the results of Sec.\ \ref{sec:sp_char} in the expression
\eqref{eq:F_pin} for the pinning force density, we find, to leading order in
$\kappa -1$, 
\begin{equation}\label{eq:F_pin_iso_result}
   F_\mathrm{pin} = 9 n_p \frac{\xas_m}{a_0} \frac{\Cbar^2}{\gamma a_0}
  (\kappa - 1)^2.
\end{equation}
The scaling $F_\mathrm{pin} \sim n_p (\xi/a_0)^2 f_p (\kappa - 1)^2$ (with
$\Cbar \xi^2/e_p \sim 1/\kappa$, up to a numerical) uniquely derives from the
scaling $\propto (\kappa - 1)^2$ of the energy jumps in \eqref{eq:d_epin^pf},
as the asymptotic trapping length $\xas_- \sim \xi$ remains finite as $\kappa
\to 1$ for the isotropic defect; this will change for the anisotropic defect.

\subsection{Relation to Landau's theory of phase transitions}\label{sec:Landau}

The expansion \eqref{eq:e_pin_expans} of the pinning energy $e_\mathrm{pin}
(\xti; \xas)$ around the inflection point $\xti_m$ of the force takes the same
form as the Landau free energy of a phase transition\cite{Koopmann_2004},
\begin{eqnarray}\label{eq:f_phi}
   f(\phi;h) &=& \frac{r_0}{2}(T/T_c-1)\phi^2 +u\phi^4 - h\phi,
\end{eqnarray}
with the straightforward transcription $\uti \leftrightarrow \phi$,
$\Cbar (1-\kappa) \leftrightarrow r_0 (T/T_c - 1)$, $\gamma /24
\leftrightarrow u$ and the conjugate field $\Cbar \uas
\leftrightarrow h$. The functional \eqref{eq:f_phi} describes a one-component
oder parameter $\phi$ driven by $h$, e.g., an Ising model with magnetization
density $\phi$ in an external magnetic field $h$. This model develops a
mean-field transition with a first-order line in the $h$--$T$ phase diagram
that terminates in a critical point at $T=T_c$ and $h=0$. The translation to strong
pinning describes a strong pinning region at large $\kappa$ that terminates
(upon decreasing $\kappa$) at $\kappa = 1$. The ferromagnetic phases with
$\phi = \pm \sqrt{r_0(1-T/T_c)/4u}$ correspond to pinned and unpinned states,
the paramagnetic phase at $T > T_c$ with $\phi = 0$ translates to the unpinned
domain at $\kappa < 1$. The spinodals associated with the hysteresis in the
first-order magnetic transition correspond to the termination of the free and
pinned branches at $\xas_\pm$; indeed, the flat inflection points appearing in
$\epin(\xti; \xas)$ at the boundaries of the bistable region $\Bas$ as discussed in
Sec.\ \ref{sec:U-B-domains} correspond to the disappearance of metastable
magnetic phases in \eqref{eq:f_phi} at the spinodals of the first-order
transition where $\partial_\phi f (\phi; h) = \partial_\phi^2 f (\phi; h) =
0$. When including correlations between defects, the unpinned phase at $\kappa
< 1$ transforms into a weakly pinned phase that continues beyond $\kappa = 1$
into the strongly pinned phase.  Including such correlations, the
strong-pinning transition at the onset of strong pinning at $\kappa = 1$
transforms into a weak-to-strong pinning crossover.

\section{Anisotropic defects}\label{sec:arb_shape}

Let us generalize the above analysis to make it fit for the ensuing discussion
of an arbitrary pinning landscape or short, pinscape.  Central to the
discussion are the unstable and bistable domains $\mathcal{U}_{\Rti}$ and
$\mathcal{B}_{\Ras}$ in tip- and asymptotic space.  The boundary of the
unstable domain $\mathcal{U}_{\Rti}$ in tip space is determined by the jump
positions of the vortex tip. The latter follows from the local differential
properties of $\epin(\Rti;\Ras)$ at fixed asymptotic coordinate $\Ras$, for
the isotropic defect, the appearence of an inflection point
$[\partial_{\xti}^2 \epin(\xti,\xas)] = 0$, see Eq.\ \eqref{eq:e_eff_jp}. In
generalizing this condition  to the anisotropic situation, we have to study
the Hessian matrix of $\epin(\Rti;\Ras)$ defined in Eq.\
\eqref{eq:en_pin_tot},
\begin{equation}\label{eq:Hessian}
   \bigl[\mathrm{Hess}\bigl[\epin(\Rti;\Ras)|_{\Ras}\bigr]\bigr]_{ij}
   =  \Cbar \delta_{ij} + \mathrm{H}_{ij}(\Rti) 
\end{equation}
with
\begin{equation}\label{eq:Hessian_e_p}
   \mathrm{H}_{ij}(\Rti) = 
   \partial_{\xti_i} \partial_{\xti_j} e_p(\Rti;\Ras)
\end{equation}
the Hessian matrix associated with the defect potential $e_p(\Rti)$. The
vortex tip jumps when the pinning landscape $\epin(\Rti;\Ras)$ at fixed $\Ras$
opens up in an unstable direction, i.e., develops an inflection point; this
happens when the lower eigenvalue $\lambda_-(\Rti) < 0$ of the Hessian matrix
$\mathrm{H}_{ij}(\Rti)$ matches up with $\Cbar$,
\begin{align}\label{eq:match_LC}
   \lambda_-(\Rti) + \Cbar = 0,
\end{align}
and strong pinning appears in the location where this happens first, say in
the point $\Rti_m$, implying that the eigenvalue $\lambda_-(\Rti)$ has a
minimum at $\Rti_m$. Furthermore, the eigenvector $\mathbf{v}_-(\Rti_m)$
associated with the eigenvalue $\lambda_-(\Rti_m)$ provides the unstable
direction in the pinscape $\epin(\Rti;\Ras)$ along which the vortex tip
escapes.

Defining the reduced curvature function
\begin{align}\label{eq:red_curv_kappa}
   \kappa(\Rti) \equiv \frac{-\lambda_-(\Rti)}{\Cbar},
\end{align}
we find the generalized Labusch parameter
\begin{align}\label{eq:gen_Lab}
   \kappa_m \equiv \kappa(\Rti_m),
\end{align}
and the Labusch criterion takes the form
\begin{align}\label{eq:gen_Lab_crit}
   \kappa_m  = 1.
\end{align}
The latter has to be read as a double condition: i) find the location $\Rti_m$
where the smaller eigenvalue $\lambda_-(\Rti)$ is negative and largest, from
which ii), one obtains the critical elasticity $\Cbar$ where strong pinning
sets in.

A useful variant of the strong pinning condition \eqref{eq:match_LC} is
provided by the representation of the determinant of the Hessian matrix,
\begin{align}\label{eq:det_Hessian}
   D(\Rti) &\equiv  \det\bigl\{\mathrm{Hess}\bigl[\epin(\Rti;\Ras)|_{\Ras}\bigr]\bigr\},
\end{align}
in terms of its eigenvalues $\lambda_\pm(\Rti)$, $D(\Rti) = [\Cbar +
\lambda_-(\Rti)] [\Cbar + \lambda_+(\Rti)]$; near onset, the second factor
$\Cbar + \lambda_+(\Rti)$ stays positive and the strong pinning onset appears
in the point $\Rti_m$ where $D(\Rti)$ has a minimum which touches zero for the
first time, i.e., the two conditions $\nabla D(\Rti)|_{\Rti_m} = 0$ and
$D(\Rti_m) = 0$ are satisfied simultaneously. The latter conditions make sure
that the minima of $\lambda_-(\Rti)$ and $D(\Rti)$ line up at $\Rti_m$.  Note
that the Hessian determinant $D(\Rti)$ does not depend on the asymptotic
coordinate $\Ras$ as it involves only second derivatives of
$\epin(\Rti;\Ras)$.

The Labusch criterion defines the situation where jumps of vortex tips appear
for the first time in the isolated point $\Rti_m$.  Increasing the pinning
strength, e.g., by decreasing the elasticity $\Cbar$ for a fixed pinning
potential $e_p({\bf R})$ (alternatively, the pinning scale $e_p$ could be
increased at fixed $\Cbar$) the condition \eqref{eq:match_LC} is satisfied on
the boundary of a finite domain and we can define the unstable domain
$\mathcal{U}_{\Rti}$ through (see also Ref.\ \onlinecite{Buchacek_PhD})
\begin{align}\label{eq:def_calU}
   \mathcal{U}_{\Rti} = \left\{ \Rti~~|~~\lambda_-(\Rti) + \Cbar \leq 0 \right\}.
\end{align}
Once the latter has been determined, the bistable domain $\mathcal{B}_{\Ras}$
follows straightforwardly from the force balance equation
\begin{align}\label{eq:gen_force_balance}
   \Cbar (\Rti - \Ras) =  {\bf f}_p(\Rti) = {\bf f}_\mathrm{pin}(\Ras),
\end{align}
i.e.,\cite{Buchacek_PhD}
\begin{align}\label{eq:def_calB}
   \mathcal{B}_{\Ras} = \left\{ \Ras = \Rti -{\bf f}_p(\Rti)/\Cbar~~|~~\Rti \in
   \mathcal{U}_{\Rti}\right \}.
\end{align}
In a last step, one then evaluates the energy jumps appearing at the boundary
of $\mathcal{B}_{\Ras}$ and proper averaging produces the pinning force density
$\mathbf{F}_\mathrm{pin}$.

Let us apply the above generalized formulation to the isotropic situation.
Choosing cylindrical coordinates $(r,\varphi)$, the Hessian matrix
$\mathrm{H}_{ij}$ is already diagonal; close to the inflection point $\rti_m$,
where $e_p'''(\rti_m) = 0$, the eigenvalues are $\lambda_-(\rti)  =
e_p''(\rti) < 0$ and $\lambda_+(\rti) = e_p'(\rti)/\rti > 0$, producing
results in line with our discussion above.

\subsection{Expansion near strong pinning onset}\label{sec:ell_expansion}
With our focus on the strong pinning transition near $\kappa(\Rti_m) = 1$, we
can obtain quantitative results using the expansion of the pinning energy
$\epin(\Rti;\Ras)$, Eq.\ \eqref{eq:en_pin_tot}, close to $\Rti_m$, cf.\ Sec.\
\ref{sec:sp_char}. Hence, we construct the Landau-type pinning energy
corresponding to \eqref{eq:f_phi} for the case of an anisotropic pinning
potential, i.e., we generalize \eqref{eq:e_pin_expans} to the two-dimensional
situation.

When generalizing the strong pinning problem to the anisotropic situation, we
are free to define local coordinate systems $(\uti,\vti)$ and $(\uas, \vas)$
in tip- and asymptotic space centered at $\Rti_m$ and $\Ras_m$, where the
latter is associated with $\Rti_m$ through the force balance equation
\eqref{eq:gen_force_balance} in the original laboratory system.  Furthermore,
we fix our axes such that the unstable direction coincides with the $u$-axis,
i.e., the eigenvector ${\bf v}_-(\Rti_m)$ associated with $\lambda_-(\Rti_m)$
points along $u$; as a result, the mixed term $\propto \uti \vti$ is absent
from the expansion.  Keeping all potentially relevant terms up to fourth order
in $\uti$ and $\vti$ in the expansion, we then have to deal with an expression
of the form 
\begin{align}\nonumber
  &e_\mathrm{pin}(\Rti; \Ras) = 
  \frac{\Cbar+\lambda_-}{2} \, \uti^2 
  + \frac{\Cbar + \lambda_+}{2}\, \vti^2 -\Cbar\,\uas \uti - \Cbar\, \vas \vti  \nonumber \\
  &\quad+\frac{a}{2}\, \uti \vti^2 + \frac{a'}{2}\, \uti^2 \vti + \frac{b'}{6}\, \uti^3 
  + \frac{b''}{6}\, \vti^3 \label{eq:e_pin_expans_ani_orig} \\ \nonumber
  &\qquad+ \frac{\alpha}{4}\, \uti^2\vti^2 + \frac{\beta}{6}\, \uti^3\vti 
  +\frac{\beta''}{6}\, \uti\vti^3 +\frac{\gamma}{24}\, \uti^4  + \frac{\gamma''}{24}\, \vti^4,
\end{align}
with $\lambda_\pm = \lambda_\pm(\Rti_m)$, 
\begin{align}\label{eq:coord_uv}
   \Rti &= \Rti_m + \delta\Rti, \quad \delta\Rti = (\uti,\vti), \\ \nonumber
   \Ras &= \Ras_m + \delta\Ras, \quad \delta\Ras = (\uas,\vas), 
\end{align}
and coefficients given by the corresponding derivatives of $e_p(\bf R)$, e.g.,
$a \equiv \partial_u\partial_v^2 e_p({\bf R})|_{\Rti_m}$, $\dots$, $\gamma''
\equiv \partial_v^4 e_p({\bf R})|_{\Rti_m}$.  As we are going to see, the
primed terms in this expansion vanish due to the condition of a minimal
Hessian determinant at the onset of strong pinning, while double-primed terms
will turn out irrelevant to leading order in the small distortions $\uti$ and
$\vti$.

The first term in \eqref{eq:e_pin_expans_ani_orig} drives the strong pinning
transition as it changes sign when $\lambda_- = -\Cbar$.  Making use of the
Labusch parameter $\kappa_m$ defined in \eqref{eq:gen_Lab}, we can replace
(see also \eqref{eq:e_pin_expans})
\begin{equation} \label{eq:one-kappa}
    \Cbar +\lambda_- \to \Cbar(1-\kappa_m).
\end{equation}
In our further considerations below, the quantity $\kappa_m - 1
\ll 1$ acts as the small parameter; it assumes the role of the distance
$1-T/T_c$ to the critical point in the Landau expansion of a thermodynamic
phase transition.

The second term in \eqref{eq:e_pin_expans_ani_orig} stabilizes the theory
along the $v$ direction as $\Cbar + \lambda_+ > 0$ close to the Labusch point,
while the sign of the cubic term $a\,\uti\vti^2/2$ determines the direction of
the instability along $x$, i.e., to the right ($a > 0$) or left ($a < 0$).  The
quartic terms $\propto \alpha, \gamma >0$ bound the pinning energy at large
distances, while the term $\propto \beta$ determines the skew angle in the shape
of the unstable domain $\Uti$, see below.  Finally, we have used the force
balance equation \eqref{eq:gen_force_balance} in the derivation of the driving
terms $\Cbar \, \uas \uti$ and $\Cbar \, \vas \vti$.

The parameters in \eqref{eq:e_pin_expans_ani_orig} are constrained by the
requirement of a minimal determinant $D(\Rti)$ at the strong
pinning onset $\Rti = \Rti_m$ and $\kappa_m = 1$, i.e., its gradient has to vanish,
\begin{equation}
   \mathbf{\nabla}_{\Rti}\,D(\Rti)\big|_{\Rti_m} = 0,
\end{equation}
and its Hessian $\mathrm{Hess}[D(\Rti)]$ has to satisfy the relations
\begin{align}\label{eq:det_min2D}
   \mathrm{det}\bigl[\mathrm{Hess}\bigl[ D(\Rti) \bigr]\bigr] \big|_{\Rti_m} &> 0,\\
   \mathrm{tr} \bigl[\mathrm{Hess}\bigl[ D(\Rti) \bigr]\bigr] \big|_{\Rti_m} &> 0.
   \label{eq:det_tr2D}
\end{align}
Making use of the expansion \eqref{eq:e_pin_expans_ani_orig}, the determinant
$D(\Rti)$ reads
\begin{equation}\label{eq:full_det}
   D(\Rti) = \big\{[\partial_{\uti}^2 \epin][\partial_{\vti}^2 \epin] - [\partial_{\uti}
\partial_{\vti} \epin]^2\big\}_{\Rti}
\end{equation}
with
\begin{align}\nonumber
   &\partial_{\uti}^2 \epin = 
   \Cbar\left(1\!-\!\kappa_m\right) + a'  \vti + b'  \uti + \alpha \vti^2\!/2 + \beta \uti \vti 
   + \gamma \uti^2\!/2,  
   \\ \nonumber
   &\partial_{\vti}^2 \epin = 
   \Cbar + \lambda_+ + a \uti + b''\vti + \alpha\uti^2/2 +\beta''\uti\vti + \gamma''\vti^2/2,
   \\ \nonumber
   &\partial_{\uti} \partial_{\vti} \epin = a\vti + a'\uti + \alpha\uti\vti + \beta\uti^2/2 
   + \beta''\vti^2/2,
\end{align}
and produces the gradient
\begin{align}\label{eq:grad_D}
  \mathbf{\nabla}_{\Rti}\,D(\Rti)\Big|_{\Rti_m} = (\Cbar + \lambda_+)(b',a'),
\end{align}
hence the primed parameters indeed vanish, $a' = 0$ and $b' = 0$.
The Hessian then takes the form
\begin{align}\label{eq:Hess_D}
  \mathrm{Hess}\bigl[ D(\Rti) \bigr]\Big|_{\Rti_m} &=  (\Cbar + \lambda_+)
  \begin{bmatrix} 
    \gamma &~~~ \beta\\
    \beta &~~~ \delta
  \end{bmatrix}
\end{align}
at the Labusch point $\kappa_m = 1$, where we have introduced the parameter
\begin{equation}\label{eq:delta}
   \delta \equiv \alpha -\frac{2a^2}{\Cbar}\frac{1}{1+ \lambda_{+}/\Cbar}.
\end{equation}
The stability conditions \eqref{eq:det_min2D} and \eqref{eq:det_tr2D}
translate, respectively, to
\begin{equation}\label{eq:detHD}
   \gamma \delta - \beta^2 > 0
\end{equation}
(implying $\delta > 0$) and
\begin{equation}\label{eq:trHD}
   \gamma + \delta > 0.
\end{equation}

The Landau-type theory \eqref{eq:e_pin_expans_ani_orig} involves the two
`order parameters' $\uti$ and $\vti$ and is driven by the dual coordinates
$\uas$ and $\vas$.  This $n=2$ theory involves a soft order parameter $\uti$
and the stiff $\vti$, allowing us to integrate out $\vti$ and reformulate the
problem as an effective one-dimensional Landau theory \eqref{eq:eff_landau_1D}
of the van der Waals kind---the way of solving the strong pinning problem near
onset in this 1D formulation is presented in Appendix \ref{sec:eff_1D_onset}.

\subsection{Unstable domain $\Uti$}\label{sec:Uti}
Next, we determine the unstable domain $\Uti$ in tip space as defined in
\eqref{eq:def_calU}. We will find that, up to quadratic order, the boundary
of $\Uti$ has the shape of an ellipse with the semiaxes lengths scaling as
$\sqrt{\kappa_m-1}$.  

\subsubsection{Jump line $\Jti$}\label{sec:Jti}

We find the unstable domain $\mathcal{U}_{\Rti}$ by determining its boundary
$\partial \Uti$ that is given by the set of jump positions $\Rjp$ making up the
jump line $\Jti$. The boundary $\partial \Uti$ is determined by the condition
$\Cbar + \lambda_- = 0$ or, equivalently, the vanishing of the determinant 
\begin{equation}\label{eq:Rjp}
   D(\Rjp) \equiv 0.
\end{equation}
The latter condition guarantees the existence of an unstable direction
parallel to the eigenvector $\mathbf{v}_-(\Rjp)$ associated with the
eigenvalue $\lambda_-(\Rjp)$ where the energy \eqref{eq:e_pin_expans_ani_orig}
turns flat, cf.\ our discussion in Sec.\ \ref{sec:U-B-domains}.  The edges of
the unstable domain $\Uti$ therefore correspond to a line of inflection points
in $\epin(\Rti;\Ras)$ along which one of the bistable tip configurations of
the force balance equation \eqref{eq:gen_force_balance} coalesces with the
unstable solution.  Near the onset of strong pinning, the unstable domain
$\Uti$ is closely confined around the point $\Rti_m$ where
$\mathbf{v}_-(\Rti_m) \parallel \hat{\mathbf{u}}$. The unstable direction
$\mathbf{v}_-(\Rjp)$ is therefore approximately homogeneous within the
unstable domain $\Uti$ and is parallel to the $u$ axis.  This fact will be of
importance later, when determining the topological properties of the unstable
domain $\Uti$.

\begin{figure}
        \includegraphics[width = 1.\columnwidth]{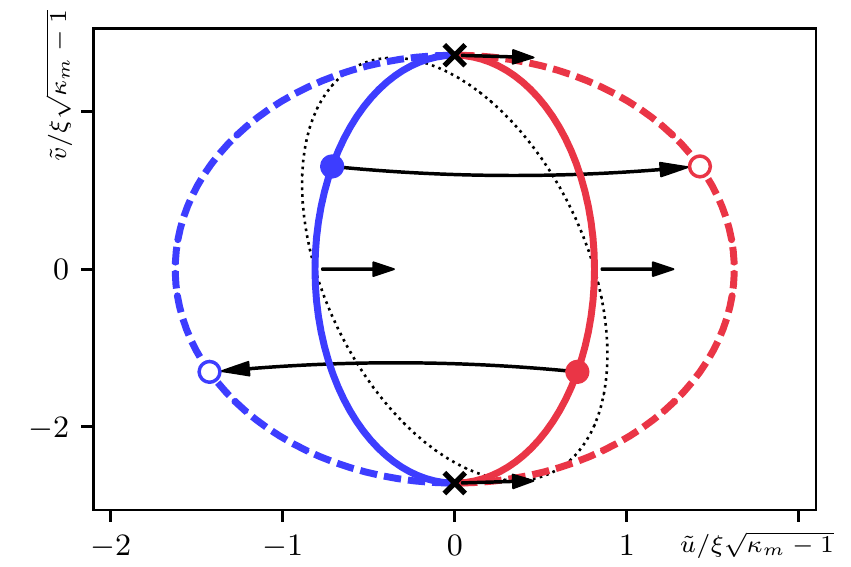}
	\caption{Jump line $\Jti$ (solid red/blue, see Eq.\
	\eqref{eq:matrix_eq_jp}) and landing line (dashed red/blue, see Eq.\
	\eqref{eq:matrix_eq_lp}) $\Lti$ in tip space $\Rti$ (in units of
	$\xi$), with the ellipse $\Jti$ representing the edge $\partial \Uti$
	of the unstable domain $\Uti$.  We choose parameters $\kappa_m - 1=
	10^{-2}$, with $\lambda_- = -0.25 \,e_p/\xi^2, \lambda_+ = 0.05
	\,e_p/\xi^2$, and $a = 0.07 \,e_p/\xi^3$, $\alpha = 0.1\, e_p/\xi^4,
	\beta = 0, \gamma = 0.75 \,e_p/\xi^4$ inspired by the choice of the
	uniaxial defect with 10 \% anisotropy in Sec.\ \ref{sec:uniax_defect};
	the dotted ellipse shows the effect of a finite skew parameter $\beta
	= 0.05\, e_p/\xi^4$ on the jump ellipse $\Jti$.  Along the edges of
	$\Uti$, one of the stable tip configurations coalesces with the
	unstable solution of \eqref{eq:gen_force_balance} and the total
	pinning energy $\epin(\Rti;\Ras)$ develops an inflection line in the
	tip coordinate $\Rti$.  Crosses correspond to the contact points
	\eqref{eq:contact_points} between the two ellipses $\Jti$ and $\Lti$.
	Blue and red colors identify different types of vortex deformations
	upon jump and landing.  Pairs of solid and open circles connected via
	long arrows are, respectively, examples of pairs of jumping- and
	landing tip positions for vortices approaching the defect from the left (top)
	and right (bottom), see Fig.\ \ref{fig:f_pin}(c) for the isotropic
	problem's counterpart. The unstable direction $\mathbf{v}_-(\Rjp)$,
	shown as short black arrows for different points on the ellipse,
	always points in the $u-$direction and are parallel to the tangent
	vector of the unstable ellipse at the contact points
	\eqref{eq:contact_points}.}
    \label{fig:ellipses}
\end{figure}

Inspection of the condition \eqref{eq:Rjp} with $D(\Rti)$ given by Eq.\
\eqref{eq:full_det} shows that the components of $\delta\Rti_\mathrm{jp}$
scale as $\sqrt{\kappa_m - 1}$: in the product $[\partial_{\uti}^2\epin]
[\partial_{\vti}^2\epin]$, the first factor involves the small constant
$\Cbar (1 - \kappa_m)$ plus quadratic terms (as $a' = 0$ and $b' = 0$), while
the second factor comes with the large constant $\Cbar + \lambda_+$ plus
corrections. The leading term in $[\partial_{\uti}\partial_{\vti}\epin]$ is
linear in $\vti$ with the remaining terms providing corrections. To leading
order, the condition of vanishing determinant then produces the quadratic form
\begin{equation}\label{eq:quadratic_form}
   [\gamma\,\uti^2 + 2\beta\,\uti\vti + \delta\, \vti^2]_{\Rjp}
   = 2\Cbar\left(\kappa_m-1\right).
\end{equation}
With $\gamma$ and $\delta$ positive, this form is associated with an elliptic
geometry of extent $\propto \sqrt{\kappa_m - 1}$.  For later convenience, we
rewrite Eq.\ \eqref{eq:quadratic_form} in matrix form
\begin{equation}\label{eq:matrix_eq_jp}
   \delta\Rjp^\mathrm{T} M_\mathrm{jp}\, \delta\Rjp = \Cbar (\kappa_m - 1)
\end{equation}
with
\begin{align}\label{eq:ellipse_jp}
      M_\mathrm{jp} &= \begin{bmatrix} \gamma/2 &~~~ \beta/2\\
                           \beta/2 &~~~ \delta/2
           \end{bmatrix}
\end{align}
and $\det M_\mathrm{jp} = (\gamma\delta -\beta^2)/4 >0$, see Eq.\ \eqref{eq:detHD}.
The jump line $\Jti$ can be expressed in the parametric form
\begin{equation}\label{eq:uti_jp}
\begin{split}
   \uti_\mathrm{jp}(|\vti| < \vti_c) &= -\frac{1}{\gamma}\Bigl[\beta\vti\\
   &\pm \sqrt{2\gamma\Cbar(\kappa_m-1)- (\gamma\delta -\beta^2) \vti^2} \Bigr],
\end{split}
\end{equation}
with
\begin{equation}\label{eq:vti_c}
   \vti_c = \sqrt{2\gamma\,\Cbar(\kappa_m - 1)/(\gamma\delta -\beta^2)}
\end{equation}
and is shown in Fig.\ \ref{fig:ellipses} for the example of an anisotropic
potential inspired by the uniaxial defect in Sec.\ \ref{sec:uniax_defect} with
10 \% anisotropy.  The associated unstable domain $\Uti$ assumes a compact
elliptic shape, with the parameter $\beta$ describing the ellipse's skew.
Comparing with the isotropic defect, this ellipse assumes the role of the ring
bounded by solid lines in Fig.\ \ref{fig:f_pin}(c), see Sec.\
\ref{sec:topology} for a discussion of its different topology.

An additional result of the above discussion concerns the terms that we need to keep
in the expansion of the pinning energy \eqref{eq:e_pin_expans_ani_orig}: indeed,
dropping corrections amounts to dropping terms with double-primed coefficients
and we find that the simplified expansion
\begin{align}\label{eq:e_pin_expans_ani}
  &e_\mathrm{pin}(\Rti; \Ras) =
  \frac{\Cbar}{2} (1 - \kappa_m) \, \uti^2
  + \frac{\Cbar + \lambda_+}{2}\, \vti^2  
  +\frac{a}{2}\, \uti \vti^2 \nonumber \\
  &\quad+\frac{\alpha}{4}\, \uti^2\vti^2
  +\frac{\beta}{6}\, \uti^3\vti
  +\frac{\gamma}{24}\, \uti^4
  -\Cbar\,\uas \uti - \Cbar\, \vas \vti
\end{align}
produces all of our desired results to leading order.

\subsubsection{Landing line $\Lti$}\label{sec:Lti}

We find the landing positions $\Rlp$ by extending the discussion of the
isotropic situation in Sec.\ \ref{sec:sp_char} to two dimensions: we shift the
origin of the expansion \eqref{eq:e_pin_expans_ani} to the jump point $\Rjp$
and find the landing point $\Rlp = \Rjp + \Delta \Rti$ by minimizing the total
energy $\epin(\Delta\Rti)$ at the landing position.  Below, we use $\Delta
\Rti$ both as a variable and as the jump distance to avoid introducing more
coordinates.

We exploit the differential properties of $\epin$ at the jump and landing
positions.  At landing, $\epin(\Rjp+\Delta\Rti)$ has a minimum, hence, the
configuration is force free, in particular along $\vti$,
\begin{align}\nonumber
   \partial_{\vti} \epin (\Rjp+\Delta \Rti) &\approx
   [\partial_{\vti}\partial_{\uti} \epin]_{\Rjp} \Delta\uti\\ \nonumber
   &\qquad + [\partial_{\vti}^2 \epin]_{\Rjp} \Delta\vti = 0,
\end{align}
from which we find that $\Delta\uti$ and $\Delta\vti$ are related via
\begin{equation}\label{eq:dv-du}
   \Delta \vti \approx -\frac{[\partial_{\vti}\partial_{\uti} \epin]_{\Rjp}}
   {[\partial_{\vti}^2 \epin]_{\Rjp}} \Delta\uti.
\end{equation}
Here, we have dropped higher order terms in the expansion, assuming that the
jump is mainly directed along the unstable $u$-direction---indeed, using the
expansion \eqref{eq:e_pin_expans_ani}, we find that
\begin{equation}\label{eq:dv-mall}
   \Delta \vti \approx -\frac{a \vjp} {\Cbar +\lambda_+}\,  \Delta\uti 
   \propto \sqrt{\kappa_m - 1}\> \Delta\uti.
\end{equation}
Note that we cannot interchange the roles of $\uti$ and $\vti$ in this force
analysis, as higher order terms in the expression for the force along $\uti$
cannot be dropped.

At the jump position $\Rjp$, the state is force-free, i.e., the derivatives
$[\partial_{\uti} \epin]_{\Rjp}$ and $[\partial_{\vti} \epin]_{\Rjp}$ vanish,
and the Hessian determinant vanishes as well. Therefore, the expansion of
$\epin(\Rjp+\Delta\Rti)$ has no linear terms and the second order terms
$[\partial_{\uti}^2 \epin]_{\Rjp} \Delta\uti^2/2 + [\partial_{\uti}
\partial_{\vti} \epin]_{\Rjp} \Delta\uti \Delta\vti + [\partial_{\vti}^2
\epin]_{\Rjp} \Delta\vti^2/2$ combined with Eq.\ \eqref{eq:dv-du} can be
expressed through the Hessian determinant, $\{[\partial_{\uti}^2
\epin][\partial_{\vti}^2 \epin] - [\partial_{\uti}\partial_{\vti}
\epin]^2\}_{\Rjp} \Delta\uti^2/2 = 0$, that vanishes as well.  Therefore, the
expansion of $\epin$ around $\Rjp$ starts at third order in $\Delta \Rti
\approx (\Delta \uti, 0)$ and takes the form (we make use of \eqref{eq:dv-mall},
dropping terms $\propto \Delta\vti$ and a constant)
\begin{equation}\label{eq:epin_at_jp}
   \epin(\Rjp + \Delta \Rti) \approx 
   \frac{1}{6} \bigl(\gamma \ujp + \beta \vjp \bigr) \Delta\uti^3
   +\frac{\gamma}{24} \Delta\uti^4.
\end{equation}
Minimizing this expression with respect to $\Delta\uti$ (as $\epin$ is minimal
at $\Rlp$), we obtain the result
\begin{equation}\label{eq:du}
   \Delta \uti \approx - 3 (\gamma\ujp + \beta\vjp)/\gamma.
\end{equation}

Making use of the quadratic form \eqref{eq:matrix_eq_jp}, we can show that the
equation for the landing position $\Rti_\mathrm{lp} = \Rti_\mathrm{jp} +
\Delta\Rti$ can be cast into a similar quadratic form (with $\delta\Rlp$
measured relative to $\Rti_m$)
\begin{equation}\label{eq:matrix_eq_lp}
   \delta\Rlp^\mathrm{T} M_\mathrm{lp}\, \delta\Rlp = \Cbar (\kappa_m - 1),
\end{equation}
but with the landing matrix now given by
\begin{equation}\label{eq:ellipse_lp}
   M_\mathrm{lp} = \frac{1}{4} M_\mathrm{jp} +
   \begin{bmatrix} 0 & 0\\
                   0 & ~~~\displaystyle{\frac{3}{4}\Bigl(\frac{\delta}{2}
                   - \frac{\beta^2}{2\gamma}\Bigr)}
   \end{bmatrix}.
\end{equation}
In the following, we will refer to the solutions of Eq.\
\eqref{eq:matrix_eq_lp} as the `landing' or `stable' ellipse $\Rlp$ and
write the jump distance in a parametric form involving the
shape $\ujp(\vti)$ in Eq.\ \eqref{eq:uti_jp} of the jumping ellipse,
\begin{align}
   &\Delta \uti(\vti) = -3\left[\gamma\, \uti_\mathrm{jp}(\vti)
   + \beta\, \vti\right]/\gamma,\label{eq:delta_rx}\\
   &\Delta \vti(\vti) = - \left[a/(\Cbar + \lambda_+)\right]\, \vti\,
   \Delta \uti(\vti)\label{eq:delta_ry}.
 \end{align}
The landing line derived from \eqref{eq:matrix_eq_lp} is displayed as a dashed
line in Fig.\ \ref{fig:ellipses}. Two tip jumps connected by an arrow are
shown for illustration, with solid dots marking the jump position
$\Rti_\mathrm{jp}$ of the tip and open dots its landing position
$\Rti_\mathrm{lp}$; they describe tip jumps for a vortex approaching the
unstable ellipse once from the left (upper pair) and another time from the
right (lower pair). The different topologies associated with jumps and landing
showing up for the isotropic defect in Fig.\ \ref{fig:f_pin}(c) (two
concentric circles) and for the generic onset in Fig.\ \ref{fig:ellipses} (two
touching ellipses) will be discussed later.

Inspecting the matrix equation \eqref{eq:matrix_eq_lp}, we can gain several
insights on the landing ellipse $\Lti$: (i) the matrix $M_\mathrm{jp}/4$ on
the right-hand side of \eqref{eq:ellipse_lp} corresponds to an ellipse with
the same geometry as for $\Jti$ but double in size, (ii) the remaining matrix with
vanishing entries in the off-diagonal and the $M_{xx}$ elements leaves the
size doubling of the stable ellipse $\Lti$ at $\vti = 0$ unchanged, and (iii)
the finite $M_{yy}$ component exactly counterbalances the doubling along the
$v-$direction encountered in (i), cf.\ the definiton \eqref{eq:ellipse_jp} of
$M_\mathrm{jp}$, up to a term proportional to the skew parameter $\beta$
accounting for deviations of the semiaxis from the $v-$axis. Altogether, the
stable ellipse $\Lti$ extends with a double width along the $u-$axis and
smoothly overlaps with the unstable ellipse at the two contact points
$\vti_{c,\pm}$.  The latter are found by imposing the condition $\Delta \uti =
\Delta \vti = 0$ in Eqs.\ \eqref{eq:delta_rx} and \eqref{eq:delta_ry}; we find
them located (relative to $\Rti_m$) at
\begin{align}
        \delta\Rti_{c,\pm} &= \pm \left(-\beta/\gamma, 1\right)\,\vti_{c},
        \label{eq:contact_points}
\end{align}
with the endpoint coordinate $\vti_c$ given in Eq.\ \eqref{eq:vti_c}, and mark
them with crosses in Fig.\ \ref{fig:ellipses}.  As anticipated, the contact
points are off-set with respect to the $v-$axis for a finite
skew parameter $\beta$. At these points, the unstable and the stable tip
configurations coincide and the vortex tip undergoes no jump. Furthermore, the
vector tangent to the jump (or landing) ellipse is parallel to the
$u-$direction at the contact points. To see that, we consider
\eqref{eq:uti_jp} and find that
\begin{align}\label{eq:tangent_y}
   \frac{\partial \uti}{\partial \vti}\Big|_{\vti\to\pm\vti_{c}} \!\!\!\!\!
   &\approx\pm\left(\sqrt{\vti_c^2 - \frac{2\gamma\,\Cbar(\kappa_m-1)}
   {\gamma\beta - \delta^2}}\right)^{-1} \!\!\! \to \pm\infty,
\end{align}
hence, the corresponding tangents $\partial_{\uti} \vti$ vanish.  

The asymptotic positions $\Ras$ where the vortex tips jump and land belong to
the boundary of the bistable region $\Bas$; for the isotropic case in Fig.\
\ref{fig:f_pin}(d) these correspond to the circles with radii $\ras_-$
(pinning) and $\ras_+$ (depinning) with jump and landing radii
$\rti_\mathrm{f-}(\ras_-)$ and $\rti_\mathrm{p-}(\ras_-)$ and
$\rti_\mathrm{p+}(\ras_+)$ and $\rti_\mathrm{f+}(\ras_+)$, respectively, see
Fig.\ \ref{fig:f_pin}(c). For the anisotropic defect, we have only a single
jump/landing event at one asymptotic position $\Ras$ that we are going to
determine in the next section.

\subsection{Bistable domain $\Bas$}\label{sec:Bas}
The set of asymptotic positions $\Ras$ corresponding to the tip positions
$\Rti_\mathrm{jp}$ along the edges of $\Uti$ forms the boundary $\partial\Bas$
of the bistable domain $\Bas$; they are related through the force-balance
equation \eqref{eq:gen_force_balance}, with every vortex tip position
$\Rti_\mathrm{jp} \in \partial \Uti$ defining an associated asymptotic
position $\Ras(\Rti_\mathrm{jp}) \in \partial\Bas$. 

\begin{figure}
        \includegraphics[width = 1.\columnwidth]{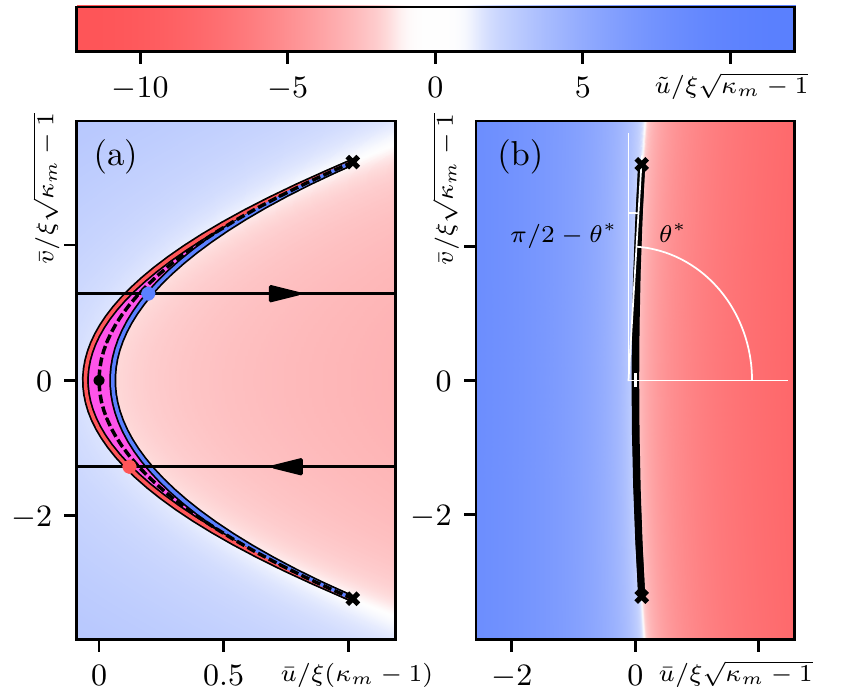}
	\caption{(a) Bistable domain $\Bas$ in asymptotic $\Ras$-space
	measured in units of $\xi$; the same parameters as in Fig.\
	\ref{fig:ellipses} have been used.  Note the different scaling of the
	axes in $\kappa_m - 1$; the right panel (b) shows $\Bas$ in
	isotropic scales. The bistable domain $\Bas$ is elongated along the
	transverse direction $\vas$ and narrow/bent along the unstable
	direction $\uas$, giving $\Bas$ its peculiar crescent-like shape. The
	branch crossing line $\Ras_0$, see \eqref{eq:x_0_line}, is shown as a
	dashed black line. Black crosses mark the cusps of $\Bas$ and are
	associated with the contact points of $\Uti$ through the force balance
	equation \eqref{eq:gen_force_balance}; they correspond to critical
	end-points in the thermodynamic Ising analogue, while the boundaries
	$\partial \Bas$ map to spinodals.  Blue and red colors identify
	different characters of vortex tip configurations as quantified
	through the `order parameter' $\uti$ of the Landau expansion (at
	$\beta = 0$), see text, while magenta is associated to the bistable
	area $\Bas$; the blue and red branches extend to the far side of the
	crescent and terminate in the blue and red colored boundaries
	$\partial\Bas^\mathrm{b}$ and $\partial\Bas^\mathrm{r}$, respectively.
	Thin horizontal lines show vortex trajectories that proceed smoothly
	in asymptotic space, see also Fig.\ \ref{fig:f_pin}(d). Blue and red
	dots mark the asymptotic positions associated with vortex tip jumps
	that happen at the exit of $\Bas$; they correspond to the pairs of tip
	positions in Fig.\ \ref{fig:ellipses}.  (b) Bistable domain $\Bas$ in
	isotropic scaled coordinates $\uas$ and $\vas$ showing the `true'
	shape of $\Bas$.  Vortices impacting on the bistable domain with an
	angle $|\theta|\leq \theta^\ast$ undergo a single jump on the far side
	of $\Bas$, with the pinning force density directed along $u$ and
	scaling as $\Fpin^{\parallel}\propto (\kappa-1)^{5/2}$. Vortices
	crossing $\Bas$ at large angles close to $\pi/2$ jump either never,
	once, or twice; at $\theta = \pi/2$ the pinning force density is
	small, $\Fpin^{\perp}\propto (\kappa-1)^{3}$, and directed along
	$v$.}
    \label{fig:bananas}
\end{figure}

At the onset of strong pinning, the bistable domain corresponds to the
isolated point $\Ras_m$, related to $\Rti_m$ through
\eqref{eq:gen_force_balance}. Beyond the Labusch point, $\Bas$ expands out of
$\Ras_m$ and its geometry is found by evaluating the force balance equation
\eqref{eq:gen_force_balance} at a given tip position $\Rti_\mathrm{jp} \in
\partial\Uti$, $\Ras(\Rti_\mathrm{jp}) = \Rti_\mathrm{jp} - \mathbf{f}_p
(\Rti_\mathrm{jp})/\Cbar \in \partial\Bas$.  Using the expansion
\eqref{eq:e_pin_expans_ani} for $\epin(\Rti;\Ras)$, this force equation can be
expressed as $\nabla_\mathbf{R} e_\mathrm{pin} (\mathbf{R};\Ras) \big|_{\Rti}
= 0$, or explicitly (we remind that we measure $\Ras = \Ras_m +(\uas,\vas)$
relative to $\Ras_m$),
\begin{align}\nonumber
   \Cbar \uas &=  \Cbar(1-\kappa_m) \uti + \frac{a}{2}\vti^2 
   + \frac{\gamma}{6}\uti^3 + \frac{\beta}{2}\uti^2 \vti 
   + \frac{\alpha}{2}\uti \vti^2,\\
   \Cbar \vas &= (\Cbar + \lambda_+) \vti + a\,\uti \vti + \frac{\beta}{6}\uti^3 
   + \frac{\alpha}{2}\uti^2 \vti.
   \label{eq:asymptotic_positions}
\end{align}
Inserting the results for the jump ellipse $\Jti$, Eq.\ \eqref{eq:uti_jp}, into
Eqs.\ \eqref{eq:asymptotic_positions}, we find the crescent-shape bistable domain
$\Bas$ shown in Fig.\ \ref{fig:bananas}; let us briefly derive the origin of this shape.

Solving \eqref{eq:asymptotic_positions} to leading order, $\Cbar
\uas^{\scriptscriptstyle (0)} \approx (a/2) \vti^2$ and $\Cbar
\vas^{\scriptscriptstyle (0)} \approx (\Cbar + \lambda_+) \vti$, we find the
parabolic approximation
\begin{align}\label{eq:parabola_x}
   \uas^{\scriptscriptstyle (0)} &\approx \frac{a}{2\Cbar}
   \frac{1}{(1 + \lambda_+/\Cbar)^2}\,\vas^{{\scriptscriptstyle (0)}\,2},
\end{align}
telling that the extent of $\Bas$ scales as $(\kappa_m - 1)$ along
$\uas$ and $\propto (\kappa_m - 1)^{1/2}$ along $\vas$, i.e., we find a flat
parabola opening towards positive $\uas$ for $a > 0$, see Fig.\ \ref{fig:bananas}.

In order to find the width of $\Bas$, we have to solve
\eqref{eq:asymptotic_positions} to the next higher order, $\uas =
\uas^{\scriptscriptstyle (0)} + \uas^{\scriptscriptstyle (1)}$; for $\beta =
0$, we find the correction
\begin{equation}\label{eq:duas}
   \uas^{\scriptscriptstyle (1)} =  (1-\kappa_m) \uti + \frac{\gamma}{6\Cbar}\uti^3
   + \frac{\alpha}{2\Cbar}\uti \vti^2
\end{equation}
that produces a $\vas \leftrightarrow -\vas$ symmetric crescent.
Inserting the two branches \eqref{eq:uti_jp} of the jump ellipse, we arrive
at the width of the crescent that scales as $(\kappa_m-1)^{3/2}$. The correction
to $\vas$ is $\propto (\kappa_m - 1)$ and we find the closed form
\begin{align}\label{eq:dvas}
    \vas &\approx [1+ (\lambda_+ + a\uti)/\Cbar]\> \vti
\end{align}
with a small antisymmetric (in $\uti$) correction. For a finite $\beta \neq
0$, the correction $\uas^{\scriptscriptstyle (1)}$ picks up an additional term
$(\beta/2\Cbar)\,\uti^2 \vti$ that breaks the $\vas \leftrightarrow -\vas$
symmetry and the crescent is distorted.

Viewing the boundary $\partial\Bas$ as a parametric curve in the variable
$\vti$ with $\uti = \uti_\mathrm{jp}(\vti)$ given by Eq.\ \eqref{eq:uti_jp},
we obtain the boundary $\partial\Bas$ in the form of two separate arcs that
define the crescent-shaped domain $\Bas$ in Fig.\ \ref{fig:bananas}(a). The
two arcs merge in two cusps at $\Ras_{c,\pm}$ that are associated to the touching
points \eqref{eq:contact_points} in dual space and derive from Eqs.\
\eqref{eq:asymptotic_positions}; measured with respect to $\Ras_m$, these cusps
are located at 
\begin{eqnarray}\label{eq:cusps}
   \delta\Ras_{c,\pm}  &=& (\uas_c,\pm \vas_c) \\ \nonumber
   &\approx& \left[\left(a/2\,\Cbar\right)\,\vti^2_c,\,
   \pm (1 + \lambda_+/\Cbar)\vti_c\right].
\end{eqnarray}

The coloring in Fig.\ \ref{fig:bananas} indicates the characters `red' and
`blue' of the vortex states; these are defined in terms of the `order
parameter' $\uti-\uti_m(\vas)$ of the Landau functional
\eqref{eq:e_pin_expans_ani} that changes sign at the branch crossing line Eq.\
\eqref{eq:x_0_line}, with the shift
\begin{equation}\label{eq:shift}
   \uti_m(\vas) = -\frac{\beta}{\gamma} \vti(\vas) 
            \approx -\frac{\beta}{\gamma} \frac{\vas}{1 + \lambda_+/\Cbar},
\end{equation}
$\uti_m(\vas) = 0$ for our symmetric case with $\beta = 0$ in Fig.\
\ref{fig:bananas}.  Going beyond the cusps (or critical points) at
$\Ras_{c,\pm}$, the two states smoothly crossover between `red' and `blue'
(indicated by the smooth blue--white--red transition), as known for the van
der Waals gas (or Ising magnet) above the critical point. Within the bistable
region $\Bas$, both `red' and `blue' states coexist and we color this region
in magenta.

The geometry of the bistable domain $\Bas$ is very different from the
ring-shaped geometry of the isotropic problem discussed in Sec.\
\ref{sec:iso_def}, see Fig.\ \ref{fig:f_pin}(d); in the discussion of the
uniaxial anisotropic defect below, we will learn how these two geometries are
interrelated.  Comparing the overall dimensions of the crescent with the
ring in Fig.\ \ref{fig:f_pin}(d), we find the following scaling behavior in
$\kappa_m -1$: while the crescent $\Bas$ grows along $\vas$ as
$(\kappa_m-1)^{1/2}$, the isotropic ring involves the characteristic size
$\xi$ of the defect, $\ras_- \sim \xi$ and hence its extension along $\vas$ is
a constant.  On the other hand, the scaling of the crescent's and the ring's
width is the same, $\propto (\kappa_m - 1)^{3/2}$.  The different scaling of
the transverse width then will be responsible for the new scaling of the
pinning force density, $\Fpin \propto (\kappa_m -1 )^{5/2}$.

\subsection{Comparison to isotropic situation}\label{sec:discussion}

Let us compare the unstable domains $\Uti$ for the isotropic and anisotropic
defects in Figs.\ \ref{fig:f_pin}(c) and \ref{fig:ellipses}, respectively.  In
the isotropic example of Sec.\ \ref{sec:iso_def}, the jump- and
landing-circles $\rti_\mathrm{jp}(\ras)$ and $\rti_{\mathrm{lp}}(\ras)$ are
connected to different phases, e.g., free (colored in blue at
$\rti_\mathrm{jp} = \rti_\mathrm{f-}$) and pinned (colored in red at
$\rti_{\mathrm{lp}} =\rti_\mathrm{p-}$) associated with $\ras_-$. Furthermore,
the topology is different, with the unstable ring domain separating the two
distinct phases, free and pinned ones.  As a result, a second pair of jump-
and landing-positions associated with the asymptotic circle $\ras_{+}$ appears
along the vortex trajectory of Fig.\ \ref{fig:f_pin}(c); these are the located
at the radii $\rti_\mathrm{jp} = \rti_\mathrm{p+}$ and $\rti_\mathrm{lp} =
\rti_\mathrm{f+}$ and describe the depinning process from the pinned branch
back to the free branch (while the previous pair at radii $\rti_\mathrm{f-}$
and $\rti_\mathrm{p-}$ describes the pinning process from the free to the
pinned branch). The pinning (at $\ras_-$) and depinning (at $\ras_+$)
processes in the asymptotic coordinates are shown in figure
\ref{fig:f_pin}(d).  The bistable area $\Bas$ with coexisting free and pinned
states has a ring-shape as well (colored in magenta, the superposition of blue
and red); the two pairs of jump and landing points in tip space have collapsed
to two pinning and depinning points in asymptotic space.

In the present situation describing the strong pinning onset for a generic
anisotropic potential, the unstable domain $\Uti$ grows out of an isolated
point (in fact, $\Rti_m$) and assumes the shape of an ellipse that is simply
connected; as a result, a vortex incident on the defect undergoes only a
single jump, see Fig.\ \ref{fig:ellipses}. The bistable domain $\Bas$ is
simply connected as well, but now features two cusps at the end-points of the
crescent, see Fig.\ \ref{fig:bananas}.  The bistability again involves two
states, but we cannot associate them with separated pinned and free
phases---we thus denote them by `blue'-type and `red'-type.  The two states
approach one another further away from the defect and are distiguishable only
in the region close to bistability; in Fig.\ \ref{fig:bananas}, this is
indicated with appropriate color coding. Note that the Landau-type expansion
underlying the coloring in Fig.\ \ref{fig:bananas} fails at large distances;
going beyond a local expansion near $\Rti_m$, the distortion of the vortex
vanishes at large distances and red/blue colors faint away to approach
`white'.

\subsection{Topology}\label{sec:topology}

The different topologies of unstable and bistable regions appearing in the
isotropic and anisotropic situations are owed to the circular symmetry of the
isotropic defect; we will recover the ring-like topology for the anisotropic
situation later when describing a uniaxially anisotropic defect at larger
values of the Labusch parameter $\kappa_m$. Indeed, such an increase in pinning
strength will induce a change in topology with two crescents facing one
another joining into a ring-like shape.

Let us discuss the consequences of the different topologies that we
encountered for the isotropic and anisotropic defects in the discussion above.
Specifically, the precise number and position of the contact points have an
elegant topological explanation. When a vortex tip touches the edges $\Rjp$ of
the unstable domain there are two characteristic directions: one is given by
the unstable eigenvector $\mathbf{v}_-(\Rjp)$ discussed in Sec.\ \ref{sec:Uti}
along which the tip will jump initially. The second is the tangent vector to
the boundary $\partial \Uti$ of the unstable domain, i.e., to the unstable
ellipse.  While the former is approximately constant and parallel to the
unstable $u$-direction along $\Rjp$, the latter winds around the ellipse
exactly once after a full turn around $\Uti$.  The contact points
$\Rti_{c,\pm}$ of the unstable and stable ellipses then coincide with those
points on the ellipse where the tangent vector are parallel and anti-parallel
to $\mathbf{v}_-$; at these points, the tip touches the unstable ellipse but
does not undergo a jump any more.  Given the different winding numbers of
$\mathbf{v}_-$ and of the tangent vector, there are exactly two points along
the circumference of $\Uti$ where the tangent vector is parallel/anti-parallel
to the $u$-direction; these are the points found in \eqref{eq:contact_points}.
This argument remains valid as long as the contour $\partial\Uti$ is not
deformed to cross/encircle the singular point of the $\mathbf{v}_-(\Rjp)$ field
residing at the defect center.

The same arguments allow us to understand the absence of contact points in the
isotropic scenario: For an isotropic potential, the winding number
$n_{\scriptscriptstyle \mathcal{U}}$ of the tangent vector around $\Uti$
remains unchanged, i.e., $n_{\scriptscriptstyle \mathcal{U}} = \pm 1$, while
the unstable direction $\mathbf{v}_-$ is pointing along the radius and thus
acquires a unit winding number as well. Indeed, the two directions, tangent
and jump, then rotate simultaneously and do not wind around each other after a
full rotation, explaining the absence of contact points in the isotropic
situation.

\subsection{Energy jumps}\label{sec:de_pin}
\begin{figure}
        \includegraphics[width = 1.0\columnwidth]{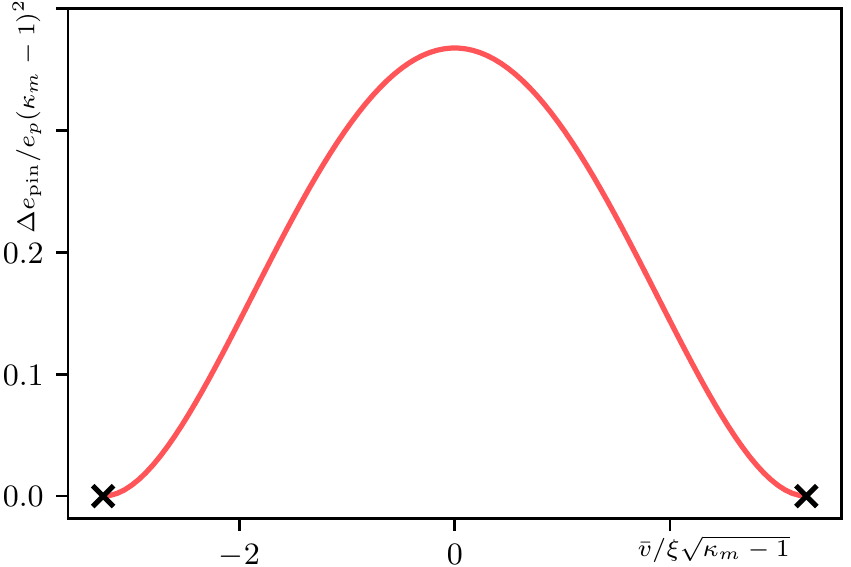}
	\caption{Energy jump $\Delta e_\mathrm{pin}$ along the edges of the
	bistable domain $\Bas$ as a function of the transverse coordinate
	$\vas$; we have used the same parameters as in Fig.\
	\ref{fig:ellipses}. The energy jump vanishes at the cusps
	$\pm\vas_{c}$, as the bistable tip configurations become identical and
	their energies turn equal.}
    \label{fig:energies}
\end{figure}

Within strong pinning theory, the energy jump $\Delta e_\mathrm{pin}$
associated with the vortex tip jump between bistable vortex configurations at
the boundaries of $\Bas$ determines the pinning force density $\Fpin$ and the
critical current $j_c$, see Eqs.\ \eqref{eq:F_pin} and
\eqref{eq:macroscopic_force_balance}.  Formally, the energy jump $\Delta
e_\mathrm{pin}$ is defined as the difference in energy $\epin(\Rti;\Ras)$ at
fixed asymptotic position $\Ras \in \partial\Bas$ between vortex
configurations with tips in the jump ($\Rjp(\Ras)$) and landing ($\Rlp(\Ras) =
\Rjp(\Ras) + \Delta\Rti$) positions,
\begin{multline}\label{eq:energy_jump_def}
   \Delta e_\mathrm{pin}(\Ras \in \partial \Bas) \equiv e_\mathrm{pin}[\Rjp(\Ras); \Ras]\\
 - e_\mathrm{pin}[\Rlp(\Ras); \Ras].
\end{multline}
In Sec.\ \ref{sec:Lti} above, we have found that the jump $\Delta\Rti$ is
mainly forward directed along $u$.  Making use of the expansion
\eqref{eq:epin_at_jp} of $\epin$ at $\Rjp$ and the result \eqref{eq:du} for
the jump distance $\Delta\uti$, we find the energy jumps $\Delta
e_\mathrm{pin}$ in tip- and asymptotic space in the form (cf.\ with the
isotropic result Eq.\ \eqref{eq:d_epin^pf}),
\begin{align}\label{eq:energy_jump}
   \Delta e_\mathrm{pin}(\Ras) &\approx \frac{\gamma}{72}\Delta\uti^4 \approx
   \left(\frac{9}{8\gamma^3}\right)
   \left[\gamma\, \ujp(\vti) + \beta\, \vti \right]^4\\
   &\approx\left(\frac{9}{8\gamma^3}\right)\left[(\gamma\delta - \beta^2)
   \left(\vti_c^2 - \vti^2\right)\right]^2
   \nonumber\\
   & \approx\left(\frac{9}{8\gamma^3}\right)
   \left[\frac{(\gamma\delta - \beta^2)}{(1+\lambda_+/\Cbar)^2}
   \left(\vas_c^2 - \vas^2\right)\right]^2. \nonumber
\end{align}
Here, we have used the parametric shape $\ujp(\vti)$ in Eq.\ \eqref{eq:uti_jp}
for the jumping ellipse as well as \eqref{eq:asymptotic_positions} to lowest
order, $\vti \approx \vas/(1 + \lambda_+ / \Cbar)$, to relate the tip and
asymptotic positions in the last equation.  The energy jump
\eqref{eq:energy_jump} scales as $(\kappa_m - 1)^2$ and is shown in
Fig.\ \ref{fig:energies}.  It depends on the $v$ coordinate of the asymptotic
(or tip)  position only and vanishes at the cusps $\Ras_{c,\pm}$, see Eq.\
\eqref{eq:cusps} (or at the touching points $\Rti_{c,\pm}$, see Eq.\
\eqref{eq:contact_points}).  To order $(\kappa_m - 1)^2$, the energy jumps are
identical at the left and right edges of the bistable domain $\Bas$.

Following the two bistable branches and the associated energy jumps between
them to the inside of $\Bas$, the latter vanish along the branch crossing line
$\Ras_0$. In the thermodynamic analogue, this line corresponds to the
first-order equilibrium transition line that is framed by the spinodal lines;
for the isotropic defect, this is the circle with radius $\ras_0 = x_0$ framed
by the spinodal circles with radii $\ras_\pm$, see Figs.\ \ref{fig:e_pin} and
\ref{fig:f_pin}(d). For the anisotropic defect with $\beta = 0$, this line is
trivially given by the centered parabola of $\Bas$, see Eq.\
\eqref{eq:parabola_x}, and hence
\begin{equation} 
   \uas_0 \approx \frac{a}{2\Cbar}\frac{1}{(1+ \lambda_+/\Cbar)^2} \vas_0^2.
   \label{eq:x_0_line}
\end{equation}
The result for a finite skew parameter $\beta \neq 0$ is given by Eq.\
\eqref{eq:x_0_line_beta} in Appendix \ref{sec:eff_1D_onset}.  

\subsection{Pinning force density}\label{sec:F_pin_anis}

The pinning force density $F_\mathrm{pin}$ is defined as the average force
density exerted on a vortex line as it moves across the superconducting
sample. For the isotropic case described in Sec.\ \ref{sec:F_pin_iso}, the
individual pinning force $\mathbf{f}_\mathrm{pin}(\Ras) = - \nabla_{\Ras}
\epin(\Ras)$, see Eq.\ \eqref{eq:f_pin}, is directed radially and the force
density $F_\mathrm{pin}$ is given by the (constant) energy jump $\Delta
e_\mathrm{pin}\propto (\kappa-1)^2$ on the edge $\partial\Bas$ of the bistable
domain and the transverse length $t_\perp \sim \xi$, hence, $\Fpin\propto
t_\perp \Delta e_\mathrm{pin}$ scales as $(\kappa-1)^2$.

For an anisotropic defect, the pinning force depends on the vortex direction
of motion $\hat{\mathbf{v}} = (\cos\theta,\sin\theta)$ relative to the axis of
the bistable region: we choose angles $-\pi/2\leq\theta\leq\pi/2$ measured
from the unstable direction $\uas$, i.e., vortices incident from the left; the
case of larger impact angles $|\theta| > \pi/2$ corresponds to vortices
incident from the right and can be reduced to the previous case by inverting
the sign of the parameter $a$ in the expansion \eqref{eq:e_pin_expans_ani},
i.e., the curvature of the parabola \eqref{eq:parabola_x}; to our leading
order analysis, the results remain the same. The pinning force is no longer
directed radially but depends on $\theta$; furthermore, the energy jump
\eqref{eq:energy_jump} is non-uniform along the boundary $\Bas$.

In spite of these complications, we can perform some simple scaling estimates
as a first step: let us assume a uniform distribution of identical anisotropic
defects, all with their unstable direction pointing along $x$. The jumps in
energy still scale as $\Delta e_\mathrm{pin} \propto (\kappa_m-1)^2$, however,
the trapping distance is no longer finite but grows from zero as $\kappa_m -
1$ increases. Due to their elongated shapes, the bistable domains $\Bas$
exhibit different extensions along the $y$ and $x$ directions, i.e., $\propto
\vas_{c} \propto \sqrt{\kappa_m - 1}$ along $y$ and $\propto \uas_{c} \propto
(\kappa_m - 1)$ along $x$, respectively.  These simple considerations then
suggest that the pinning force density exhibits a scaling $\Fpin \propto
(\kappa_m-1)^\mu$ with $\mu > 2$, different from the setup with isotropic
defects.  Even more, vortices moving along the $x$ or $y$ directions,
respectively, will experience different forces $\Fpin^{\parallel}$ and
$\Fpin^{\perp}$ scaling as
\begin{equation}\label{eq:fpin_scaling}
   \Fpin^{\parallel}\propto (\kappa_m-1)^{5/2}, \quad \Fpin^{\perp}\propto (\kappa_m-1)^{3}
\end{equation}
near the onset of strong pinning. While such uniform anisotropic defects could
be created artificially, a more realistic scenario will involve defects that
are randomly oriented and an additional averaging over angles $\theta$ has to
be performed; this will be done at the end of this section.

We first determine the magnitude and orientation of the pinning force density
$\mathbf{F}_\mathrm{pin}(\theta)$ as a function of the vortex impact angle
$\theta$ for randomly positioned but uniformly oriented (along $x$) defects of
density $n_p$.  The pinning force density is given by the average over
relative positions between vortices and defects (with a minus sign following
convention; $\Vas$ denotes the vortex lattice unit cell),
\begin{eqnarray}\label{eq:formal_pinning_force}
   &&\mathbf{F}_\mathrm{pin}(\theta) = -n_p 
   \int_{\Vas\setminus\Bas} \!\!  \frac{\mathrm{d}^2\Ras}{a_0^2}\, 
   \mathbf{f}_\mathrm{pin}(\Ras)  \\ \nonumber 
   &&\quad - 
   n_p \int_{\Bas} \!\!\! \frac{\mathrm{d}^2 \Ras}{a_0^2} \left[p_\mathrm{b}(\Ras;\theta)\,
   \mathbf{f}^\mathrm{b}_\mathrm{pin}(\Ras) + p_\mathrm{r}(\Ras;\theta)\,
   \mathbf{f}^\mathrm{r}_\mathrm{pin}(\Ras)\right].
\end{eqnarray}
Outside of the bistable domain, i.e., in $\Vas\setminus\Bas$, a single stable
vortex tip configuration exists and the pinning force
$\mathbf{f}_\mathrm{pin}(\Ras)$ is uniquely defined.  Inside $\Bas$, the
branch occupation functions $p_\mathrm{b,r}(\Ras;\theta)$ are associated with
the tip positions appertaining to the `blue' and the `red' vortex
configurations with different tip positions $\Rti^\mathrm{b,r}(\Ras)$, cf.\
Figs.\ \ref{fig:ellipses} and \ref{fig:bananas}. The pinning forces
$\mathbf{f}^\mathrm{b,r}_\mathrm{pin}(\Ras)$ are evaluated for the
corresponding vortex tip positions and are defined as
\begin{equation}
  \mathbf{f}^\mathrm{b,r}_\mathrm{pin}(\Ras) = -\mathbf{\nabla}_{\Ras} 
  e_\mathrm{pin}[\Rti^\mathrm{b,r}(\Ras);\Ras].
\end{equation}

Let us now study how vortex lines populate the bistable domain as a function
of the impact angle $\theta$.  Examining Fig.\ \ref{fig:bananas}, we can
distinguish between two different angular regimes: a \emph{frontal}-impact
regime at angles away from $\pi/2$, $|\theta| \leq \theta^\ast$, where
all the vortices that cross the bistable domain undergo exactly one jump on
the far edge of $\Bas$, see the blue dot and blue boundary
$\partial\Bas^\mathrm{b}$ in Fig.\ \ref{fig:bananas}; and a \emph{transverse}
regime for angles $\theta^\ast \leq |\theta| \leq \pi/2$, where
vortices crossing the bistable domain undergo either no jump, one or two.  The
angle $\theta^\ast$ is given by the (outer) tangent of the bistable domain at
the cusps $\Ras_{c,\pm}$; making use of the lowest order approximation
\eqref{eq:parabola_x} of the crescent's geometry, we find that
\begin{align}
   \tan (\theta^\ast) &= 
   \frac{\partial \vas^{\scriptscriptstyle (0)}}
        {\partial \uas^{\scriptscriptstyle (0)}} \Big|_{\vas_c}
   = \frac{(\Cbar + \lambda_+)}{a}
   \sqrt{\frac{\gamma\delta -\beta^2}{2\gamma\Cbar(\kappa_m -1)}},
\end{align}
implying that $\pi/2 - \theta^\ast \propto \sqrt{\kappa_m - 1}$ is small,
\begin{align}
   \theta^\ast \approx \pi/2 - \frac{a}{(\Cbar + \lambda_+)}
   \sqrt{\frac{2\gamma\Cbar(\kappa_m -1)}{\gamma\delta -\beta^2}}.
\end{align}

\subsubsection{Impact angles $|\theta| < \theta^\ast$}\label{sec:F_par}

For a frontal impact with $|\theta| < \theta^\ast$, 
vortices occupy the `blue' branch and remain there throughout 
the bistable domain $\Bas$ until its termination on the
far edge $\partial\Bas^\mathrm{b}$, see Fig.\ \ref{fig:bananas}, implying that
$p_\mathrm{b}(\Ras\in\Bas) = 1$ and $p_\mathrm{r}(\Ras\in\Bas) = 0$,
independent of $\theta$. As a consequence, the pinning force
$\mathbf{F}_\mathrm{pin}$ does not depend an the impact angle and
is given by the expression
\begin{equation*}
   \mathbf{F}^{<}_\mathrm{pin} = -n_p \! \int_{\Vas\setminus\Bas} \!\!\!\!
   \frac{\mathrm{d}^2\Ras}{a_0^2}\, \mathbf{f}_\mathrm{pin}(\Ras)
   - n_p \! \int_{\Bas} \!\!\!\! \frac{\mathrm{d}^2\Ras}{a_0^2}\,
   \mathbf{f}^\mathrm{b}_\mathrm{pin}(\Ras).
\end{equation*}
Next, Gauss' formula tells us that for a function $e(\mathbf{x})$, we can
transform
\begin{equation}\label{eq:gauss_theorem}
   \int_\mathcal{V} \mathrm{d}^n x \,\mathbf{\nabla} e(\mathbf{x}) 
   = \int_{\partial\mathcal{V}}\mathrm{d}^{n-1}\, \mathbf{S}_\perp \,e(\mathbf{x}),
\end{equation}
with the surface element $\mathrm{d}^{n-1}\, \mathbf{S}_\perp$ oriented
perpendicular to the surface and pointing outside of the domain $\mathcal{V}$.
In applying \eqref{eq:gauss_theorem} to the first integral of
$\mathbf{F}^{<}_\mathrm{pin}$, we can drop the contribution from the
outer boundary $\partial\Vas$ since we assume a compact defect potential.  The
remaining contribution from the crescent's boundary $\partial\Bas$ joins up
with the second integral but with an opposite sign, as the two terms involve
the same surface but with opposite orientations.  Altogether, we then arrive
at the expression
\begin{multline}\label{eq:intermediate_flux_fpin}
   \mathbf{F}^{<}_\mathrm{pin} = n_p \int_{\partial \Bas^{\mathrm{b}}}
   \frac{\mathrm{d}\, \mathbf{S}_\perp}{a_0^2}
   \left(e^\mathrm{b}_\mathrm{pin}(\Ras) - e_\mathrm{pin}(\Ras)\right)\\ 
   + n_p \int_{\partial \Bas^{\mathrm{r}}} \frac{\mathrm{d}\, \mathbf{S}_\perp}{a_0^2}
   \left(e^\mathrm{b}_\mathrm{pin}(\Ras) - e_\mathrm{pin}(\Ras)\right),
\end{multline}
where we have separated the left and right borders $\partial \Bas^{\mathrm{r,b}}$
of the bistable domain.  Due to continuity, the stable vortex energy
$e_\mathrm{pin}(\Ras)$ will be equal to $e_\mathrm{pin}^\mathrm{b}(\Ras)$ on
the left border $\partial \Bas^{\mathrm{r}}$ and equal to
$e_\mathrm{pin}^\mathrm{r}(\Ras)$ on the right border $\partial
\Bas^{\mathrm{b}}$.  The expression \eqref{eq:intermediate_flux_fpin} for
$\mathbf{F}^{<}_\mathrm{pin}$ then reduces to
\begin{align}\label{eq:fpin_frontal}
   \mathbf{F}^{<}_\mathrm{pin} &= n_p \int_{\partial \Bas^{\mathrm{b}}} 
   \frac{\mathrm{d}\, \mathbf{S}_\perp}{a_0^2} \left(e^\mathrm{b}_\mathrm{pin}(\Ras) 
   - e^\mathrm{r}_\mathrm{pin}(\Ras)\right)\nonumber\\
    &= n_p \int_{-\vas_{c}}^{\vas_{c}}\frac{\mathrm{d}\vas}{a_0}\,
    \frac{\Delta e_\mathrm{pin}(\vas)}{a_0} \left[1,-\partial{\uas}/\partial{\vas}\right]
   \nonumber\\
    &=n_p \left[\frac{2\vas_{c}}{a_0}\frac{\langle\Delta e_\mathrm{pin}\rangle}{a_0},\, 0\right]
    \equiv [F^{\parallel}_\mathrm{pin}, 0]
\end{align}
with $\langle\Delta e_\mathrm{pin}\rangle$ the average energy jump evaluated
along the $v$-direction. The force $\mathbf{F}^{<}_\mathrm{pin}$ is
aligned with the unstable directed along $u$, with the $v$-component vanishing
due to the antisymmetry in $\vas \leftrightarrow-\vas$ of the derivative
$\partial{\uas} /\partial{\vas}$, and is independent on $\theta$ for $|\theta|
< \theta^*$.

\subsubsection{Impact angle $|\theta| = \pi/2$}\label{sec:F_perp}

Second, let us find the pinning force density
$\mathbf{F}^{\pi/2}_\mathrm{pin}$ for vortices moving along the (positive)
$v$-direction, $\theta = \pi/2$. As follows from Fig.\ \ref{fig:bananas},
vortices occupy the blue branch and jump to the red one upon hitting the lower
half of the boundary $\partial\Bas^\mathrm{b}$; vortices that enter $\Bas$ but
do not cross $\partial\Bas^\mathrm{b}$ undergo no jump and hence do not
contribute to $\mathbf{F}^{\pi/2}_\mathrm{pin}$. As vortices in the red branch
proceed upwards, they jump back to the blue branch upon crossing the red
boundary $\partial\Bas^\mathrm{r}$. While jumps appear on all of the lower
half of $\partial\Bas^\mathrm{b}$, a piece of the upper boundary
$\partial\Bas^\mathrm{r}$ that contributes with a second jump is cut away (as
vortices to the left of $\uas^{\scriptscriptstyle (0)} +
\uas^{\scriptscriptstyle (1)}$ do not change branch from blue to red). The
length $\Delta\vas$ of this interval scales as $\Delta\vas/\vas_c \propto
(\kappa_m - 1)^{1/4}$; ignoring this small jump-free region, we determine
$\mathbf{F}^{\pi/2}_\mathrm{pin}$ assuming that vortices contributing to
$\mathbf{F}^{\pi/2}_\mathrm{pin}$ undergo a sequence of two jumps, from blue
to red on the lower half $\partial\Bas^\mathrm{b<}$ and back from red to blue
on the upper half $\partial\Bas^\mathrm{r>}$ of the boundary $\partial\Bas$.
Repeating the above analysis, we find that the $u$-components in
$\mathbf{F}^{\pi/2}_\mathrm{pin}$ arising from the blue and red boundaries now
cancel, while the $v$-components add up,
\begin{align}\label{eq:fpin_perp}
   \mathbf{F}^{\pi/2}_\mathrm{pin}
   &=n_p \int_{\partial \Bas^{\mathrm{b<}}} \frac{\mathrm{d}\, \mathbf{S}_\perp}{a_0^2} 
   \left(e^\mathrm{b}_\mathrm{pin}(\Ras) - e^\mathrm{r}_\mathrm{pin}(\Ras)\right)\nonumber\\
   &+n_p \int_{\partial \Bas^{\mathrm{r>}}} \frac{\mathrm{d}\, \mathbf{S}_\perp}{a_0^2}
   \left(e^\mathrm{r}_\mathrm{pin}(\Ras) - e^\mathrm{b}_\mathrm{pin}(\Ras)\right)\nonumber\\
   &= 2 n_p \int_0^{\vas_{c}} \frac{\mathrm{d}\vas}{a_0}\,
   \frac{\Delta e_\mathrm{pin}(\vas)}{a_0} \left[0,\partial{\uas}/\partial{\vas}\right]\\
   \nonumber
    &= n_p \left[0,\frac{2\vas_{c}}{a_0}\frac{\langle\Delta e_\mathrm{pin}
    \partial_{\vas} \uas \rangle}{a_0}\right] \equiv [0,F^{\perp}_\mathrm{pin}].
\end{align}

\begin{figure}[t]
        \includegraphics[width = 1.\columnwidth]{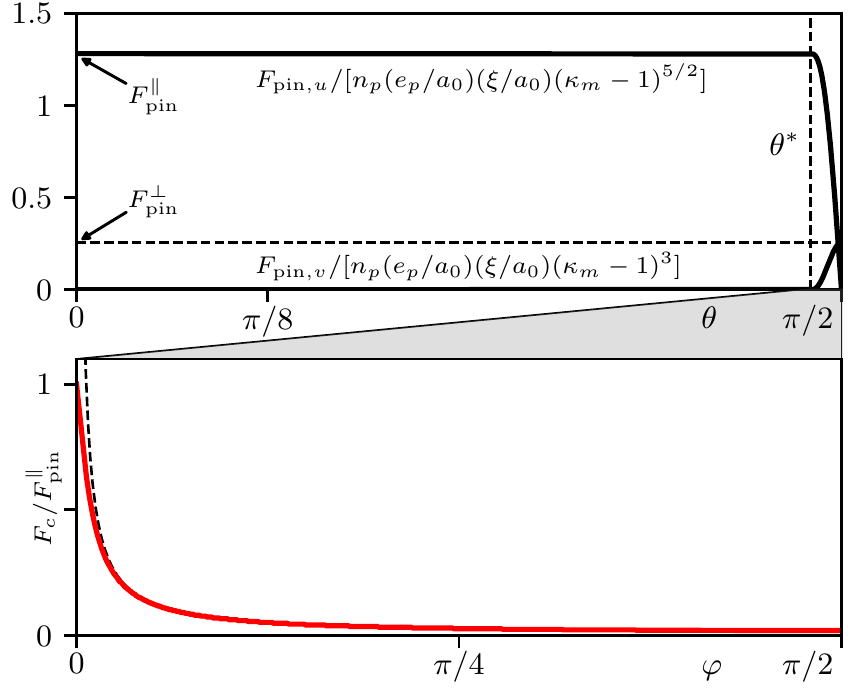}
	\caption{Top: scaled pinning force densities $F_{\mathrm{pin},u}$ and
	$F_{\mathrm{pin},v}$ versus impact angle $\theta$; we have used the
	same parameters as in Fig.\ \ref{fig:ellipses}. The longitudinal
	(along $u$) force $F_{\mathrm{pin},u}$ remains constant and equal to
	$F_\mathrm{pin}^\parallel$ for all angles $|\theta| < \theta^\ast$,
	while the transverse (along $v$) component $F_{\mathrm{pin},v}$
	vanishes in this regime.  The longitudinal force drops and vanishes
	over the narrow interval $\theta^\ast < |\theta| < \pi/2$, while the
	transverse force $F_{\mathrm{pin},v}$ increases up to
	$F_\mathrm{pin}^\perp$.  Bottom: critical force density $F_c$
	(directed along the Lorentz force $\mathbf{F}_{\rm \scriptscriptstyle
	L} = \mathbf{j} \wedge\mathbf{B}/c$) versus angle $\varphi$ of the
	Lorentz force; the dashed line shows the upper bound $F_c <
	F_\mathrm{pin}^\perp/ \sin(\varphi)$.}
    \label{fig:forces_angle}
\end{figure}

Making use of the result \eqref{eq:energy_jump} for $\Delta
e_\mathrm{pin}(\vas)$ in \eqref{eq:fpin_frontal}, we find explicit expressions
for the pinning force densities for impacts parallel and perpendicular to the
unstable direction $u$,
\begin{align}\label{eq:fpin_explicit_par}
   F_\mathrm{pin}^{\parallel} &\approx \left(\frac{9n_p}{8\,a_0^2\gamma^3}\right)
   \!\int_{-\vas_c}^{\vas_c} \!\!\!\!\!\! \mathrm{d}\vas 
   \left[\frac{\gamma\delta -\beta^2}{(1+\lambda_+/\Cbar)^2} 
   \left(\vas_c^2 - \vas^2\right)\right]^2
   \\ \nonumber
   &=\frac{24}{5} n_p \frac{\sqrt{2\Cbar/\gamma}}{a_0}
   \frac{\Cbar^2}{\gamma a_0} \frac{\gamma (1 + \lambda_+/\Cbar)}
   {\sqrt{\gamma\delta -\beta^2}} (\kappa_m - 1)^{5/2}
\end{align}
and
\begin{align}\label{eq:fpin_explicit_perp}
   F_\mathrm{pin}^\perp &\approx 
   3 \frac{\Cbar^2}{\gamma a_0} \frac{\gamma a/a_0}{\gamma\delta - \beta^2}(\kappa_m - 1)^3,
\end{align}
that confirm the scaling estimates of Eq.\ \eqref{eq:fpin_scaling}.  Here, we
have made use of the definition \eqref{eq:cusps} of $\vas_{c}$ and have
brought the final result into a form similar to the isotropic result
\eqref{eq:F_pin_iso_result} (with the length $\sqrt{\Cbar/\gamma}$ and the
force $\Cbar^2/\gamma a_0$, equal to $\xi/\sqrt{3}\kappa$ and $e_p/12\kappa^2$
for a Lorentzian potential).  The result \eqref{eq:fpin_explicit_par} provides
the pinning force density $\mathbf{F}_\mathrm{pin} =
[F_\mathrm{pin}^{\parallel},0]$ for all impact angles $|\theta| \leq
\theta^\ast$ (note that \eqref{eq:fpin_explicit_par} depends on the curvature
$a$ of the crescent via $\delta$, Eq.\ \eqref{eq:delta}, that involves $a^2$
only, but higher-order corrections will introduce an asymmetry between left-
and right moving vortices). Within the interval $\theta^\ast < \theta <
\pi/2$, the longitudinal force $F_{\mathrm{pin},u}$ along $u$ decays to zero
and the transverse force $F_{\mathrm{pin},v}$ along $v$ becomes finite,
assuming the value \eqref{eq:fpin_explicit_perp} at $\theta = \pi/2$. The two
force components have been evaluated numerically over the entire angular
regime and the results are shown in Fig.\ \ref{fig:forces_angle}: when moving
away from the angle $\theta = \pi/2$, the transition from the blue to the red
boundary is moving upwards, with the relevant boundary turning fully blue at
$\theta = \theta^\ast$, thus smoothly transforming \eqref{eq:fpin_perp} into
\eqref{eq:fpin_frontal} (we have adopted the approximation of dropping the
jump-free interval $\Delta \vas$ that moves up and becomes smaller as $\theta$
decreases from $\pi/2$ to $\theta^\ast$).

\subsubsection{Anisotropic critical force density $\mathbf{F}_c$}\label{sec:F_c}

When the vortex system is subjected to a current density $\mathbf{j}$, the
associated Lorentz force $\mathbf{F}_{\rm \scriptscriptstyle L}(\varphi) =
\mathbf{j} \wedge \mathbf{B}/c$ directed along $\varphi$ pushes the vortices
across the defects. When $\mathbf{F}_{\rm \scriptscriptstyle L}$ is directed
along $u$, we have $\mathbf{F}_\mathrm{pin} = [F_\mathrm{pin}^\parallel,0]$
and the vortex system gets immobilized at force densities $F_{\rm
\scriptscriptstyle L} < F_c = F_\mathrm{pin}^\parallel$ (or associated current
densities $\mathbf{j}_c$). When $\mathbf{F}_{\rm \scriptscriptstyle L}$ is
directed away from $u$, the driving component along $v$ has to be compensated
by a finite pinning force $F_{\mathrm{pin},v}$ that appears only for angles
$\theta^\ast < \theta < \pi/2$. Hence, the angles of force and motion,
$\varphi$ associated with the Lorentz force $\mathbf{F}_{\rm
\scriptscriptstyle L}(\varphi)$ and $\theta$ providing the direction of the
pinning force $\mathbf{F}_\mathrm{pin}(\theta)$, are different. We find them,
along with the critical force density $\mathbf{F}_c(\varphi)$, by solving the
dynamical force equation \eqref{eq:macroscopic_force_balance} at vanishing
velocity $\mathbf{v} = 0$,
  \begin{equation}\label{eq:Lor-pin-force}
   \mathbf{F}_c(\varphi) = \mathbf{F}_\mathrm{pin}(\theta)
\end{equation}
resulting in a critical force density
  \begin{equation}\label{eq:Fc}
   F_c(\varphi) = \sqrt{F_{\mathrm{pin},u}^2(\theta) + F_{\mathrm{pin},v}^2(\theta)}
\end{equation}
with angles $\varphi$ and $\theta$ related via
  \begin{equation}\label{eq:varphi}
   \tan \varphi = \frac{F_{\mathrm{pin},u}(\theta)}{F_{\mathrm{pin},v}(\theta)}.
\end{equation}
Since $F_{\mathrm{pin},u}(\theta< \theta^\ast) = 0$, the entire interval
$\theta < \theta^\ast$ is compressed to $\varphi = 0$ and it is the narrow
regime $\theta^\ast < \theta < \pi/2$ that determines the angular
characteristic of the critical force density $F_c(\varphi)$.  The critical
force density $F_c(\varphi)$ is peaked at $\varphi = 0$ as shown in Fig.\
\ref{fig:forces_angle} (with a correspondingly sharp peak in $j_c$ at right
angles). Combing Eqs.\ \eqref{eq:Fc} and \eqref{eq:varphi}, we can derive a
simple expression bounding the function $F_c(\varphi)$,
  \begin{equation}\label{eq:bound_Fc}
   F_c(\varphi) = F_{\mathrm{pin},v}(\theta)\sqrt{1+\cot^2(\varphi)} \leq
   \frac{F_\mathrm{pin}^\perp}{\sin(\varphi)},
\end{equation}
that traces $F_c(\varphi)$ over a wide angular region, see the dashed line in
Fig.\ \ref{fig:forces_angle}. At small values of $\varphi$ we cannot ignore
the angular dependence in $F_{\mathrm{pin},v}(\theta)$ any more that finally
cuts off the divergence $\propto 1/\sin(\varphi)$ at the value $F_c(\varphi
\to 0) \to F_\mathrm{pin}^\parallel$.

\subsubsection{Isotropized pinning force density $F_\mathrm{pin}$}\label{sec:F_ang-av}

In a last step, we assume an ensemble of equal anisotropic defects that are
uniformly distributed in space and randomly oriented. In this situation, we
have to perform an additional average over the instability directions
$\hat{\mathbf{u}}_i$ associated with the different defects $i = 1, \dots N$.
Neglecting the modification of $\mathbf{F}_\mathrm{pin}(\theta)$ away from
$[F_\mathrm{pin}^\parallel,0]$ in the small angular regions $\theta^\ast <
|\theta| < \pi/2$, we find that the force along any direction
$\hat{\mathbf{R}}$ has the magnitude
\begin{eqnarray}\label{eq:av_force}
   F_\mathrm{pin} &\approx& \frac{1}{N}\sum_{i=1}^N |(F_\mathrm{pin}^\parallel 
   \hat{\mathbf{u}}_i) \cdot \hat{\mathbf{R}}|
   \\ \nonumber
   &\approx& F^{\parallel}_\mathrm{pin} \int_{-\pi/2}^{\pi/2}
   \frac{\mathrm{d}\theta}{\pi} \, \cos\theta = \frac{2}{\pi} \Fpin^\parallel.
\end{eqnarray}
As a result of the averaging over the angular directions, the pinning force
density is now effectively isotropic and directed against the velocity
$\mathbf{v}$ of the vortex motion.

\section{Uniaxial defect}\label{sec:uniax_defect}
In Sec.\ \ref{sec:arb_shape}, we have analyzed the onset of strong pinning for
an arbitrary potential and have determined the shape of the unstable and
bistable domains $\Uti$ and $\Bas$---with their elliptic and crescent forms,
they look quite different from their ring-shaped counterparts for the
isotropic defect in Figs.\ \ref{fig:f_pin}(c) and (d). In this section, we
discuss the situation for a weakly anisotropic defect with a small uniaxial
deformation quantified by the small parameter $\epsilon$ in order to
understand how our previous findings, the results for the isotropic defect and
those describing the strong-pinning onset, relate to one another. 

Our weakly deformed defect is described by equipotential lines that are nearly
circular but slightly elongated along $y$, implying that pinning is strongest
in the $x$-direction.  We will find that the unstable (bistable) domain $\Uti$
($\Bas$) for the uniaxially anisotropic defect starts out with two ellipses
(crescents) on the $x$-axis as $\kappa_m$ crosses unity. With increasing
pinning strength, i.e., $\kappa_m$, these ellipses (crescents) grow and deform
to follow the equipotential lines, with the end-points approaching one another
until they merge on the $\pm y$-axis. These merger points, we denote them as
$\Rti_s$ and $\Ras_s$, define a second class of important points (besides the
onset points $\Rti_m$ and $\Ras_m$) in the buildup of the strong pinning
landscape: while the onset points $\Rti_m$ are defined as minima of the
Hessian determinant $D(\Rti)$, the merger points $\Rti_s$ turn out to be
associated with saddle points of $D(\Rti)$.  Pushing across the merger of the
deformed ellipses (crescents) by further increasing the Labusch parameter
$\kappa_m$, the unstable (bistable) domains $\Uti$ ($\Bas$) undergo a change
in topology, from two separated areas to a ring-like geometry as it appears
for the isotropic defect, see Figs.\ \ref{fig:f_pin}(c) and (d), thus
explaining the interrelation of our results for isotropic and anisotropic
defects.

With this analysis, we thus show how the strong pinning landscape for the
weakly uniaxial defect will finally assume the shape and topology of the
isotropic defect as the pinning strength $\kappa_m$ overcomes the anisotropy
$\epsilon$. Second, this discussion will introduce the merger points $\Rti_s$ as a
second type of characteristic points of strong pinning landscapes that we will
further study in section \ref{sec:hyp_expansion} using a Landau-type expansion as
done in section \ref{sec:ell_expansion} above; we will find that the geometry of
the merger points $\Rti_s$ is associated with hyperbolas, as that of the onset points
was associated with ellipses.

Our uniaxially anisotropic defect is described by the stretched (along the
$y$-axis) Lorentzian
\begin{equation}\label{eq:uniax_potential_formal}
   e_p(\xti,\yti) = -e_p\left(1+\frac{\xti^2}{2\xi^2} 
   + \frac{\yti^2}{2\xi^2\left(1 + \epsilon\right)^2}\right)^{-1},
\end{equation}
with equipotential lines described by ellipses
\begin{equation}\label{eq:equipotential_lines}
   \frac{\xti^2}{\xi^2} + \frac{\yti^2}{\xi^2\left(1 + \epsilon\right)^2} = \text{const},
\end{equation} 
and the small parameter $0 < \epsilon \ll 1$ quantifying the degree of
anisotropy. At fixed radius $\rti^2 = \xti^2 + \yti^2$, the potential
\eqref{eq:uniax_potential_formal} assumes maxima in energy and in negative
curvature on the $x-$axis, and corresponding minima on the $y-$axis.  Along
both axes, the pinning force is directed radially towards the origin
and the Labusch criterion \eqref{eq:gen_Lab} for strong pinning is determined
solely by the curvature along the radial direction.  At the onset of strong
pinning, the unstable and bistable domains then first emerge along the $x-$axis
at the points $\Rti_m=(\pm\sqrt{2}\xi,0)$ and $\Ras_m = (\pm 2\sqrt{2}\xi,0)$ 
when
\begin{equation}
  \kappa_m = \frac{e_p}{4\Cbar\xi^2} = 1.
\end{equation}
Upon increasing the pinning strength $\kappa_m$, e.g., via softening of the
vortex lattice as described by a decrease in $\Cbar$, the unstable and
bistable domains $\Uti$ and $\Bas$ expand away from these points, and
eventually merge along the $y-$axis at $\Rti_s = (0, \pm
\sqrt{2}\xi(1+\epsilon))$, $\Ras_s = (0, \pm 2\sqrt{2}\xi(1+\epsilon))$ when
\begin{equation}\label{eq:uniax_merging}
   \kappa_s = \frac{e_p}{4\Cbar\xi^2(1+\epsilon)^2} = \frac{\kappa_m}{(1+\epsilon)^2} = 1,
\end{equation}
i.e., for $\kappa_m = (1 + \epsilon)^2$.  The evolution of the strong pinning
landscape from onset to merging takes place in the interval $\kappa_m
\in [1, (1+ \epsilon)^2]$; pushing $\kappa_m$ beyond this interval,
we will analyze the change in topology and appearance of non-simply connected
unstable and bistable domains after the merging.

The quantity determining the shape of the unstable domain $\Uti$ is the
Hessian determinant $D(\Rti)$ of the total vortex energy $\epin(\Rti;\Ras)$,
see Eqs.\ \eqref{eq:det_Hessian} and \eqref{eq:en_pin_tot}, respectively.  At
onset, the minimum of $D(\Rti)$ touches zero for the first time; with
increasing $\kappa_m$, this minimum drops below zero and the condition
$D(\Rti) = 0$ determines the unstable ellipse that expands in $\Rti$-space.
Viewing the function $D(\Rti)$ as a height function of a landscape in the
$\Rti$ plane, this corresponds to filling this landscape, e.g., with water, up
to the height level $D = 0$ with the resulting lake representing the unstable
domain. In the present uniaxially symmetric case, a pair of unstable ellipses
grow simultaneously, bend around the equipotential line near the radius $\sim
\sqrt{2}\xi$ and finally touch upon merging on the $y$-axis. In our geometric
interpretation, this corresponds to the merging of the two (water-filled)
valleys that happens in a saddle-point of the function $D(\Rti)$ at the height
$D = 0$. Hence, the merger point $\Rti_s$ correspond to saddles in $D(\Rti)$
with
\begin{equation}\label{eq:det_saddle2Da}
   D(\Rti_s) = 0,\quad \mathbf{\nabla}_{\Rti}\,D(\mathbf{R})\big|_{\Rti_s} = 0,
\end{equation}
and
\begin{equation}\label{eq:det_saddle2Db}
   \mathrm{det}\bigl[\mathrm{Hess}\bigl[ D(\Rti) \bigr]\bigr] \big|_{\Rti_s} < 0,
\end{equation}
cf.\ Eq.\ \eqref{eq:det_min2D}. 

In our calculation of $D(\Rti)$, we exploit that the Hessian in
\eqref{eq:det_Hessian} does not depend on the asymptotic position $\Ras$ and
we can set it to zero,
\begin{align}
   D(\Rti) 
   &=\det\bigl\{\mathrm{Hess}[\Cbar\rti^2/2 + e_{p}^{\scriptscriptstyle (i)}(\rti) 
   + \delta e_p(\Rti)]\bigr\},
\end{align}
where we have split off the anisotropic correction $\delta
e_p(\Rti) = e_p(\Rti) - e_p^{\scriptscriptstyle (i)}(\rti)$ away from the
isotropic potential $e_p^{\scriptscriptstyle (i)}(\rti)$ with $\epsilon = 0$.
In the following, we perform a perturbative analysis around the isotropic
limit valid in the limit of weak anisotropy $\epsilon \ll 1$; this motivates
our use of polar (tip) coordinates $\rti$ and $\pti$.

The isotropic contribution $\mathrm{H}^{\scriptscriptstyle (i)}$ to the
Hessian matrix $\mathrm{H}$ is diagonal with components
\begin{align}\nonumber
   \mathrm{H}^{\scriptscriptstyle (i)}_{\tilde{R}\tilde{R}}(\rti)
   &\equiv \partial_{\rti}^2 [\Cbar\rti^2/2 + e_{p}^{\scriptscriptstyle (i)}(\rti)]\\
   \label{eq:uniax_Hrr}
   &=\Cbar + \partial_{\rti}^2 e_{p}^{\scriptscriptstyle (i)}(\rti)
\end{align}
and 
\begin{align} \nonumber
   \mathrm{H}^{\scriptscriptstyle (i)}_{\pti\pti}(\rti)
   &\equiv(\rti^{-2}\partial^2_{\pti\pti} + \tilde{R}^{-1}\partial_{\rti})
   [\Cbar\rti^2/2 + e_p^{\scriptscriptstyle (i)}(\rti)] \\
   &= \Cbar - f_p^{\scriptscriptstyle (i)}(\rti)/\rti.
   \label{eq:uniax_Hpp}
\end{align}
The radial component $\mathrm{H}^{\scriptscriptstyle
(i)}_{\tilde{R}\tilde{R}}\propto(\kappa_m - 1)$ vanishes at onset, while
$\mathrm{H}^{\scriptscriptstyle (i)}_{\pti\pti}$ remains finite, positive, and
approximately constant.

\begin{figure}[t]
        \includegraphics[width = 1.\columnwidth]{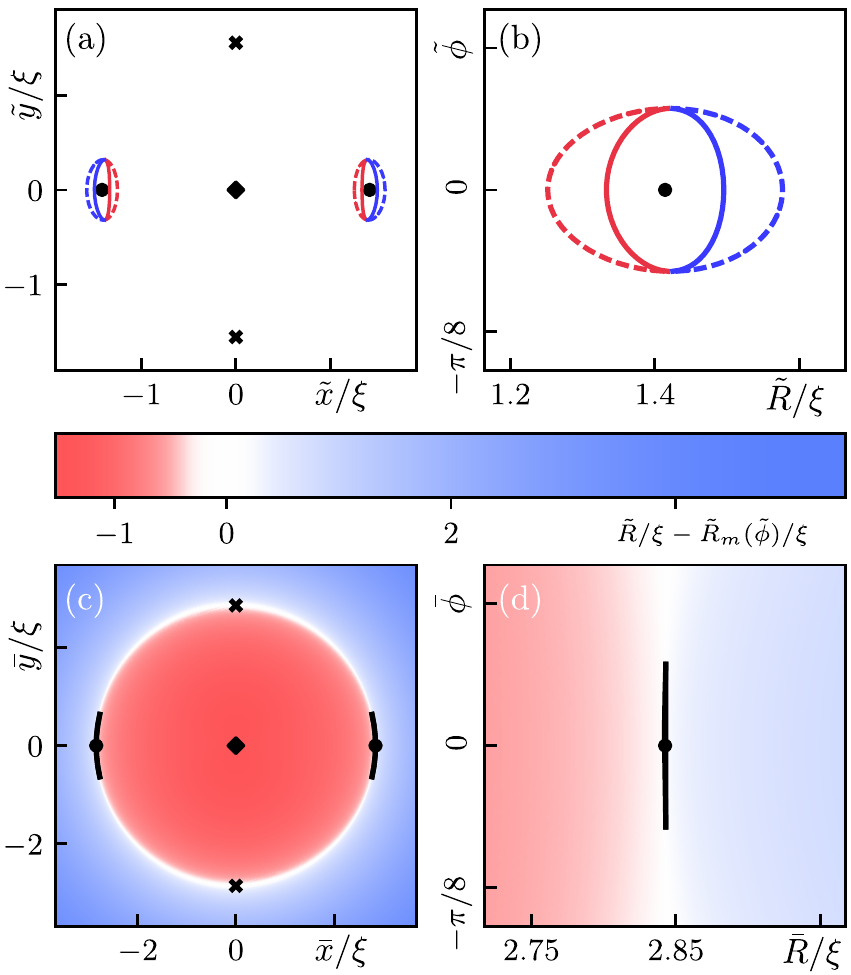}
        \caption{Unstable and bistable domains close to the onset of strong
        pinning for a uniaxial defect \eqref{eq:uniax_potential_formal}
        centered at the origin, with $\epsilon = 0.1$ and $\kappa_m - 1
        =0.01$. The pinning potential is steepest at angles $\pti = 0,\, \pi$
        and least steep at $\pti =\pm\pi/2$, hence strong pinning is realized
        first in a small interval around $\pti = 0,\, \pi$ (solid black dots)
        where $\kappa_m(\pti) \geq 1$. (a) The unstable domain $\Uti$ in tip
        space is bounded by red/blue solid lines (jump lines $\Jti$, see Eq.\
        \eqref{eq:uniax_Rjp}); dashed lines mark the associated landing lines
        $\Lti$, see \eqref{eq:uniax_Rlp}.  (b) Focus on the unstable domain near
        $\pti = 0$ in polar coordinates $\tilde{R}$ and $\pti$. The jumping
        (solid) and landing (dashed) lines have the approximate shape of
        ellipses, see Eq.\ \eqref{eq:ellipse-small}, in agreement with our
        analysis of Sec.\ \ref{sec:Uti}.  (c) The bistable domain $\Bas$ in
        asymptotic space involves symmetric crescents centered at $\pas = 0,\,
        \pi$ and a narrow width $\propto (\kappa_m(\pas) - 1)^{3/2}$, see Eq.\
        \eqref{eq:uniax_R_bas}, in agreement with the analysis of Sec.\
        \ref{sec:Bas}.  (d) Focus on the bistable domain at $\pas = 0$ in
        polar coordinates $\bar{R}$ and $\pas$.  Red/blue colors indicate
        different vortex configurations as quantified through the order
        parameter $\rti - \rti_m(\pti)$.}
    \label{fig:quasi-ring-ellipse}
\end{figure}

The anisotropic component $\delta e_p(\Rti)$ introduces corrections
$\propto\epsilon$; these significantly modify the radial entry of the full
Hessian while leaving its azimutal component $\mathrm{H}_{\pti\pti}$
approximately unchanged; the off-diagonal entries of the full Hessian scale as
$\epsilon$ and hence contribute in second order of $\epsilon$ to $D(\Rti)$.
As a result, the sign change in the determinant
\begin{equation}
   D(\Rti) \approx \mathrm{H}_{\tilde{R}\tilde{R}}(\Rti)
                   \mathrm{H}_{\pti\pti}(\rti)
   + \mathcal{O}\left(\epsilon^2\right),
\end{equation}
is determined by 
\begin{equation}
   \mathrm{H}_{\tilde{R}\tilde{R}}(\Rti) 
   = \mathrm{H}^{\scriptscriptstyle (i)}_{\tilde{R}\tilde{R}}(\rti) 
   + \partial^2_{\rti}\delta e_p(\Rti)
\end{equation}
for radii close to $\rti_m$ with $\delta\rti = \rti - \rti_m \approx
\mathcal{O}(\sqrt{\kappa_m - 1})$.  We expand the potential
\eqref{eq:uniax_potential_formal} around the isotropic part
$e_p^{\scriptscriptstyle (i)}(\rti)$,
\begin{equation}\label{eq:dep_quad}
   \delta e_p(\Rti) \approx -\epsilon\,[\partial_{\rti}
   e_p^{\scriptscriptstyle (i)}(\rti)]\rti\sin^2\pti,
\end{equation}
and additionally expand both $e_p^{\scriptscriptstyle (i)}(\rti)$ and $\delta
e_p(\Rti)$ around $\rti_m$, keeping terms $\propto \epsilon\,
\sqrt{(\kappa_m-1)}$. The radial entry of the anisotropic Hessian matrix then
assumes the form
\begin{multline}
   \mathrm{H}_{\tilde{R}\tilde{R}}(\Rti)\approx \Cbar \, [1-\kappa_m(\pti)] \\
   + \gamma \, [ \delta\rti^2/2 - \epsilon\,\sin^2{\pti}\,\rti_m \delta\rti]
\end{multline}
with $\gamma = \partial^4_{\rti}e_p^{\scriptscriptstyle (i)}(\rti)|_{\rti_m}$
and the angle-dependent Labusch parameter
\begin{equation}\label{eq:kappa_quad}
   \kappa_m(\pti) \equiv \frac{\max_{\rti}[-\partial_{\rti}^2 e_p(\rti,\pti)|_{\pti}]}{\Cbar} 
   = \kappa_m - 2\epsilon\sin^2\pti.
\end{equation}
The edges of the unstable region $\Uti$ then can be obtained by imposing the
condition $\mathrm{H}_{\tilde{R}\tilde{R}}(\Rti) = 0$ and the solution to the
corresponding quadratic equation define the jump positions
$\tilde{R}_\mathrm{jp}(\pti)$ (or boundaries $\partial\Uti$)
\begin{align}\label{eq:uniax_Rjp}
   \tilde{R}_\mathrm{jp}(\pti) \approx \rti_m(\pti) \pm \delta \tilde{R}(\pti).
\end{align}
These are centered around the (`large') ellipse defined by
\begin{equation}\label{eq:ellipse-large}
   \rti_m(\pti) = \rti_m (1 +\epsilon\sin^2\pti)
\end{equation}
and separated by (cf.\ Eq.\ \eqref{eq:ujp})
\begin{align}\label{eq:uniax_drmax}
   2\, \delta \tilde{R}(\pti) = \sqrt{\frac{8\Cbar}{\gamma}(\kappa_m(\pti) - 1)}
\end{align}
along the radius. Making use of the form \eqref{eq:kappa_quad} of
$\kappa_m(\pti)$ and assuming a small value of $\kappa_m > 1$ near onset, we
obtain the jump line in the form of a (`small') ellipse centered at $[\pm
\rti_m,0]$,
\begin{equation}\label{eq:ellipse-small}
   \gamma\, \delta \tilde{R}^2 + \epsilon \Cbar\,\pti^2 = \Cbar(\kappa_m - 1).
\end{equation}
\begin{figure}[t]
        \includegraphics[width = 1.\columnwidth]{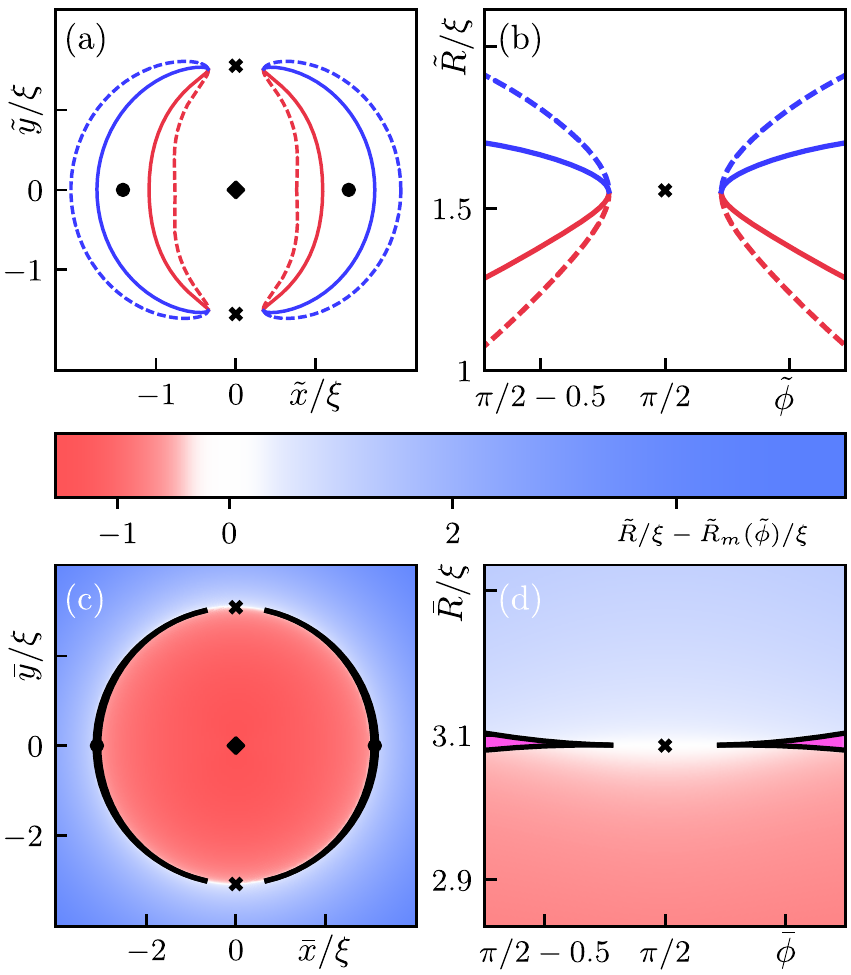}
        \caption{Unstable and bistable domains before merging for a uniaxial
        defect \eqref{eq:uniax_potential_formal} centered at the origin, with
        $\epsilon = 0.1$ and $1 - \kappa_s  \approx 0.01$. Strong pinning is
        realized everywhere but in a small interval around $\pti = \pm\pi/2$
        where $\kappa_m(\pti) < 1$. (a) The unstable domain $\Uti$ in the tip
        plane is bounded by the solid red/blue jump lines $\Jti$, see Eq.\
        \eqref{eq:uniax_Rjp}  and involves two strongly bent ellipses
        originating from angles $\pti = 0,\, \pi$ (black dots) and approaching
        one another close to $\pti =\pm\pi/2$ (black crosses); red/blue dashed
        lines are landing points as given by Eqs.\ \eqref{eq:uniax_Rlp}.  (b)
        Focus (in polar coordinates $\tilde{R},\, \pti$) on the tips of the
        unstable domain near $\pti = \pi/2$.  (c) The bistable domain $\Bas$
        in the asymptotic space consists of thin symmetric crescents (colored
        in magenta) originating from $\pas = 0,\, \pi$, with the delimiting
        black solid lines given by Eq.\ \eqref{eq:uniax_R_bas}. (d) Focus on
        the cusps of the bistable domain close to $\pas = \pi/2$ in polar
        coordinates $\bar{R},\,\pas$.  Red/blue colors indicate different
        vortex configurations as quantified through the order parameter $\rti
        - \rti_m(\pas)$.}
    \label{fig:quasi-ring-hyperbola}
\end{figure}

Hence, we find that the anisotropic results are obtained from the isotropic
ones by replacing the circle $\rti_m$ by the ellipse $\Rti_m(\pti)$ and
substituting $\kappa \to \kappa_m(\pti)$ in the width \eqref{eq:ujp}, see
Figs.\ \ref{fig:quasi-ring-ellipse}(a) and (b) evaluated for small values
$\kappa_m - 1 = 0.01$ and $\epsilon = 0.1$.  

Analogously, the boundaries of the bistable domain $\Bas$ can be found by
applying the same substitutions to the result \eqref{eq:bs_pos}, see Figs.\
\ref{fig:quasi-ring-ellipse}(c) and (d),
\begin{align}\label{eq:uniax_R_bas}
   \bar{R}(\pas) \approx \ras_m(\pas) \pm \delta \bar{R}(\pas) 
\end{align}
with $\ras_m(\pas) = \ras_m (1 +\epsilon\sin^2\pas)$ and the width
\begin{align}\label{eq:uniax_dras}
  2\, \delta \bar{R}(\pas) = \frac{2}{3}\sqrt{\frac{8\Cbar}{\gamma}}(\kappa_m(\pti) - 1)^{3/2}.
\end{align}

The landing line $\mathcal{L}_{\Rti}$ is given by (see Eq.\ \eqref{eq:j_dist}
and note that the jump point is shifted by $\ujp$ away from $\xti_m$, see Eq.\
\eqref{eq:jp_pos}) 
\begin{align}\label{eq:uniax_Rlp}
   \tilde{R}_\mathrm{lp}(\pti) \approx \rti_m(\pti) \mp 2\,\delta \tilde{R}(\pti).
\end{align}

An additional complication is the finite angular extension of the unstable and
bistable domains $\Uti$ and $\Bas$; these are limited by the condition
$\kappa_m (\phi_{\max})=1$, providing us with the constraint
\begin{equation}
  \pti_{\max}  = \pas_{\max} \approx \pm \sqrt{\frac{\kappa_m-1} {2\epsilon}}
\end{equation}
near the strong pinning onset with $(\kappa_m-1)\ll\epsilon$. The resulting
domains $\Uti$ have characteristic extensions of scale
$\propto\sqrt{\kappa_m-1}$, see Fig.\ \ref{fig:quasi-ring-ellipse}.

\begin{figure}[t]
        \includegraphics[width = 1.\columnwidth]{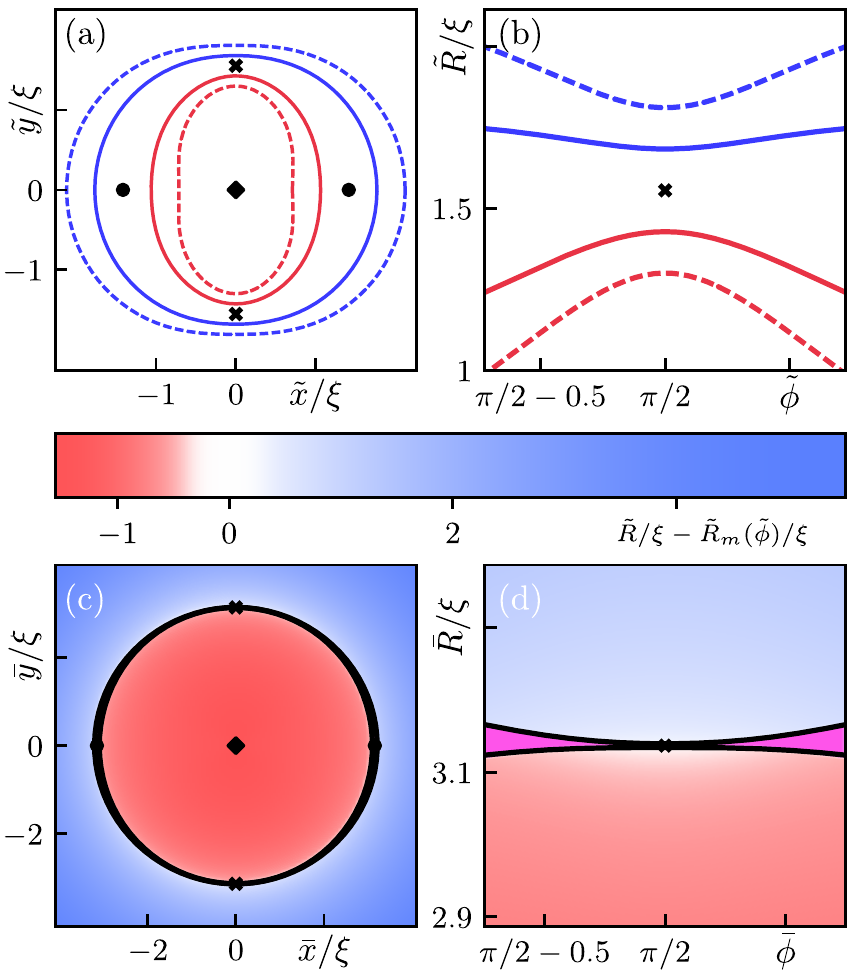}
        \caption{Unstable and bistable domains for a uniaxial defect
        \eqref{eq:uniax_potential_formal} after merging, with $\epsilon = 0.1$
        and $\kappa_s - 1 \approx 0.01$. (a) The unstable domain $\Uti$ in tip
        plane is enclosed between the jump lines $\Jti$ (solid red/blue, see Eq.\
        \eqref{eq:uniax_Rjp}) and takes the shape of a deformed ring with a
        wider (narrower) width at strongest (weakest) pinning near the solid
        dots (crosses).  Red/blue dashed lines mark the landing positions
        $\Lti$ of the vortex tips and are given by Eq.\ \eqref{eq:uniax_Rlp}.
        (b) Focus on the narrowing in the unstable domain close to the merger
        points (crosses) at $\pti = \pi/2$ in the polar coordinates
        $\tilde{R}, \, \pti$.  (c) The bistable domain $\Bas$ in asymptotic
        space is a narrow ring (colored in magenta) thicker (thinner) at points of
        strongest (weakest) pinning near $\pas = 0,\ \pi$ ($\pas = \pm\pi/2$);
        black lines correspond to Eq.\ \eqref{eq:uniax_R_bas}.  (d) Focus on
        the constriction in the bistable domain close to $\pas = \pi/2$ in
        polar coordinates $\bar{R}, \, \pas$. Red/blue colors indicate
        different vortex configurations as quantified through the order parameter
        $\rti - \rti_m(\pas)$.}
    \label{fig:quasi-ring-hyperbola-merged}
\end{figure}

Close to merging (marked by crosses in the figure) at $\phi = \pm\pi/2$, we
define the deviation $\delta\phi  = \pi/2 - \phi$ with $\delta\phi \ll 1$, and
imposing the condition $\kappa_m(\phi_{\max})=1$, we find
\begin{equation}
  \delta\pti_{\max} = \delta\pas_{\max} \approx \sqrt{1 -\frac{\kappa_m-1} {2\epsilon}}
  \approx \sqrt{\frac{1 - \kappa_s}{2\epsilon}}.
\end{equation}
The corresponding geometries of $\Uti$ and $\Bas$ are shown in Fig.\
\ref{fig:quasi-ring-hyperbola} for $1 -\kappa_s\approx 0.01$ and $\epsilon =
0.1$.  Finally, $\delta\pti_{\max}$ vanishes at merging for $\kappa_s = 1$ (or
$\kappa_m -1 \approx 2\epsilon$), in agreement, to order $\epsilon$, with the
exact result \eqref{eq:uniax_merging}.

Pushing the Labusch parameter beyond the merger with $\kappa_s > 1$ or
$\kappa_m > (1+\epsilon)^2 \approx 1 + 2\epsilon$, the unstable and bistable
regimes $\Uti$ and $\Bas$ change their topology: they develop a (non-simply
connected) ring-like geometry with separated inner and outer edges that are a
finite distance apart in the radial direction at all angles $\pti$ and $\pas$.
The situation after the merger is shown in Fig.\
\ref{fig:quasi-ring-hyperbola-merged} for $\kappa_s - 1 \approx 0.01$ and
$\epsilon = 0.1$, with the merging points $\Rti_s$ and $\Ras_s$ marked by
crosses.

The merging of the unstable domains at the saddle point $\Rti_s$ is a general
feature of irregular pinning potentials.  In the next section, we will analyze
the behavior of the unstable domains close to a saddle point $\Rti_s$ of the
Hessian determinant $D(\Rti)$ and obtain a universal description of their
geometry close to this point. We will see that the geometry associated with
this merger is of a hyperbolic type described by $\gamma \uti^2 + \delta
\vti^2 = 2\Cbar (\kappa_s -1)$, $\gamma >0$ and $\delta < 0$ (assuming no
skew). The change in topology then is driven by the sign change in $\kappa_s -
1$: before merging, $\kappa_s < 1$, the hyperbola is open along the
unstable (radial) direction $\uti$, thus separating the two unstable regions,
while after merging, $\kappa_s > 1$, the hyperbola is open along the
transverse direction $\vti$, with the ensuing passage defining the single,
non-simply connected, ring-like unstable region.

\section{Merger points}\label{sec:merger}
The merging of unstable and bistable domains is a general feature of irregular
pinning potentials that is relevant beyond the simple example of a weakly
anisotropic uniaxial defect discussed above.  Indeed, while the exact
geometries of $\Uti$ and $\Bas$ depend on the precise shape of the pinning
potential, their behavior close to merging is universal. Below, we will study
this universal behavior by generalizing the expansions of Sec.\
\ref{sec:arb_shape} to saddle points $\Rti_s$ of the determinant $D(\Rti)$.
As with the onset of strong pinning, the merger of two domains induces a
change in topology in the unstable and bistable domains; we will discuss these
topological aspects of onsets and mergers in Secs.\ \ref{sec:topology_hyp} and
\ref{sec:2D_landscape} below.

\subsection{Expansion near merger}\label{sec:hyp_expansion}
Following the strategy of Sec.\ \ref{sec:arb_shape}, we expand the energy
functional around a saddle point $\Rti_s$ of the determinant $D(\Rti)$ in
order to obtain closed expressions for the unstable and bistable domains at
merging.  In doing so, we again define local coordinate systems $(\uti,\vti)$
and $(\uas,\vas)$ in tip- and asymptotic space centered at $\Rti_s$ and
$\Ras_s$, where the latter is associated with $\Rti_s$ through the force
balance equation \eqref{eq:gen_force_balance} in the original laboratory
system.  Furthermore, we fix our axes such that $D(\Rti_s)$ is a local maximum
along the (unstable) $u$- and a local minimum along the (stable) $v$-direction
of the saddle; the mixed term $\propto \uti\vti$ is absent from the expansion
(as the Hessian matrix is symmetric).  Furthermore, the vanishing slopes at
the saddle point, see \eqref{eq:det_saddle2Da}, imply the absence of terms
$\propto \uti^3$ and $\propto \uti^2\vti$ in the expansion and dropping
higher-order terms (corresponding to double-primed terms in
\eqref{eq:e_pin_expans_ani_orig}), we arrive to the expression
\begin{align}\label{eq:e_pin_expans_hyp}
  &e_\mathrm{pin}(\Rti; \Ras) = 
  \frac{\Cbar}{2} (1-\kappa_s) \, \uti^2 
  + \frac{\Cbar + \lambda_{+,s}}{2}\, \vti^2 +\frac{a_s}{2}\, \uti \vti^2  \nonumber \\
  &\quad+\frac{\alpha_s}{4}\, \uti^2\vti^2
  +\frac{\beta_s}{6}\, \uti^3\vti
  +\frac{\gamma_s}{24}\, \uti^4
  -\Cbar\uas\uti - \Cbar\vas \vti,
\end{align}
with $\kappa_s \equiv -\lambda_-(\Rti_s)/\Cbar,\ \lambda_{+,s} \equiv
\lambda_+(\Rti_s)$ and the remaining coefficients defined in analogy to Eq.\
\eqref{eq:e_pin_expans_ani}.

The most important term in the expansion \eqref{eq:e_pin_expans_hyp} is the
curvature term $\Cbar (1-\kappa_s)\, \uti^2 /2$ along the unstable direction
$u$.  As before in Sec.\ \ref{sec:Uti}, see Eq.\ \eqref{eq:e_pin_expans_ani},
the coefficient $(1 - \kappa_s)$ changes sign at some value of the pinning
strength and will serve as the small parameter in our considerations. The
higher-order terms in the expansion \eqref{eq:e_pin_expans_hyp} are
constrained by the saddle condition \eqref{eq:det_saddle2Db}, implying that
(cf.\ \eqref{eq:Hess_D} and \eqref{eq:detHD})
\begin{equation}\label{eq:det_saddle2D_explicit}
   \gamma_s\delta_s - \beta_s^2 < 0
\end{equation}
with
\begin{equation}\label{eq:delta_s}
   \delta_s \equiv \alpha_s - \frac{2 a_s^2}{\Cbar + \lambda_{+,s}}
\end{equation}
(for the saddle point there is no condition on the trace of the Hessian).  The
mapping of the two-dimensional pinning energy \eqref{eq:e_pin_expans_hyp} to
an effective one-dimensional Landau theory \eqref{eq:eff_landau_1D_hyp} of the
van der Waals kind is discussed in Appendix \ref{sec:eff_1D_merging}, both
before and after merging.

\subsection{Unstable domain $\Uti$}\label{sec:Uti_merger}
\begin{figure}[t]
        \includegraphics[width = 1.\columnwidth]{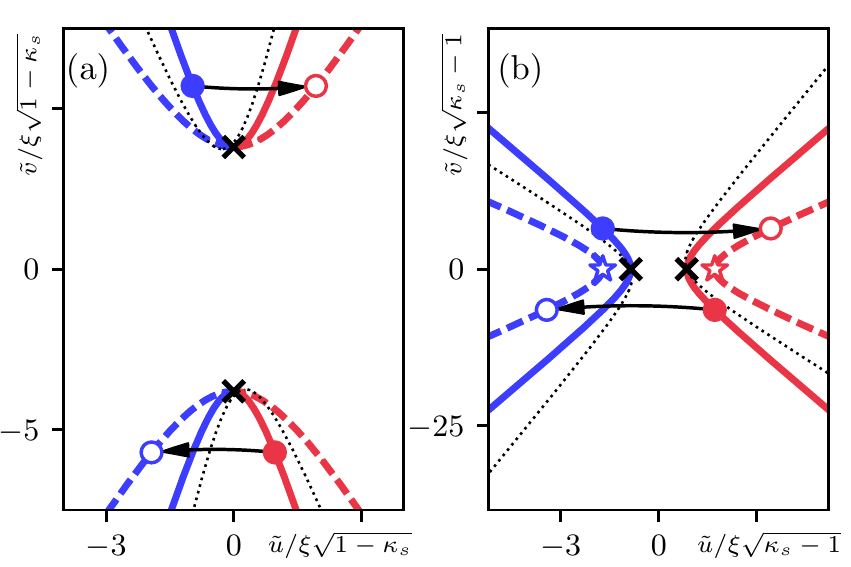}
	\caption{Jump lines $\Jti$ (solid red/blue) and landing lines $\Lti$
	(dashed red/blue) in tip space $\Rti$ (in units of $\xi$), with the
	hyperbola $\Jti$ defining the edge $\partial \Uti$ of the unstable
	domain $\Uti$, before (a) and after (b) merging, for $1- \kappa_s= \pm
	0.01$.  Parameters are $\lambda_{-,s} = -0.25 \,e_p/\xi^2,
	\lambda_{+,s} = 0$, and $a_s \approx 0.035\,e_p/\xi^3$, $\alpha_s =
	-0.025\,e_p/\xi^4$, $\beta_s = 0$, $\gamma_s \approx 0.68
	\,e_p/\xi^4$.  A finite skew parameter $\beta_s = 0.025 e_p/\xi^4$
	tilts the hyperbola away from the axes (dotted curves).  Crosses
	correspond to the vertices \eqref{eq:hyp_vertices} and
	\eqref{eq:hyp_vertices_merged} of the hyperbola before and after
	merging.  Pairs of solid and open circles connected via long arrows
	are examples of pairs of jumping- and landing tip positions. After
	merging, see (b), the unstable domain $\Uti$ is connected along the
	$\vti$-axis, dividing the tip coordinate plane into two separate
	regions. The jumping and landing hyperbolas coincide at their vertices
	before merging, see (a), but not thereafter, see (b), where the
	jumping and landing hyperbolas are separated (vertices on $\Lti$ are
	marked with open red/blue stars) and no contact point is present.
	Note the rotation by 90 degrees of the unstable direction with
	respect to Figs.\ \ref{fig:quasi-ring-hyperbola}(b) and
	\ref{fig:quasi-ring-hyperbola-merged}(b).}
    \label{fig:hyp-duo}
\end{figure}

\subsubsection{Jump line $\Jti$}\label{sec:hyp_Jti}

The boundary of the unstable domain $\Uti$ is determined by the jump condition
$D(\Rsjp) = 0$.  Making use of the expansion \eqref{eq:e_pin_expans_hyp} and
keeping only terms quadratic in $\uti,\vti$, the edges $\delta\Rsjp = (\usjp,\vsjp)$
of $\Uti$ (measured relative to $\Rti_s$) are given by the solutions of the quadratic
form (cf.\ \eqref{eq:quadratic_form})
\begin{equation}\label{eq:quadratic_form_hyp}
   [\gamma_s\,\uti^2 + 2 \beta_s\,\uti\vti + \delta_s\, \vti^2]_{\Rsjp} 
   = 2 \Cbar(\kappa_s - 1).
\end{equation}
Equation \eqref{eq:quadratic_form_hyp} describes a hyperbola (centered at
$\Rti_s$) as its associated determinant is negative, see Eq.\
\eqref{eq:det_saddle2D_explicit}. Again, \eqref{eq:quadratic_form_hyp} can be
cast in the form of a matrix equation
\begin{equation}\label{eq:matrix_eq_jp_hyp}
   \delta\Rsjp^T M_{s, \mathrm{jp}}\delta\Rsjp = \Cbar(\kappa_s - 1),
\end{equation}
with $M_{s,\mathrm{jp}}$ given by
\begin{align}\label{eq:hyperbola_jp}
      M_{s,\mathrm{jp}} &= \begin{bmatrix} \gamma_s/2 &~~~ \beta_s/2\\
                           \beta_s/2 &~~~ \delta_s/2
           \end{bmatrix}
\end{align}
with $\det M_{s,\mathrm{jp}} = (\gamma_s\delta_s - \beta_s^2)/4 < 0$.  As
shown in Fig.\ \ref{fig:hyp-duo}, the geometry of the unstable domain $\Uti$
changes drastically when $1- \kappa_s$ changes sign. Before merging, i.e., for
$1 - \kappa_s > 0$, the unstable domain (top and bottom regions in Fig.\
\ref{fig:hyp-duo}(a)) is disconnected along the stable $v$-direction and the
two red/blue branches of the hyperbola \eqref{eq:quadratic_form_hyp} describe
the tips of $\Uti$.  When $\kappa_s$ goes to unity, the tips of the unstable
domain merge at the saddle point $\Rti_s$.  After merging, the unstable domain
extends continuously from the top to the bottom in Fig.\ \ref{fig:hyp-duo}(b)
with a finite width along the unstable $u$-direction, similarly to the
isotropic case shown in Fig.\ \ref{fig:f_pin}(c).  Correspondingly, the two
(red and blue) branches of the hyperbola \eqref{eq:quadratic_form_hyp} now
describe the edges of $\Uti$.

Solving the quadratic equation \eqref{eq:quadratic_form_hyp} before merging,
i.e., $1 - \kappa_s >0$, we find solutions $\usjp(\vti)$ away from a gap
along the stable $v$-direction,
\begin{multline}\label{eq:uti_jp_hyp}
   \usjp(|\vti| \geq \vti_{s,c}) = -\frac{1}{\gamma_s}\Bigl[\beta_s \vti\\
   \pm \sqrt{2\gamma_s\Cbar(\kappa_s-1)- (\gamma_s\delta_s - \beta_s^2)\vti^2} \Bigr],
\end{multline}
i.e., Eq.\ \eqref{eq:uti_jp_hyp} has real solutions in the (unbounded)
interval $|\vti| \geq \vti_{s,c}$, with
\begin{equation}\label{eq:vti_sc}
   \vti_{s,c} = \sqrt{2\gamma_s\Cbar(1-\kappa_s)/|\gamma_s\delta_s - \beta_s^2|}.
\end{equation}
For the uniaxial defect \eqref{eq:uniax_potential_formal} before merging, this
gap corresponds to a splitting of $\Uti$ along the stable angular direction,
producing two separated domains as shown in Fig.\
\ref{fig:quasi-ring-hyperbola}(a).  The coordinates
$\left(\usjp(\pm\vti_{s,c}), \pm \vti_{s,c}\right)$ give the positions of the
vertices $\delta\Rti^<_{s,c,\pm}$ (relative to $\Rti_s$) of the hyperbola before
merging,
\begin{align}\label{eq:hyp_vertices}
        \delta\Rti^<_{s,c,\pm} &= \pm \left(-\beta_s/\gamma_s, 1\right)\,\vti_{s,c}.
\end{align}
These are marked as black crosses in Fig.\ \ref{fig:hyp-duo}(a) (note the
rotation in the geometry as compared with Fig.\
\ref{fig:quasi-ring-hyperbola}(a)).  We denote the distance between these
vertices  by  $\delta v^<$, defining a gap of width $\propto
\sqrt{1-\kappa_s}$ given by
\begin{equation}
   \delta v^< = 2|\delta\Rti^<_{s,c,\pm}| = 2 \sqrt{\left(\gamma_s+\frac{\beta_s^2}{\gamma_s}\right)
   \frac{\Cbar(1-\kappa_s)}{|\gamma_s\delta_s - \beta_s^2|}}.
\end{equation}

After merging, i.e., for $\kappa_s  - 1 > 0$, the (local) topology of $\Uti$
has changed as the gap along $v$ closes and reopens along the unstable
$u$-direction; as a result, the two separated domains of $\Uti$ have merged.
The two branches of the hyperbola derived from \eqref{eq:quadratic_form_hyp}
are now parametrized as
\begin{multline}\label{eq:vti_jp_hyp}
   \vsjp(|\uti| \geq \uti_{s,e}) = -\frac{1}{\delta_s}\Bigl[\beta_s\uti \\
   \pm \sqrt{2\delta_s\Cbar(\kappa_s-1)- (\gamma_s\delta_s - \beta_s^2)\uti^2} \Bigr],
\end{multline}
with 
\begin{equation}
   \uti_{s,e} = \sqrt{2\delta_s\Cbar(\kappa_s - 1)/|\gamma_s\delta_s - \beta_s^2|}.
\end{equation}
The corresponding unstable domain is shown in Fig.\ \ref{fig:hyp-duo}(b).  For
the uniaxial defect \eqref{eq:uniax_potential_formal} after merging, this gap
now corresponds to the finite width of $\Uti$ along the radial direction, as
shown in Fig.\ \ref{fig:quasi-ring-hyperbola-merged}(a).  The coordinates
$\left(\pm\uti_{s,e}, \vsjp(\pm \uti_{s,e})\right)$ for the vertices
$\Rti^>_{s,e,\pm}$ read
\begin{align}\label{eq:hyp_vertices_merged}
   \delta \Rti^>_{s,e,\pm} 
   &= \pm\left(1, -\frac{\beta_s}{\delta_s}\right)\,\uti_{s,e}
\end{align}
and correspond to the points of closest approach in the branches of the
hyperbola \eqref{eq:quadratic_form_hyp}; these are again marked as black
crosses in Fig.\ \ref{fig:hyp-duo}(b) but are no longer associated with
critical points (we index these extremal points by `e').  Their distance
$\delta u^>$ is given by
\begin{equation}
   \delta u^>=2|\delta\Rti^>_{s,e,\pm}|
   = 2\sqrt{\left(\delta_s+\frac{\beta_s^2}{\delta_s}\right)
   \frac{\Cbar(\kappa_s - 1)}{|\gamma_s\delta_s - \beta_s^2|}},
\end{equation}
i.e., the smallest width in $\Uti$ grows as $\propto\sqrt{\kappa_s - 1}$. 

As discussed above and shown in Fig.\ \ref{fig:hyp-duo}, the solutions of the
quadratic form \eqref{eq:quadratic_form_hyp} before and after merging are
unbounded for every value of $\kappa_s - 1$.  As a consequence, neglecting the
higher order terms in the determinant $D(\Rti)$ is valid only in a narrow 
neighborhood of the saddle $\Rti_s$, where the boundaries of $\Uti$ have the
shape of a hyperbola. Away from the saddle, these higher order terms are
relevant in determining the specific shape of the unstable and bistable domain,
e.g., the ring-like structures of $\Uti$ and $\Bas$ in Figs.\
\ref{fig:quasi-ring-hyperbola} and \ref{fig:quasi-ring-hyperbola-merged}.

\subsubsection{Landing line $\Lti$}\label{sec:hyp_Lti}

To find the second bistable vortex tip configuration $\Rslp$ associated to the
edges of $\Bas$ before and after merging, we repeat the steps of Sec.\
\ref{sec:Lti}. For the jump vector $\Delta\Rti_s = \Rslp -\Rsjp$, we find the
result 
\begin{align}
   &\Delta \uti_s(\vti) = -3\left(\gamma_s\, \usjp(\vti)
   + \beta_s\, \vti\right)/\gamma_s,\label{eq:delta_rxh}\\
   &\Delta \vti_s(\vti) = - \left[a_s/(\Cbar + \lambda_{s,+})\right]\vti\,
   \Delta \uti_s(\vti), \label{eq:delta_ryh}
 \end{align}
cf.\ Eqs.\ \eqref{eq:delta_rx} and \eqref{eq:delta_ry} above. Here, we make
use of the parametrization for the jump coordinate $\usjp(\vti)$ in
\eqref{eq:uti_jp_hyp} before merging; after merging, the above result is still
valid but should be expressed in terms of the parametrization $\vsjp(\uti)$ in
Eq.\ \eqref{eq:vti_jp_hyp}.

The landing positions $\Rslp = \Rsjp + \Delta\Rti_s$ arrange along the
branches $\Lti$ of a hyperbola in $\Rti$-space that are described by the
matrix equation
\begin{equation}\label{eq:matrix_eq_lp_hyp}
   \delta\Rslp^\mathrm{T} M_{s,\mathrm{lp}}\, \delta\Rslp = \Cbar (\kappa_s - 1),
\end{equation}
with the landing matrix now given by
\begin{equation}\label{eq:hyperbola_lp}
   M_\mathrm{s,lp} = \frac{1}{4} M_{s,\mathrm{jp}} +
   \begin{bmatrix} 0 & 0\\
                   0 & ~~~\displaystyle{\frac{3}{4}\Bigl(\frac{\delta_s}{2}
   - \frac{\beta_s^2}{2\gamma_s}\Bigr)}
   \end{bmatrix}
\end{equation}
with $\det M_\mathrm{s,lp} = (\gamma_s\delta_s -\beta_s^2)/16 < 0$.
Before merging, the vertices of the landing and jumping hyperbolas coincide
and the jump \eqref{eq:delta_rxh}--\eqref{eq:delta_ryh} vanishes at these
points.  Moreover, as for the contact points \eqref{eq:contact_points} close
to onset of strong pinning, the tangent to the jumping and landing hyperbolas
at the vertices is parallel to the $u$-direction, as is visible in Fig.\
\ref{fig:hyp-duo}(a).

For $\kappa_s = 1$, the tips of $\Uti$ merge and both the jumping and landing
hyperbolas coincide at $\Rti_s$.  After merging, i.e., for $\kappa_s - 1 > 0$,
the condition $\Delta \uti_s = \Delta \vti_s = 0$ cannot be realized along the
hyperbola \eqref{eq:quadratic_form_hyp} and the jumping and landing lines
separate completely; as a result, both the jumping distance $\Delta\Rti_s$ as
well as the jump in energy $\Delta \epin$ are always finite (see also Appendix
\ref{sec:eff_1D_merging}).  Indeed, after merging the landing hyperbola
\eqref{eq:matrix_eq_lp_hyp} has vertices
\begin{equation}
   \delta\Rti_{s,v,\pm}
   = \pm\left(1, -\frac{\gamma_s \beta_s}{(4\gamma_s \delta_s-3\beta_s^2)}\right)
   \,\uti_{s,v},
\end{equation}
with
\begin{equation}
   \uti_{s,v} = \sqrt{\frac{2\Cbar(\kappa_s-1) (4\gamma_s\delta_s-3\beta_s^2)}
   {\gamma_s(\gamma_s\delta_s -\beta_s^2)}}
\end{equation}
different from the jumping hyperbola in \eqref{eq:hyp_vertices_merged}.  At
these points, the stable and unstable hyperbolas are tangent to the
$v$-direction, as is visible in Fig.\ \ref{fig:hyp-duo}(b).

In section Sec.\ \ref{sec:topology_hyp} below, we will take a step back from
the local analysis of the unstable domain $\Uti$ close to a saddle point
$\Rti_s$ and consider the evolution of its geometry across the merging
transition from a global perspective using specific examples.  Elaborating on
the analysis of Sec.\ \ref{sec:topology}, we will provide a simple argument
explaining the absence of contact points between jump and landing lines after
merging. Furthermore, we discuss the two possible roles of mergers as changing
the number of components of $\Uti$ or changing the connectivity of $\Uti$
between simply and non-simply connected areas.  Before doing so, we discuss
the behavior of the bistable region $\Bas$ close to merging.

\subsection{Bistable domain $\Bas$}\label{sec:hyp_Bas}

\begin{figure}[t]
        \includegraphics[width = 1.\columnwidth]{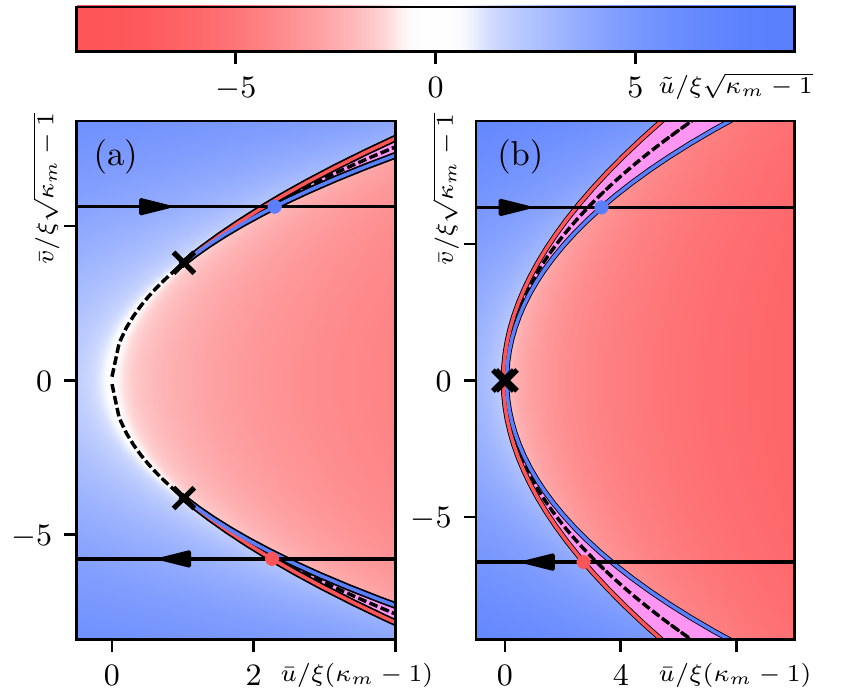}
	\caption{Bistable domain $\Bas$ in asymptotic space $\Ras$ before (a)
	and after (b) merging, for $1 - \kappa_s=\pm0.01$ and parameters as in
	Fig.\ \ref{fig:hyp-duo}.  (a) Before merging, the bistable domain $\Bas$
	consists of two parts, corresponding to the two unstable regions $\Uti$ in
	Fig.\ \ref{fig:hyp-duo}(a). These terminate in the cusps at
	$\Ras_{s,c,\pm}^<$ that approach one another along the dashed parabola
	\eqref{eq:x_0_line_hyp} to merge at $\kappa_s = 1$. Red/blue colors
	indicate different vortex configurations as quantified through the
	order parameter $\uti - \uti_m(\vas)$, while magenta is associated to
	the bistable region $\Bas$.  Colored dots mark the asymptotic positions
	associated to the pairs of jump positions in Fig.\
	\ref{fig:hyp-duo}(a).  (b) After merging, the bistable domain is
	continuously connected; the cusps/critical points have vanished and
	the dashed parabola turns into the branch cutting line.  The black
	crosses now mark the positions of strongest pinching of $\Bas$, the
	colored dots mark the asymptotic positions associated to the pairs of
	tip positions in Fig.\ \ref{fig:hyp-duo}(b).}
    \label{fig:hyp-bananas-duo}
\end{figure}

The set of asymptotic positions corresponding to $\Uti$ before and after
merging, i.e., the bistable domain $\Bas$, can be found by systematically
repeating the steps in Sec.\ \ref{sec:Bas}. Applying the force balance
equation $\nabla_\mathbf{R} e_\mathrm{pin}(\mathbf{R};\Ras)\Big|_{\Rti}=0$ to
the energy expansion \eqref{eq:e_pin_expans_hyp}, we find the counterpart of
Eqs.\ \eqref{eq:asymptotic_positions},
\begin{align}\nonumber
   \Cbar \uas &=  \Cbar(1-\kappa_s) \uti + \frac{a_s}{2}\vti^2 
   + \frac{\gamma_s}{6}\uti^3 + \frac{\beta_s}{2}\uti^2 \vti 
   + \frac{\alpha_s}{2}\uti \vti^2,\\
   \Cbar \vas &= (\Cbar + \lambda_{s,+}) \vti + a_s\,\uti \vti + \frac{\beta_s}{6}\uti^3 
   + \frac{\alpha_s}{2}\uti^2 \vti,
   \label{eq:asymptotic_positions_hyp}
\end{align}
relating tip and asymptotic positions close to merging. As for the unstable
domain, the topology of $\Bas$ depends on the sign of $1- \kappa_s$.  The
bistable domain $\Bas$ before merging is shown in Fig.\
\ref{fig:hyp-bananas-duo}(a) for $1 - \kappa_s = 0.01$. It
consists of two parts, corresponding to the two pieces of $\Uti$ for $1 - \kappa_s
> 0$, that terminate at the cusps $\Ras_{s,c,\pm}^<$. The latter are related
to the vertices $\Rti_{s,c,\pm}^<$ of the jumping hyperbola through the force
balance equation \eqref{eq:asymptotic_positions_hyp},
\begin{equation}\label{eq:cusps_hyp}
   \delta\Ras_{s,c,\pm}^< \approx \left[\left(a_s/2\,\Cbar\right)\,\vti^2_{s,c},\, 
   \pm\left(1 + \lambda_{s,+}/\Cbar\right)\vti_{s,c}\right].
\end{equation}
For finite values of $(1-\kappa_s)$, the cusps are separated by a distance
$2|\delta{\Ras}_{s,c,\pm}^<|\approx 2\left(1 + \lambda_{s,+} /\Cbar\right) \vti_{s,c} \propto
\sqrt{1-\kappa_s}$. They approach one another along the parabola 
\begin{equation}
   \uas_{s,0} \approx \frac{a}{2\Cbar}\frac{1}{(1+ \lambda_+/\Cbar)^2} \vas_{s,0}^2,
   \label{eq:x_0_line_hyp}
\end{equation}
see the black dashed line in Fig.\ \ref{fig:hyp-bananas-duo}, with
higher-order corrections appearing at finite skew $\beta \neq 0$. 
After merging, this line lies within $\Bas$ and defines the branch
crossing line, cf.\ Eq.\ \eqref{eq:x_0_line}. 

After merging, when $\kappa_s - 1 > 0$, the cusps have vanished and the edges
have rearranged to define a connected bistable region, see Fig.\
\ref{fig:hyp-bananas-duo}(b).  The extremal points of the two edges are found
by evaluating the force balance equation \eqref{eq:asymptotic_positions_hyp}
at the vertices $\Rti_{s,e,\pm}^>$, Eq.\ \eqref{eq:hyp_vertices_merged}, to
lowest order,
\begin{equation}\label{eq:cusps_hyp_merged}
   \delta\Ras_{s,e,\pm}^> \approx \frac{\beta_s}{\delta_s}
   \left[\frac{a_s}{2\,\Cbar}\frac{\beta_s}{\delta_s}\,\uti_{s,e}^2,\, 
   \mp\left(1 + \frac{\lambda_{s,+}}{\Cbar}\right)\,\uti_{s,e}\right].
\end{equation}
For finite values of $(\kappa_s - 1)$, these points are separated by a
distance $2|\delta\Ras_{s,e,\pm}^>|\approx 2\left(1 + \lambda_{s,+} / \Cbar\right)
(\beta_s/\delta_s)\uti_{s,e}\propto\sqrt{\kappa_s -1}$. Note that the extremal
points $\Ras_{s,e,\pm}^>$ are no longer associated to cusps or critical points
as these have disappeared in the merging process. When the skew parameter
vanishes as in Fig.\ \ref{fig:hyp-bananas-duo}, $\beta_s = 0$, higher-order
terms in $(\kappa_s - 1)$ in the force-balance equation
\eqref{eq:asymptotic_positions_hyp} become relevant in determining the
positions $\Ras_{s,e,\pm}^>$, separating them along the unstable
$u$-direction. In this case, we obtain a different scaling for their distance,
i.e., $|\delta\Ras_{s,e,\pm}^>| \propto\left(1-\kappa_s\right)^{3/2}$.

\subsection{Topological aspect of mergers}\label{sec:topology_hyp}

\begin{figure}
        \includegraphics[width = 1.\columnwidth]{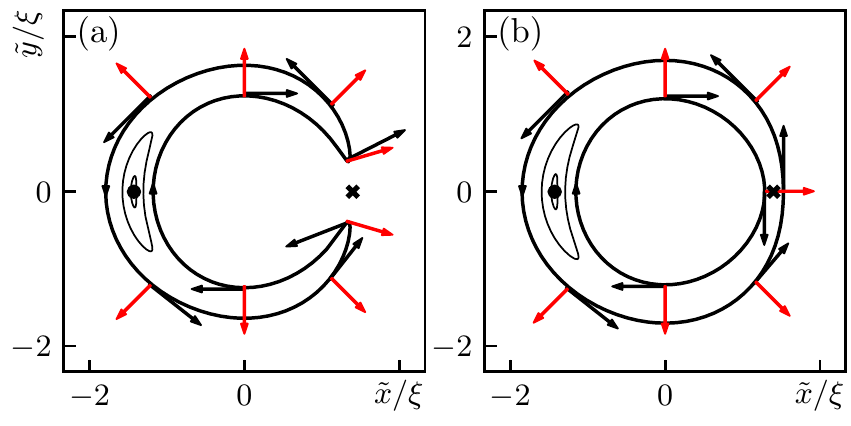}
 \caption{Left: Unstable region $\Uti$ for a defect with dipolar asymmetry.
        Upon the onset of strong pinning, an unstable ellipse appears to the
        left of the defect center (black solid dot).  With increasing pinning
        strength (decreasing $\Cbar$) the ellipse grows and deforms into a
        horseshoe geometry. The unstable eigenvector field $\mathbf{v}_-$ (red
        arrows) points radially outward away from the defect center.  The
        tangent field to the boundary $\partial\Uti$ (black arrows) follows
        the unstable direction at an angle of $\pi/2$ over most of
        $\partial\Uti$, with the exception of the two turning points where the
        tangent rotates by $\pi$ with respect to $\mathbf{v}_-$, producing a
        relative winding of $2\pi$. Right: After the merger of the turning
        points the unstable region $\Uti$ changes topology and assumes the
        shape of a ring. The windings of the tangent field with respect to the
        eigenvector-field $\mathbf{v}_-$ vanish separately for both
        boundaries of $\Uti$.
        }
    \label{fig:top_horseshoe}
\end{figure}

In order to discuss the topological aspect of a merger, it is convenient to
consider some specific examples. In Sec.\ \ref{sec:uniax_defect}, we have
analyzed the case of a uniaxial defect with a quadrupolar anisotropy $\delta
e_p \propto \epsilon \sin^2\pti$ in the pinning potential, see \eqref{eq:dep_quad},
that produced a degenerate onset at symmetric points $[\pm\xti_m,0]$. Here, we
choose again a weakly anisotropic defect centered in the origin but with a dipolar
deformation $\delta e_p \propto \epsilon \cos\pti$ that results in an angle-dependent
Labusch parameter
\begin{equation}\label{eq:kappa_dip}
   \kappa_m(\pti) = \kappa_m - \epsilon\cos\pti,
\end{equation}
see Eq.\ \eqref{eq:kappa_quad}. The strong pinning onset of such a defect then
appears in an isolated point on the negative $x$-axis, with the unstable
ellipse $\Uti$ deforming with increasing $\kappa_m$ into a horseshoe that is
open on the positive $x$-axis---the closing of the horseshoe to produce a
ring, see Fig.\ \ref{fig:top_horseshoe}, then corresponds to the local merger
shown in Fig.\ \ref{fig:hyp-duo}. With this example in mind, we can repeat the
discussion in Sec.\ \ref{sec:topology}: The unstable eigenvector
$\mathbf{v}_-(\mathbf{R}_\mathrm{jp})$ points radially outwards from the
origin over the entire horseshoe, including the merging region at positive
$x$.  On the other hand, the tangent to the boundary $\partial\Uti$ rotates
forward and back along the horseshoe as shown in Fig.\ \ref{fig:top_horseshoe}
(we attribute a direction to $\partial\Uti$ with the convention of following
the boundary with the unstable region on the left); in fact, over most of the
boundary, the tangent is simply orthogonal to $\mathbf{v}_-$, with both
vectors rotating together when going along $\partial\Uti$.  At the ends of the
horseshoe, however, the tangent locally aligns parallel (anti-parallel) to
$\mathbf{v}_-$ and the two vectors rotate (anti-clockwise) with respect to one
another, with the total winding equal to $2\pi$.  After the merger, this
winding has disappeared, with the resulting ring exhibiting no winding in the
tangent fields on the inner/outer boundary; as a result, the contact points
between the jump and landing lines have disappeared.

Furthermore, the merger changes the topology of $\Uti$ from the
simply-connected horseshoe to the non-simply connected ring, while the number
of components in $\Uti$ has not changed.  Note that the change in the relative
winding is not due to crossing the singularity of the vector field
$\mathbf{v}_-$ as alluded to in Sec.\ \ref{sec:topology}---rather, it is the
merger of the horseshoe tips that rearranges the boundaries of $\Uti$ and make
them encircle the singularity.

\begin{figure}[t]
        \includegraphics[width = 1.\columnwidth]{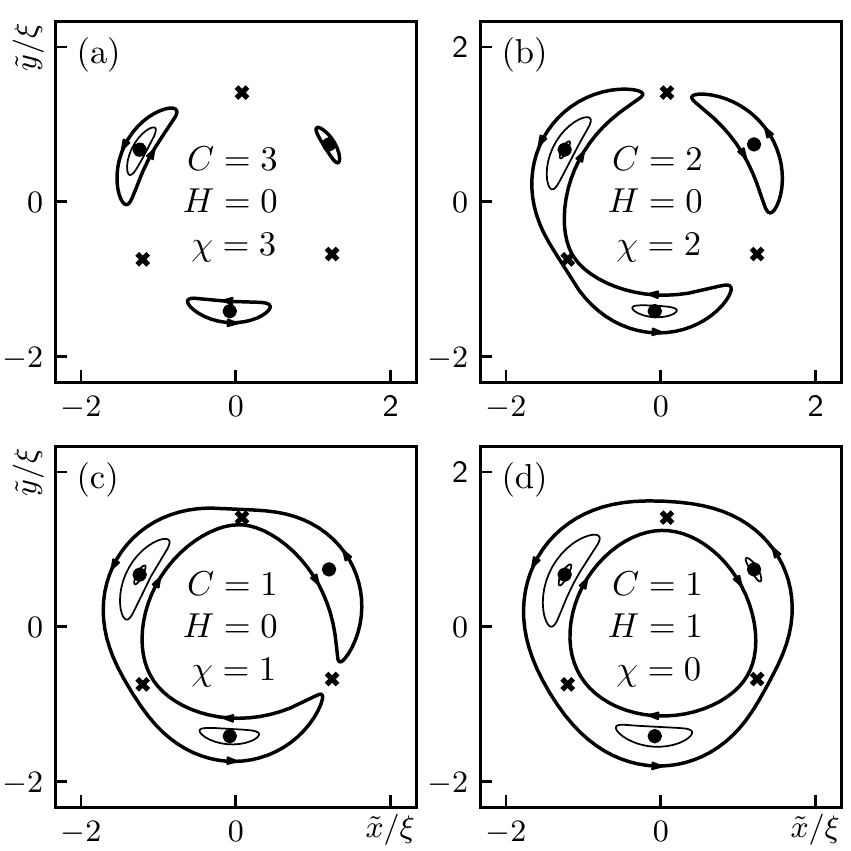}
 \caption{The unstable domain $\Uti$ starting out with $C = 3$ components in
	(a) changes topology in three steps: after the first (b) and second
	(c) mergers the number of components $C$ has changed from three in (a)
	to two in (b) to one in (c), leading to a horseshoe shape of $\Uti$.
	The third merger closes the horseshoe to produce the ring
	geometry in (d) characterized by the coefficients $C = 1$ and $H = 1$ ($H$
	denotes the number of `holes' in $\Uti$); the Euler characteristic
	$\chi = C - H$ changes by unity in every merger. }
    \label{fig:top_three}
\end{figure}

In the above example, we have discussed a merger that changes the
connectedness of $\Uti$. On the other hand, as we are going to show, a merger
might leave the connectedness of $\Uti$ unchanged, while modifying the number
of components, i.e., the number of disconnected parts, in $\Uti$.  Let us
again consider a specific example in the form of an anisotropic defect with a
warped well shape, producing several (in general subsequent) onsets and
mergers; in Fig.\ \ref{fig:top_three}, we consider a situation with three
onset points and subsequent individual mergers. After the onset, the three
ellipses define an unstable region $\Uti$ with three disconnected parts that
are simply-connected each. This configuration is characterized by its number
of components measuring $C = 3$.  As two of the three ellipses merge, the
number of components of $\Uti$ reduces to $C = 2$, the next merger generates a
horseshoe that is still simply-connected with $C = 1$. The final merger
produces a ring; while the number of components remains unchanged, $C = 1$,
the unstable area assumes a non-simply connected shape with a `hole'; we
associate the index $H = 1$ with the appearance of this hole within $\Uti$. In
physics terms, the last merger producing a hole in $\Uti$ is associated with
the appearance of a pinned state; the unstable ring separates stable tip
positions that are associated with pinned and free vortex configurations
residing at small and large radii, respectively.

Defining the (topological) characteristic $\chi \equiv C - H$, we see that
$\chi$ changes by unity at every onset and merger, either through an increase
(for an onset) or decrease (for a merger) in the number of components $C \to
C\pm 1$, or through the appearance of a hole (in a merger) $H \to H+1$.
Indeed, the quantity $\chi$ is known as the Euler characteristic of a manifold
and describes its global topological properties; it generalizes the well known
Euler characteristic of a polyhedron to surfaces and manifolds\cite{Nakahara_2003}, see Sec.\
\ref{sec:2D_landscape} below. Finally, Morse theory \cite{NashSen_2011} 
connects the Euler characteristic with the local differential properties
(minima, maxima, saddles) of that manifold, hence establishing a connection
between local onsets and mergers (at minima and saddles of $D(\Rti)$) and the
global properties of $\Uti$ such as the appearance of new pinned states. In
Sec.\ \ref{sec:2D_landscape} below, we consider the general case of a random
pinning landscape in two dimensions and discuss the connection between local
differential and global topological properties of $\Uti$ in the light of Morse
theory---the topology of bistable domains $\Bas$ then follows trivially.

\section{$\Uti$ of a two-dimensional pinscape}\label{sec:2D_landscape}

\begin{figure}[t]
        \includegraphics[width = 1.\columnwidth]{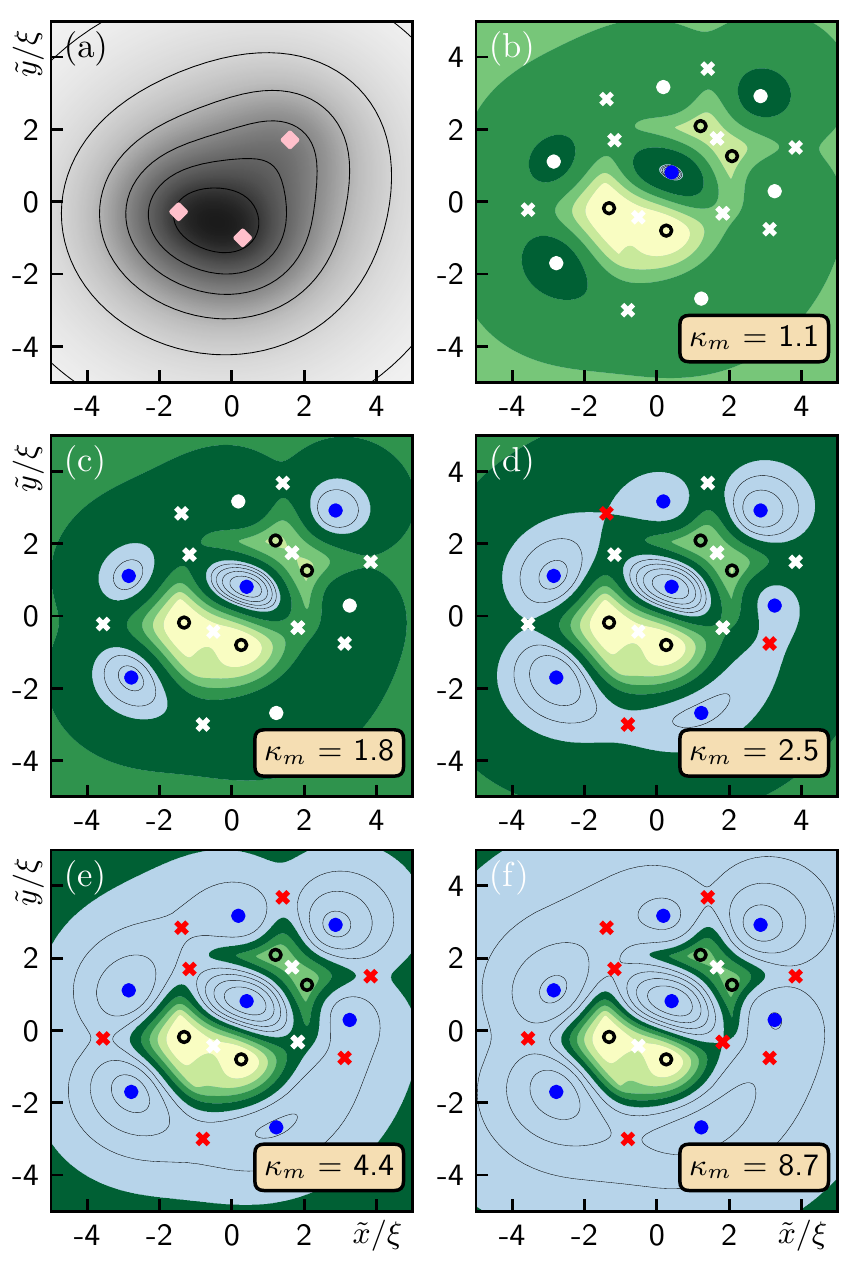}
	\caption{(a) Grayscale image of the pinning potential landscape
	$e_p(\Rti)$, with the three diamonds marking the positions of the
	defects.  (b)--(f) Shifted curvature function $\Lambda_{\Cbar}(\Rti)$
	versus tip position $\Rti$ for increasing values of $\kappa_m$
	(decreasing $\Cbar$) as we proceed from (b) to (f).  We make use of
	the topographic interpretation with positive values of
	$\Lambda_{\Cbar}$ marked as landmass (greenish colors, with low/high
	elevation in dark/light green) and negative values of
	$\Lambda_{\Cbar}$ constituting $\Uti$ in flat light blue (height
	levels are shown by thin black lines). The pinscape in (a) produces a
	curvature landscape with $7$ minima (solid dots), $4$ maxima (open
	dots), and $10$ saddles (crosses). Several unstable regions $\Uti$
	appear (solid dots turn blue) and merge (crosses turn red) to change
	the topology of $\Uti$.  The Euler characteristic $\chi(\Uti) = m - s
	+ M = 1 - 0 + 0 = 1$ in (b) changes to $\chi(\Uti) = 4$ in (c) and
	(d), drops to $\chi(\Uti) = 0$ in (e) and $\chi(\Uti) = -1$ in (f); indeed,
	$\Uti$ in (f) has one component $C = 1$ and two holes $H = 2$,
	reproducing $\chi(\Uti) = C - H = -1$.
}
    \label{fig:3_defects}
\end{figure}

\begin{figure}[t]
        \includegraphics[width = 1.\columnwidth]{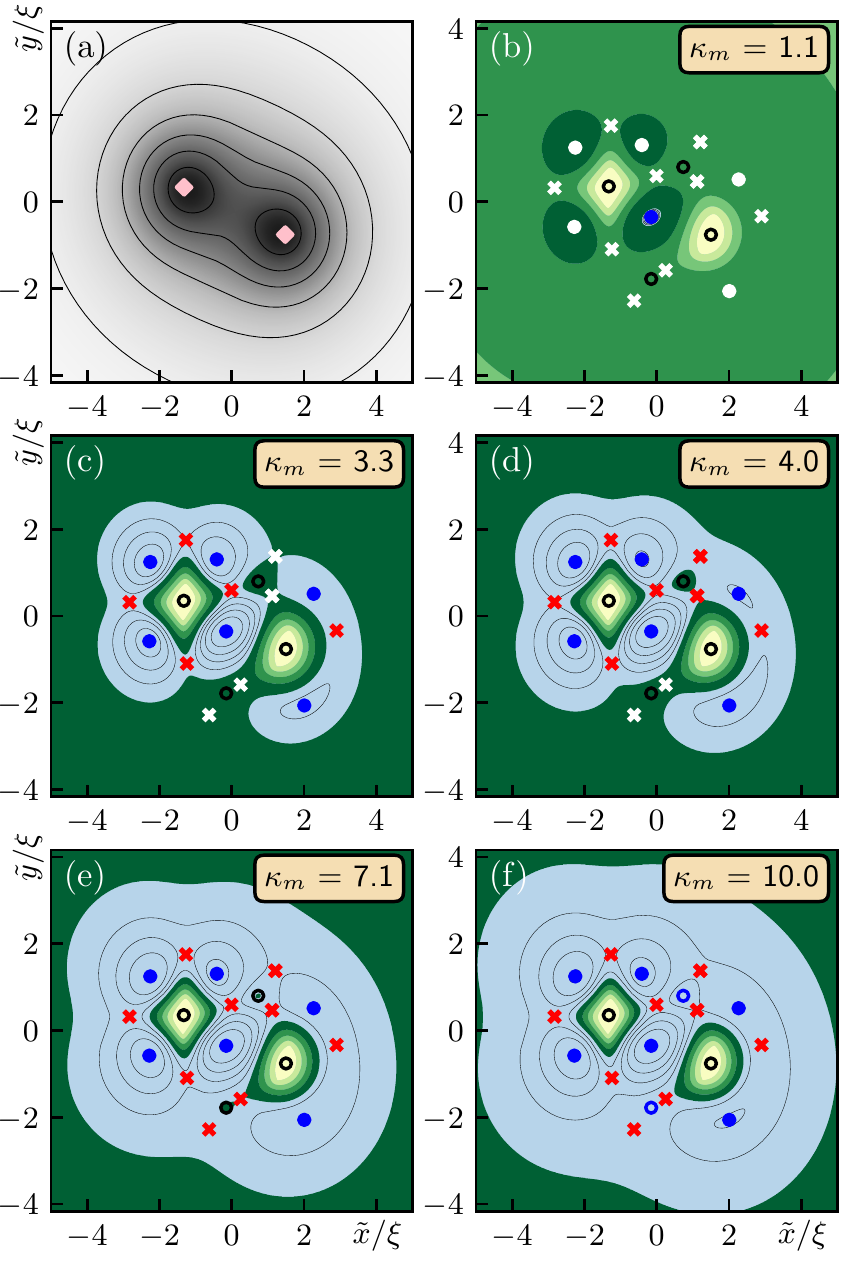}
 \caption{(a) Grayscale image of the pinning potential landscape
        $e_p(\Rti)$, with the two diamonds marking the positions of the
        defects.  (b)--(f) Shifted curvature function $\Lambda_{\Cbar}(\Rti)$
        (in topographic coloring, see caption of Fig.\ \ref{fig:3_defects})
        versus tip position $\Rti$ for increasing values of $\kappa_m$ as we
        proceed from (b) to (f).  The pinscape in (a) produces a curvature
        landscape with $6$ minima (solid dots), $4$ maxima (open dots), and
        $9$ saddles (crosses).  Upon increasing $\kappa_m$, several unstable
        regions $\Uti$ appear (solid dots turn blue) and merge (crosses turn
        red) to change the topology of $\Uti$.  The Euler characteristic
        $\chi(\Uti) = m - s + M = 1 = C$ in (b), remains $\chi(\Uti) = 1$ in
        (c), but with $C = 2$ and $H = 1$, changes to $\chi(\Uti) = -1$ in
        (d), and $\chi(\Uti) = -3$ with one component $C = 1$ and four holes
        $H = 4$ in (e).  In going from (e) to (f) two of the maxima (black
        open dots turn blue) drop below zero, producing a characteristic
        $\chi(\Uti) = 6 - 9 + 2 = -1$; indeed, $\Uti$ in (f) has one component
        $C = 1$ and two holes $H = 2$, reproducing $\chi(\Uti) = C - H = -1$.}
    \label{fig:2_defects_maxima}
\end{figure}

We consider a two-dimensional pinning landscape $e_p(\mathbf{R})$, e.g., as
produced by a superposition of several (anisotropic Lorentzian) defects
residing in the $z = 0$ plane. In the figures \ref{fig:3_defects} and
\ref{fig:2_defects_maxima}, we analyse two specific cases with $n = 3$ and $n
= 2$ defects as given in Eq.\ \eqref{eq:uniax_potential_formal} with $\epsilon
= 0.1$ and positions listed in Tables \ref{table1} and \ref{table2}; these
produce unstable landscapes $\Uti$ of considerable complexity already, see
Figs.\ \ref{fig:3_defects}(a) and \ref{fig:2_defects_maxima}(a). Our defects
are compact with $e_p(\mathbf{R}) \to 0$ vanishing at $R \to \infty$; as a
result, $\epin$ becomes flat at infinity.  Note that a dense assembly of
uniformly distributed individual defects produces a random Gaussian pinning
landscape, as has been shown in Ref.\ \onlinecite{Willa_2022}.

Here, we are interested in the evolution of the unstable and bistable domains
$\Uti$ and $\Bas$ associated with the 2D pinning landscape $\epin$; we focus
on the unstable domain $\Uti$, with the properties of the bistable domain
$\Bas$ following straightforwardly from the solution of the force balance
equation \eqref{eq:force_balance}. Unlike the analysis above that is centered
on special points of $\Uti$, ellipses near onset and hyperbolas near mergers,
here, we are interested in the global properties of the unstable region
produced by a generic (though still two-dimensional) pinscape.

As discussed in Sec.\ \ref{sec:arb_shape} above, the unstable region $\Uti$
associated with strong pinning is determined by the condition $D(\Rti) = 0$ of
vanishing Hessian determinant, more precisely, by the competition between the
lowest eigenvalue $\lambda_-(\Rti)$ of the Hessian matrix $\mathrm{H}_{ij}$ of
the pinning potential $e_p(\mathbf{R})$ and the effective elasticity $\Cbar$,
see Eq.\ \eqref{eq:def_calU}. In order to avoid the interference with the
second eigenvalue $\lambda_+(\Rti)$ of the Hessian matrix, we consider the
shifted (by $\Cbar$) curvature function
\begin{equation}\label{eq:def_Lambda}
  \Lambda_{\Cbar}(\Rti) \equiv \Cbar + \lambda_-(\Rti),
\end{equation}
i.e., the relevant factor of the determinant $D(\Rti) = [\Cbar +
\lambda_-(\Rti)] [\Cbar + \lambda_-(\Rti)]$.  The condition
\begin{equation}\label{eq:vanish_Lambda}
  \Lambda_{\Cbar}(\Rti) = 0
\end{equation}
then determines the boundaries of $\Uti$.

\begin{table}
\caption{\label{table1} Positions and relative weights of 3 uniaxially
anisotropic Lorentzian defects in Fig.\ \ref{fig:3_defects} as given by Eq.\
\eqref{eq:uniax_potential_formal}.}
  \vskip 3 pt
  \begin{tabular}{l | c c c}
   & $~x/\xi$ & ~~$y/\xi$ & ~~weight\\
            \hline
  defect~\#1~~ & $1.14$ & $1.07$ & 0.65\\
  defect~\#2~~ & $-0.98$ & $-0.19$ & 1\\
  defect~\#3~~ & $0.20$ & $-0.67$ & 1 \\
\end{tabular}
\end{table}
\begin{table}
\caption{\label{table2} Positions and relative weights of 2 uniaxially
anisotropic Lorentzian defects in Fig.\ \ref{fig:2_defects_maxima} as given by
Eq.\ \eqref{eq:uniax_potential_formal}.}
  \vskip 3 pt
  \begin{tabular}{l | c c c}
  & $~x/\xi$ & ~~$y/\xi$& ~~weight \\
            \hline
  defect~\#1~~ & $-1.32$ & $0.33$  & 1\\
  defect~\#2~~ & $1.48$ & $-0.76$ & 1\\
\end{tabular}
\end{table}

The above problem can be mapped to the problem of cutting a surface, where
$\Lambda_{\Cbar}(\Rti)$ is interpreted as a height-function over
$\mathbb{R}^2$ that is cut at zero level; the elasticity $\Cbar$ then plays
the role of a shift parameter that moves the function $\lambda_-(\Rti)$
downwards in height with decreasing $\Cbar$ (that corresponds to increasing
the relative pinning strength of the pinscape in physical terms). As $\Cbar$
is decreased to compensate the absolute {\it minimum} of $\lambda_-(\Rti) <
0$, $\Cbar + \lambda_-(\Rti) = 0$, strong pinning sets in locally at $\Rti_m$
for the first time in the form of an unstable ellipse $\Uti$, see Fig.\
\ref{fig:3_defects}(b) for our specific example with three defects; the
Labusch parameter $\kappa(\Rti)$ evaluated at the point $\Rti_m$ defines
$\kappa_m$, the parameter tuned in Fig.\ \ref{fig:3_defects}.  Decreasing
$\Cbar$ further, this ellipse grows and deforms, while other local {\it
minima} of $\lambda_-(\Rti)$ produce new disconnected parts of $\Uti$, a
situation illustrated in Fig.\ \ref{fig:3_defects}(c) where four `ellipses'
have appeared around (local) minima (blue filled dots).  A further increase in
pinning strength (decrease in $\Cbar$) continuous to deform these `ellipses'
and adds three new ones.  As the first {\it saddle} drops below the zero level
(red cross), two components merge and the number of components decreases; in
Fig.\ \ref{fig:3_defects}(d), we have three below-zero saddles and only
four components remain, $C = 4$.  In Fig.\ \ref{fig:3_defects}(e) four
further mergers have reduced $C$ to 1 as the corresponding {\it saddles} drop
below zero level. This produces a single non-simply connected component, i.e.,
$C = 1$ and a hole, increasing the number of holes $H$ from zero to one.
The last merger leading to (f) finally leaves $C = 1$ but cuts the stable
region inside the ring into two, increasing the number of holes to $H = 2$.

This sequence of onsets and mergers is conveniently described in the
topographic language introduced in section \ref{sec:uniax_defect} that
interprets stable tip regions as land mass (green with bright regions
indicating higher mountains in Fig.\ \ref{fig:3_defects}) and unstable regions
as lakes (flat blue with (below-water) height levels indicated by thin black
lines), with the height $\Lambda_{\Cbar} = 0$ defining the water level.  The
sequence (b) to (f) then shows the flooding of the landscape as pinning
increases ($\Cbar$ decreasing), with white dot minima turning blue at strong
pinning onsets and white cross saddles turning red at mergings; maxima in the
landscape are shown as black open circles. Note that we distinguish critical
points (minima, saddles) residing below (blue and red) and above (white) water
level.  Similarly, a (local) maximum above sea level (black open dot) turns
into a blue open dot as it drops belop sea level; such an event is missing in
Fig.\ \ref{fig:3_defects} but can be produced with other configurations of
defects, see Fig.\ \ref{fig:2_defects_maxima} where the curvature landscape
for two defects is shown.

The above discussion relates the local differential properties of the function
$\Lambda_{\Cbar}(\Rti) < 0$, minima and saddles, to the global topological
properties of $\Uti$, its number of components $C(\Uti)$ and holes $H(\Uti)$.
This connection between local and global properties is conveniently discussed
within Morse theory \cite{NashSen_2011}. Before presenting a general
mathematical formulation, let us discuss a simple heuristic argument producing
the result relevant in the present context; in doing so, we make use of the
above topographic language.

Starting with the {\it minima} of the function $\Lambda_{\Cbar}(\Rti)$, a new
disconnected component appears in $\Uti$ whenever the minimum drops below sea
level as $\Cbar$ is decreased, that produces an increase $C \to C+1$. With the
further decrease of $\Cbar$, these disconnected regions expand and merge
pairwise whenever a {\it saddle} point of $\Lambda_{\Cbar}(\Rti)$ goes below
sea level, thereby inducing a change in the topology of $\Uti$ by either
reducing the number of components $C \to C-1$ (keeping $H$ constant) or
leaving it unchanged (changing $H \to H + 1$), see, e.g., the example with the
horseshoe closing up on itself in Sec.\ \ref{sec:topology_hyp}. The below
sea-level minima and saddles of $\Lambda_{\Cbar}(\Rti)$ can naturally be
identified with the vertices and edges of a graph; the edges in the graph then
define the boundaries of the graph's faces (the same way as the vertices are
the boundaries of the edges).  For a connected graph, Euler's formula then
tells us that the number $V$ of vertices, $E$ of edges, and $F$ of faces are
constrained via $V - E + F = 1$ (not counting the outer face extending to
infinity) and a graph with $C$ components satisfies the relation $C = V - E +
F$ as follows from simple addition.

We have already identified minima and saddles of $\Lambda_{\Cbar}(\Rti) < 0$
with vertices and edges of a graph; denoting the number of below sea-level
minima and saddles by $m$ and $s$, we have $V = m$ and $E = s$. It remains to
express the number $F$ of faces in terms of critical points of the surface
$\Lambda_{\Cbar}(\Rti) < 0$. Indeed, the faces of our graph are associated
with maxima of the function $\Lambda_{\Cbar}(\Rti)$: following the boundaries
of a face, we cross the corresponding saddles with the function
$\Lambda_{\Cbar}(\Rti)$ curving upwards away from the edges, implying that the
faces of our graph include maxima of $\Lambda_{\Cbar}(\Rti)$.  These maxima
manifest in two possible ways: either the face contains a single below
sea-level maximum or a single above sea-level landscape.  The above sea-level
landscape comprises at least one maximum but possibly also includes other
extremal points that we cannot analyse with our knowledge of the below
sea-level function $\Lambda_{\Cbar}(\Rti) < 0$ only; we therefore call the
above sea-level landscape a (single) hole. The appearance of a {\it single}
maximum or hole is owed to the fact that faces are not split by a below
sea-level saddle as these have already been accounted for in setting up the
graph.

Let us denote the number of (below sea-level) maxima by $M$ and the number of
holes by $H$, then $F = H + M$.  Combining this last expression with Euler's
formula and regrouping topological coefficients $C(\Uti)$ and $H(\Uti)$ on one
side and extremal points $m[\Lambda_{\Cbar}(\Rti)]$, $s[\Lambda_{\Cbar}(\Rti)]$,
and $M[\Lambda_{\Cbar}(\Rti)]$ on the other, we arrive at the Euler
characteristic $\chi \equiv C- H$ and its representation through local
differential properties,
\begin{equation}\label{eq:def_chi_Euler_Morse}
  \chi(\Uti)  \equiv [C - H]_{\Uti} = [m - s + M]_{\Lambda_{\Cbar}(\Rti) < 0}.
\end{equation}

The result \eqref{eq:def_chi_Euler_Morse} follows rigorously from the
Euler-Poincar\'e theorem\cite{Nakahara_2003, NashSen_2011} in combination with
Morse's theorem \cite{NashSen_2011}, with the former expressing the Euler
characteristic $\chi(\Uti)$ through the so-called Betti numbers $b_i(\Uti)$,
\begin{equation}\label{eq:def_chi_Betti}
  \chi(\Uti)  \equiv \sum_{i=0}^{2} (-1)^i b_i(\Uti),
\end{equation}
where the $i$-th Betti number $b_i(\Uti) = \mathrm{Dim}[H_i(\Uti)]$ is given
by the dimension or rank of the $i$-th (singular) homology group $H_i(\Uti)$.
In colloquial terms, the Betti numbers $b_i$ count the number of `holes' in
the manifold with different dimensions $i$: the zeroth Betti number gives the
number of components $b_0 = C$ of $\Uti$, the first Betti number $b_1 = H$
counts the holes, and the second Betti number refers to cavities, here $b_2 =
0$ for our open manifold. Hence, we find that the Euler characteristic is
given by the number of components and holes in $\Uti$,
\begin{equation}\label{eq:chi_CH}
  \chi(\Uti)  = C(\Uti) - H(\Uti),
\end{equation}
in agreement with the discussion in Sec.\ \ref{sec:topology_hyp} and
\eqref{eq:def_chi_Euler_Morse}.

Morse theory\cite{NashSen_2011} then provides a connection between the
topological properties of the manifold $\Uti$ and the local differential
properties of the surface $\Lambda_{\Cbar}(\Rti) < 0$ defining it: with $C_i$
the number of critical points with index $i$ of the surface
$\Lambda_{\Cbar}(\Rti) < 0$ (the index $i$ counts the number of negative
eigenvalues of the Hessian matrix evaluated at the critical point), the Euler
characteristic $\chi(\Uti)$ relates the manifold's topology to the number and
properties of critical points,
\begin{equation}\label{eq:chi_Morse_C}
  \chi(\Uti)  = \sum_{i=0}^{2} (-1)^i C_i(\Lambda_{\Cbar} <0).
\end{equation}
For our 2D manifold the coefficients $C_i$ count the minima $C_0 = m$, the
number of saddles $C_1 = s$, and $C_2 = M$ refers to the number of maxima,
hence,
\begin{equation}\label{eq:chi_Morse_msM}
  \chi(\Uti)  = [m - s + M]_{\Lambda_{\Cbar} < 0}
\end{equation}
and the combination with \eqref{eq:chi_CH} produces the result
\eqref{eq:def_chi_Euler_Morse} anticipated above.

Summarizing, knowing the number of critical points $m$, $M$, and $s$ of the
seascape, i.e., its {\it local differential properties}, we can determine the
global topological aspects of the pinning landscape via the evaluation of the
Euler characteristic $\chi(\Uti)$ with the help of Eq.\
\eqref{eq:chi_Morse_msM}.  The latter then informs us about the number $C$ of
unstable domains in $\Uti$ where locally pinned states appear and the number
of holes $H$ in $\Uti$ where globally distinct pinned states show up.
Furthermore, the outer boundaries of the lakes, of which we have $C$
components, are to be associated with instabilities of the free vortex state,
while inner boundaries (or boundaries of holes, which count $H$ elements) tell
about instabilities of pinned states, hence the Betti numbers $C$ and $H$
count different types of instabilities. It would then have been nice to
determine the separate topological coefficients $C$ and $H$
individually---unfortunately, $\chi(\Uti)$ as derived from local differential
properties provides us only with the difference $C-H$ between locally and
globally pinned areas and not their individual values. Nevertheless, using
Morse theory, we could connect our discussion of local differential properties
of the pinning landscape in Secs.\ \ref{sec:ell_expansion} and
\ref{sec:hyp_expansion} with the global pinning properties of the pinning
energy landscape as expressed through the topology of the unstable domain
$\Uti$.

Regarding our previous examples, the isotropic and uniaxial defects, we remark
that for the latter the two simultaneous mergers on the $y$-axis produce a
reduction in $C = 2 \to 1$ and an increase of $H = 0 \to 1$ and hence a jump
from $\chi = 2$ to $\chi = 0$ in one step, as expected for two
simultaneous mergers. The symmetry of the isotropic defect produces a
(degenerate) critical line at $\rti_m$ rather than a critical point; adding a
small perturbation $\propto x^3$ breaks this symmetry and produces the
horseshoe geometry discussed in Sec.\ \ref{sec:topology_hyp} above that is
amenable to the standard analysis. 

A last remark is in place about the topological properties in dual space,
i.e., of bistable regions $\Bas$. Here, the mergers produce another
interesting phenomenon as viewed from the perspective of its thermodynamic
analogue.  Indeed, the merger of deformed ellipses in tip-space corresponds to
the merger of cusps in asymptotic space, what translates to the vanishing of
critical points and a smooth continuation of the first-order critical and
spinodal lines in the thermodynamic analogue, see also Sec.\
\ref{sec:hyp_Bas}. We are not aware of a physical example in thermodynamics
that produces such a merger and disappearance of critical points.

\section{Summary and outlook}\label{sec:summary}
Strong pinning theory is a quantitative theory describing vortex pinning in
the dilute defect limit where this complex many-body system can be reduced to
an effective single-pin--single-vortex problem. The accuracy offered by this
theory then allows for a realistic description of the shape of the pinning
potential $e_p(\mathbf{R})$ associated with the defects.  While previous work
focused on the simplest case of isotropic defects, here, we have generalized
the strong pinning theory to the description of arbitrary anisotropic pinning
potentials. Surprisingly, going from an isotropic to an anisotropic defect has
quite astonishing consequences for the physics of strong pinning---this
reminds about other physical examples where the removal of symmetries or
degeneracies produces new effects.

While the strong pinning problem is quite a complex one requiring the use of
numerical tools in general, we have identified several generic features that
provide the essential physics of the problem and that are amenable to an
analytic treatment. Specifically, these are the points of strong pinning onset
and the merger points, around which the local expansions of the pinning
potential $\epin(\Rti;\Ras)$ in the tip coordinate $\Rti$ allow us to find all
the characteristics of strong pinning. In particular, we identify the
instability region $\Uti$ in the vortex tip space (with coordinates $\Rti$)
and the bistable region $\Bas$ in the space of asymptotic vortex positions
$\Ras$ as the main geometric objects that determine the critical pinning force
density $\Fpin$, from which the critical current density $j_c$, the
technologically most relevant quantity of the superconductor, follows
straightforwardly. While the relevance of the bistable region $\Bas$ was
recognized in the past \cite{Labusch_1969,LarkinOvch_1979,Koopmann_2004}, the
important role played by the unstable region $\Uti$ went unnoticed so far.

When going from an isotropic defect to an anisotropic one, the strong pinning
onset changes dramatically: while the unstable region $\Uti$ grows out of a
circle of radius $\sim \xi$ and assumes the shape of a ring at $\kappa > 1$
for the isotropic situation, for an anisotropic defect the onset appears in a
point $\Rti_m$ and grows in the shape of an ellipse with increasing $\kappa_m
> 1$; the location where this onset appears is given by the Hessian of
$\epin$, specifically, the point $\Rti_m$ where its determinant touches zero
first, $\det\{\mathrm{Hess}[\epin(\Rti;\Ras)|_{\Ras}]\}_{\Rti_m} = 0$. The boundary of this
ellipse defines the jump positions $\Jti$ associated with the strong pinning
instabilities; when combined with the landing ellipse $\Lti$, these two
ellipses determine the jump distance $\delta \uti$ of the vortex tip, from
which follows the jump in the pinning energy $\Delta \epin \propto \delta\uti^4$,
which in turn determines $\Fpin$ and $j_c$. 

The bistable region $\Bas$ in asymptotic vortex space comes into play when
calculating the average critical force density $\Fpin$ opposing the vortex
motion: while the vortex tip undergoes a complex trajectory including jumps,
the vortex motion in asymptotic space $\Ras$ is described by a straight line.
As this trivial trajectory in $\Ras$-space traverses the bistable region
$\Bas$, the vortex tip jumps upon exiting $\Bas$, that produces the jump
$\Delta \epin$ and hence $\Fpin$.  Again, the shape of $\Bas$ changes when
going from the isotropic to the anisotropic defect, assuming a ring of finite
width around a circle of radius $\sim\xi$ in the former case, while growing in
the form of a crescent out of a point for the anisotropic defect. 

The new geometries associated with $\Uti$ and $\Bas$ then produce a
qualitative change in the scaling behavior of the pinning force density $\Fpin
\propto (\kappa_m - 1)^{\mu}$ near onset, with the exponent $\mu$ changing
from $\mu = 2$ to $\mu = 5/2$ when going from the isotropic to the anisotropic
defect. This change is due to the change in the scaling of the geometric size
of $\Bas$, with the replacement of the fixed radius $\sim \xi$ of the ring by
the growing size of the crescent $\sim \xi (\kappa_m-1)^{1/2}$ [the exponent
$\mu$ assumes a value $\mu =3$ for trajectories cutting the crescent along its
short dimension of size $\xi (\kappa_m-1)$]. Furthermore, for directed
defects, the pinning force density $\Fpin(\theta)$ depends on the impact angle
$\theta$ relative to the unstable direction $u$ and is aligned with $u$,
except for a small angular regime close to $\theta = \pi/2$. This results in a
pronounced anisotropy in the critical current density $j_c$ in the vicinity of
the strong pinning onset.

A fundamental difference between the strong pinning onsets in the isotropic
and in the anisotropic case are the geometries of the unstable $\Uti$ and
bistable $\Bas$ regions: these are non-simply connected for the isotropic case
(rings) but simply connected for the anisotropic defect (ellipse and
crescent). The resolution of this fundamental difference is provided by the
second type of special points, the mergers. Indeed, for a general anisotropic
defect, the strong pinning onset appears in a multitude of points, with
unstable and bistable regions growing with $\kappa_m > 1$ and finally merging
into larger areas.  Two examples illustrate this behavior particularly well,
the uniaxial defects with a quadrupolar and a dipolar deformation, see Secs.\
\ref{sec:uniax_defect} and \ref{sec:topology_hyp}. In the first case,
symmetric onset points on the $x$ axis produce two ellipses/crescents that
grow, approach one another, and finally merge in a ring-shaped geometry that
is non-simply connected. In the case of a dipolar deformation, we have seen
$\Uti$ grow out of a single point with its ellipse expanding and deforming
around a circle, assuming a horseshoe geometry, that finally undergoes a
merging of the two tips to produce again a ring; similar happens when multiple
$\Uti$ domains grow and merge as in Figs.\ \ref{fig:top_three} (a warped
defect) and \ref{fig:2_defects_maxima}(c) (a 2D pinning landscape where four
unstable domains have merged to enclose an `island').

These merger points are once more amenable to an analytic study using a proper
expansion of $\epin(\Rti;\Ras)$ in $\Rti$ around the merger point $\Rti_s$,
the latter again defined by the local differential properties of the
determinant $\det\{\mathrm{Hess}[\epin(\Rti;\Ras)|_{\Ras}]\}$, this time not a
minimum but a saddle.  Rather than elliptic as at onset, at merger points the
geometry is hyperbolic, with the sign change associated with increasing
$\kappa_s \equiv \kappa(\Rti_s)$ across unity producing a reconnection of the jump-
and landing lines $\Jti$ and $\Lti$.

While the expansions of $\epin(\Rti;\Ras)$ are describing the local pinning
landscape near onset and merging (and thus produce generic results), the study
of the {\it combined set} of onset- and merger-points describe the global
topological properties of $\Uti$ as discussed in Sec.\ \ref{sec:2D_landscape}:
every new (nondegenerate) onset increases the number of components $C$ in
$\Uti$, while every merger either decreases $C$ or increases $H$, the number
of `holes' or `islands' (or nontrivial loops in a non-simply connected region)
in the pinning landscape. It is the `last' merging producing a non-simply
connected domain that properly defines a new pinned state; in our examples
these are the closing of the two deformed ellipses in the uniaxial defect with
quadrupolar deformation and the closing of the horseshoe in the defect with a
dipolar deformation. Formally, the relation between the local differential
properties of the curvature function $\Lambda_{\Cbar}(\Rti) = \Cbar +
\lambda_-(\Rti)$ [with $\lambda_-(\Rti)$ the lower eigenvalue of the Hessian
of $e_p(\Rti)$], its minima, saddles, and maxima, are related to the global
topological properties of $\Uti$ as described by its Euler characteristic
$\chi = C - H$ through Morse theory, see Eq.\ \eqref{eq:def_chi_Euler_Morse}.
Such topological structures have recently attracted quite some interest, e.g.,
in the context of Fermi surface topologies and topological Lifshitz
transitions \cite{Volovik_2017, Kane_2022}.

The physics around the onset points as expressed through an expansion of
$\epin(\Rti;\Ras)$ resembles a Landau theory with $\Rti$ playing the role of
an order parameter and $\Ras$ the dual variable corresponding to a driving
field---here, $\Ras$ drives the vortex lattice across the defect and $\Rti$
describes the deformation of the pinned vortex. The endpoints of the crescent
$\Bas$ correspond to critical end points as they appear in the Landau theory
of a first-order transition line, e.g., the Ising model in an external field
or the van der Waals gas. The boundary lines of $\Bas$ correspond to spinodal
lines where phases become unstable, e.g., the termination of
overheated/undercooled phases in the van der Waals gas. The existence of
critical end points tells that `phases', here in the form of different pinning
branches, are smoothly connected when going around the critical point, similar
as in the gas--liquid transition of the van der Waals gas. As the `last'
critical point vanishes in a merger, a well defined new phase, here a new
pinned branch, appears.

Perspectives for future theoretical work include the study of correlations
between anisotropic defects (see Ref.\ \onlinecite{Buchacek_2020} addressing
isotropic defects) or the inclusion of thermal fluctuations, i.e., creep (see
Refs.\ \onlinecite{Buchacek_2019} and \onlinecite{Gaggioli_2022}).
Furthermore, our discussion of the extended pinscape in Sec.\
\ref{sec:2D_landscape} has been limited to a two-dimensional pinning
potential. In reality, defects are distributed in all three dimensions that
considerable complicates the corresponding analysis of a full
three-dimensional disordered pinning potential, with the prospect of
interesting new results.

On the experimental side, there are several possible applications for our
study of anisotropic defects.  For a generic anisotropic defect, the inversion
symmetry may be broken. In this case, the pinning force along opposite
directions is different in magnitude, as different jumps are associated to the
boundaries of the bistable region $\Bas$ away from onset, i.e., at
sufficiently large values of $\kappa_m$. Reversing the current, the different
critical forces then result in a ratchet effect \cite{Villegas_2003,
SouzaSilva_2006}. This leads to a rectification of an ac current and hence a
superconducting diode effect.  While for randomly oriented defects the pinning
force is averaged and the symmetry is statistically restored, for specially
oriented defects, the diode effect will survive.  Indeed, introducing
nanoholes into the material, vortex pinning was enhanced \cite{Wang_2013,
Kwok_2016} and a diode effect has been observed recently \cite{Lyu_2021}.
Generalizing strong pinning theory to this type of defects then may help in
the design of superconducting metamaterials with interesting functionalities.
Furthermore, vortex imaging has always provided fascinating insights into vortex
physics.  Recently, the SQUID-on-tip technique has been successful in mapping
out a 2D pinning landscape in a film \cite{Embon_2015} (including the
observation of vortex jumps) that has inspired a new characterization of the
pinscape through its Hessian analysis \cite{Willa_2022}; the adaptation of this
current-driven purely 2D setup to the 3D situation described in the present
paper is an interesting challenge.

Finally, we recap the main benefits of this work in a nutshell: For one, we
have established a detailed connection of the strong pinning transition with a
the concept of first-order phase transitions in thermodynamics, with the main
practical result that the scaling of the pinning force density $\Fpin \propto
(\kappa_m - 1)^\mu$ comes with an exponent $\mu = 5/2$ when working with
generic defects of arbitrary shapes. Second, we have found a mechanism, the
breaking of a defect's inversion symmetry, that produces rachets and a diode
effect in superconducting material.  Third, we have uncovered the geometric
structure and its topological features that is underlying strong pinning
theory, including a proper understanding of the appearance of distinguished
pinned states. While understanding these geometric structures seems to be of
rather fundamental/scholarly interest at present, future work may establish
further practical consequences that can be used in the development of
superconducting materials with specific functional properties.

\section*{Acknowledgments}
We thank Tom\'a\v{s} Bzdu\v{s}ek, Gian Michele Graf, and Roland Willa for
discussions and acknowledge financial support of the Swiss National Science
Foundation, Division II.

\appendix

\section{Effective $1$D Landau theory}\label{sec:eff_1D}
The Landau-type pinning energies \eqref{eq:e_pin_expans} and
\eqref{eq:e_pin_expans_hyp} for the vector order parameter $(\uti,\vti)$
involves a soft variable $\uti$ with a vanishing quadratic term $\propto
(1-\kappa_m)\>\uti^2$, as well as a stiff one, $\vti$, characterized by a
finite elasticity.  By eliminating the stiff direction $\vti$, we can arrive
at a 1D Landau expansion for the order parameter $\uti$ that provides us with
the desired results for the unstable and bistable domains $\Uti$ and $\Bas$
near onset and merging in a very efficient manner.

\subsection{Close to onset}\label{sec:eff_1D_onset} 
We start with the two-dimensional Landau-type energy functional \eqref{eq:e_pin_expans_ani}
\begin{align}\nonumber
  &e_\mathrm{pin}(\Rti; \Ras) =
  \frac{\Cbar\left(1 -\kappa_m\right)}{2} \, \uti^2
  + \frac{\Cbar + \lambda_+}{2}\, \vti^2 +\frac{a}{2}\, \uti \vti^2\\
  &\quad +\frac{\alpha}{4}\, \uti^2\vti^2
  +\frac{\beta}{6}\, \uti^3\vti
  +\frac{\gamma}{24}\, \uti^4
  -\Cbar\,\uas \uti - \Cbar\, \vas \vti
  \label{eq:e_pin_expans_ani_repeat}
\end{align}
written in terms of the tip coordinates $\uti,\vti$ measured relative to
$\Rti_m$, the position of the minimal determinant $D(\Rti)$ at strong
pinning onset, and with $\uti$ and $\vti$ aligned with the stable and unstable
directions, respectively.  The expansion \eqref{eq:e_pin_expans_ani_repeat} is
anisotropic: the quadratic (elastic) coefficient along the unstable
$\uti$-direction vanishes at the onset of strong pinning, while the one along
the stable $\vti$-direction stays positive and large, allowing us to
`integrate out' the latter. The asymptotic coordinates $\uas,\> \vas$ assume
the role of the driving (conjugate) fields for the tip positions (or order
parameters) $\uti,\> \vti$; the latter then are determined by the force
equations $\partial_{\Rti}e_\mathrm{pin}(\Rti; \Ras) = 0$,
\begin{align}\label{eq:asymptotic_positions_repeat_x}
   \Cbar \uas &=  \Cbar(1-\kappa) \uti + \frac{a}{2}\vti^2
   + \frac{\gamma}{6}\uti^3 + \frac{\beta}{2}\uti^2 \vti
   + \frac{\alpha}{2}\uti \vti^2,\\
   \Cbar \vas &= (\Cbar + \lambda_+) \vti + a\,\uti \vti + \frac{\beta}{6}\uti^3
   + \frac{\alpha}{2}\uti^2 \vti,
   \label{eq:asymptotic_positions_repeat_y}
\end{align}
see Eq.\ \eqref{eq:asymptotic_positions}, with $\delta \Ras = (\uas,\vas)$
measured relative to $\Ras_m$. Inspection of Eqs.\
\eqref{eq:asymptotic_positions_repeat_x} and
\eqref{eq:asymptotic_positions_repeat_y} shows that near the strong pinning
onset, the Ansatz  $\uti,~\vti,~\vas \propto \sqrt{\kappa_m - 1}$ and $\uas
\propto (\kappa_m - 1)$ produces a consistent solution. Solving the second
equation \eqref{eq:asymptotic_positions_repeat_y} for the stiff degree of
freedom $\vti$, we then find that
\begin{equation}\label{eq:vti_from_vas}
  \vti \approx \frac{\Cbar\vas}{\Cbar + \lambda_+ \! + a\uti}
  \approx\frac{\vas}{1\! +\! \lambda_+/\Cbar}
   \Bigl(1 - \frac{a/\Cbar}{1 \! + \! \lambda_+/\Cbar}\>\uti\Bigr)
\end{equation}
which is precise to order $(\kappa_m -1)$.  Inserting $\vti$ back into the
force-balance equation \eqref{eq:asymptotic_positions_repeat_x} for the
unstable component $\uti$, we find a cubic equation for $\uti$ (precise to
order $(\kappa_m -1)^{3/2}$) that is driven by a combination of $\uas$ and
$\vas^2$,
\begin{multline}\label{eq:eff_1D_force-balance}
  \Cbar \uas - \frac{(a/2)\, \vas^2}{(1+\lambda_+/\Cbar)^2}
  \approx \left[\Cbar(1-\kappa_m) + \frac{(\delta/2)\,\vas^2}
   {(1+\lambda_+/\Cbar)^2}\right] \uti \\
   +\frac{(\beta/2)\,\vas}{(1+\lambda_+/\Cbar)} \uti^2 + \frac{\gamma}{6}\uti^3.
\end{multline}
Upon integration, we finally arrive at the effective one-dimensional Landau
expansion for the 1D order parameter $\uti$ that is precise to order
$(\kappa_m -1)^2$ (up to an irrelevant shift $\propto \vas^2$),
\begin{equation}\label{eq:eff_landau_1D}
   e_\mathrm{pin}^\mathrm{eff}(\uti; \uas,\vas)
   = \frac{r(\vas)}{2}\uti^2 + \frac{w(\vas)}{6}\uti^3 + \frac{\gamma}{24}\uti^4 - h(\uas,\vas) \uti,
\end{equation}
with the coefficients $r,\> w$, and $h$ defined as
\begin{align}\begin{split}\label{eq:eff_1D_coeffs}
   r(\vas) &=\left[\Cbar(1-\kappa_m) + \frac{\delta}{2}\frac{\vas^2}
   {(1+\lambda_+/\Cbar)^2}\right],\\
   w(\vas) &= \beta\frac{\vas}{(1+\lambda_+/\Cbar)},\\
   h(\uas,\vas) &= \Cbar\uas - \frac{a}{2}\frac{\vas^2}{(1+\lambda_+/\Cbar)^2}.
\end{split}\end{align}
The Landau-type energy function \eqref{eq:eff_landau_1D} belongs to the van
der Waals (gas-liquid) universality class; its first-order transition line
maps to the branch crossing line in the strong pinning problem, its spinodals
correspond to the arcs of the crescent defining the bistable region $\Bas$,
and its critical points map to the two cusps of $\Bas$, i.e., in the strong
pinning problem, the spinodals end in {\it two} critical points. The cubic term
$w\uti^3/6$ is determined by the skew parameter $\beta$; in the absence of
such a skew, i.e., for a $\pm\vti$-symmetric unstable ellipse $\Uti$, we have
$\beta = 0$ and our problem assumes an Ising-type $\mathbb{Z}_2$ symmetry.

Let us begin with the determination of the critical coefficients $r_c$, $w_c$,
and $h_c$. These are found by setting the first three derivatives of
$e_\mathrm{pin}^\mathrm{eff}(\uti)$ to zero [two spinodals (implying 
$\partial_{\uti} e_\mathrm{pin}^\mathrm{eff} =0$ and $\partial^2_{\uti} 
e_\mathrm{pin}^\mathrm{eff} =0$) coalescing into a single
point ($\to \partial^3_{\uti} e_\mathrm{pin}^\mathrm{eff} =0$)]. Setting the
cubic derivative to zero, we find the order parameter 
\begin{equation}\label{eq:uti_c}
   \uti_c = -w_c/\gamma \approx -(\beta/\gamma)\vti_c,
\end{equation}
where we have used Eq.\ \eqref{eq:eff_1D_coeffs} and the transformation $\vas
\leftrightarrow \vti$ in \eqref{eq:vti_from_vas} to leading order.  

The vanishing of the second derivative relates the critical coefficients $r_c$
and $w_c$,
\begin{equation}\label{eq:r_c}
   r_c = w_c^2/2\gamma,
\end{equation}
(where we have made use of $\uti_c$). Inserting the dependencies $r(\vas)$ and
$w(\vas)$, see Eq.\ \eqref{eq:eff_1D_coeffs}, we find that
\begin{equation}\label{eq:vas_c}
   \frac{\vas_c^2}{(1+\lambda_+/\Cbar)^2}
   = \frac{\gamma\Cbar(\kappa_m-1)}{2 \det M_\mathrm{jp}},
\end{equation}
with $\det M_\mathrm{jp} = (\gamma\delta-\beta^2)/4$. Using again Eq.\
\eqref{eq:vti_from_vas} to leading order, we find that 
\begin{equation}\label{eq:vti_c_rep}
   \vti_c \approx  \sqrt{\frac{2\gamma\Cbar(\kappa_m - 1)}{\gamma\delta-\beta^2}},
\end{equation}
cf.\ Eq.\ \eqref{eq:vti_c}.  The critical endpoints of the 1D Landau theory
then correspond to the touching points \eqref{eq:contact_points} of the
unstable domain $\Uti$
\begin{equation}\label{eq:eff_1D_crit_point}
  \delta \Rti_{c,\pm} = \pm \left(-\beta/\gamma, 1\right)\,\vti_{c},
\end{equation}
found before, see Eq.\ \eqref{eq:contact_points} with \eqref{eq:vti_c}.

Finally, the vanishing of the first derivative defines the critical drive
\begin{equation}\label{eq:h_c}
   h_c = [r\uti + w\uti^2/2+\gamma\uti^3/6]_c = -\frac{w_c^3}{6\gamma^2}. 
\end{equation}
Making use of the coefficients \eqref{eq:eff_1D_coeffs}, this translates to
the critical drive $\uas_c$
\begin{equation}\label{eq:crescent_centre}
   \uas_c = (a/2\Cbar)\vti_c^2 -\frac{w_c^3}{6\Cbar\gamma^2}
\end{equation}
and its combination with the result for $\vas_c$ tells us that the critical
drives match up, to leading order, with the cusps \eqref{eq:cusps} 
of the bistable domain at $\Ras_{c,\pm}$,
\begin{eqnarray}\label{eq:cusps_rep}
   \delta \Ras_{c,\pm}  &=& (\uas_c,\pm \vas_c) \\ \nonumber
   &\approx& \left[\left(a/2\Cbar\right)\,\vti^2_c,\,
   \pm (1 + \lambda_+/\Cbar)\vti_c\right].
\end{eqnarray}

Next, we find the entire boundary of the unstable region $\Uti$ that is
defined as the points where local minima and maxima of
$e_\mathrm{pin}^\mathrm{eff}$ coalesce, i.e., where
$\partial^2_ue_\mathrm{pin}^\mathrm{eff}=0$,
\begin{equation}\label{eq:eff_1D_unstable_edge}
   r + w\ujp + \frac{\gamma}{2}\ujp^2 = 0.
\end{equation}
Making use of the Landau coefficients \eqref{eq:eff_1D_coeffs} as well as the
relation between $\vti$ and $\vas$ in \eqref{eq:vti_from_vas}, we recover the
equation \eqref{eq:quadratic_form} for the ellipse (we drop corrections
$\propto\left(\kappa_m - 1\right)^{3/2}$)
\begin{equation}\label{eq:ellipse_A}
   \gamma\ujp^2 + 2\beta\ujp\vjp + \delta \vjp^2
   \approx 2\Cbar(\kappa_m-1).
\end{equation}

In order to find the shape of the bistable region $\Bas$, we exploit the fact
that for fixed drives $\uas$ and $\vas$, the bistable and the unstable vortex
tip configurations are local extrema of $e_\mathrm{pin}^\mathrm{eff}$,
implying that $\partial_{\uti} e_\mathrm{pin}^\mathrm{eff} = 0$ and hence
\begin{equation}\label{eq:eff_1D_cubic}
   r\uti + \frac{w}{2}\uti^2 + \frac{\gamma}{6}\uti^3 = h,
\end{equation}
what corresponds to the force-balance equation \eqref{eq:eff_1D_force-balance}
expressed in terms of the coefficients \eqref{eq:eff_1D_coeffs}.  The cubic
equation \eqref{eq:eff_1D_cubic} with its left side $\propto (\kappa_m
-1)^{3/2}$ depends on $\uas$ through the drive $h$. According to
\eqref{eq:eff_1D_coeffs}, the two terms in the drive are of order $(\kappa_m -
1)$ and hence have to cancel one another to lowest order. As a result, we find
that the bistable domain is centered around the parabola
\begin{equation}\label{eq:eff_1D_parabola}
   \uas = \frac{a}{2\Cbar}\frac{\vas^2}{(1+\lambda_+/\Cbar)^2},
\end{equation}
that matches up with Eq.\ \eqref{eq:parabola_x} found in Sec.\
\ref{sec:arb_shape}. Finding the precise form of the bistable region $\Bas$,
we have to solve Eq.\ \eqref{eq:eff_1D_cubic} to cubic order in
$\sqrt{\kappa_m - 1}$ with the help of an expansion around the center parabola
\eqref{eq:eff_1D_parabola}, what amounts to repeating the analysis leading to
the results \eqref{eq:duas} and \eqref{eq:dvas} in Sec.\ \ref{sec:Bas}.

Finally, we find the landing line $\mathcal{L}_{\Rti}$ defined as the second
bistable tip position at fixed $\uas$ and $\vas$. We make use of the cubic
equation \eqref{eq:eff_1D_cubic} and represent it in the factorized form
(with the inflection point at $\ujp$ having multiplicity two)
\begin{equation}\label{eq:cubic_factorized}
(\uti - \ujp)^2(\uti - \ulp) = 0,
\end{equation}
and $\ulp$ the landing position of the tip introduced in Sec.\ \ref{sec:Lti}.
A somewhat tedious but straightforward calculation shows that the stable
solution $\ulp$ satisfies the quadratic equation
\begin{equation}\label{eq:ulp_quadratic}
   r - \frac{3}{8}\frac{w^2}{\gamma} + \frac{w}{4}\ulp + \frac{\gamma}{8}\ulp^2=0
\end{equation}
and thus arranges along the ellipse
\begin{equation}\label{eq:land_ellipse}
   \frac{\gamma}{8}\ulp^2 + \frac{\beta}{4}\ulp\vlp +\left(\frac{\delta}{2} 
   - \frac{3}{8}\frac{\beta^2}{\gamma}\right)\vlp^2=\Cbar(\kappa_m - 1)
\end{equation}
when expressed in the original two-dimensional tip space; this coincides with
the original result \eqref{eq:matrix_eq_lp}. 

In a last step, we may go over to an Ising-type Landau expansion by measuring
the order parameter $\uas$ with reference to the skewed line
\begin{equation}\label{eq:skewed_u}
   \uti_m(\vas) = \left(-\frac{\beta}{\gamma}\right)
   \frac{\vas}{(1+\lambda_+/\Cbar)},
\end{equation}
i.e.,
\begin{equation}
 \uti' = \uti - \uti_m(\vas).
 \end{equation}
The 1D effective Landau expansion now reads, with precision to order $(\kappa_m -1)^2$,
\begin{equation}\label{eq:eff_landau_1D_shifted}
   e_\mathrm{pin}^\mathrm{eff}(\uti'; \uas,\vas )
   = \frac{r'}{2}\uti'^2 + \frac{\gamma}{24}\uti'^4 - h' \uti',
\end{equation}
with the new coefficients
\begin{align}\label{eq:eff_1D_coeffs_shifted}
   r' = r -\frac{w^2}{2\gamma}, \quad h'= h - \frac{w^3}{3\gamma^2} + \frac{rw}{\gamma}.
\end{align}
The condition $h'=0$ now defines the equilibrium state of the thermodynamic
problem that translates into the branch crossing line where the bistable
vortex tip positions have equal energy. Using the definitions
\eqref{eq:eff_1D_coeffs} and \eqref{eq:eff_1D_coeffs_shifted} for $h$ and
$h'$, we find that the branch crossing line $\uas_0(\vas_0)$ in the original
two-dimensional asymptotic space reads
\begin{multline}\label{eq:x_0_line_beta}
   \uas_0 = \frac{a}{2\Cbar}\frac{\vas_0^2}{(1+\lambda_+/\Cbar)^2} 
   -\frac{\beta}{\gamma}\biggl[
   (\kappa_m -1)\frac{\vas_0}{1+\lambda_+/\Cbar} \\
   +\left(\frac{\delta}{2} - \frac{\beta^2}{3\gamma}\right)
   \frac{1}{\Cbar} \frac{\vas_0^3}{(1+\lambda_+/\Cbar)^3}\biggr],
\end{multline}
extending the result \eqref{eq:x_0_line} from Sec.\ \ref{sec:arb_shape} to
finite values of $\beta$ with an additional term
$\propto\left(\kappa_m-1\right)^{3/2}$.

\subsection{Close to merging}\label{sec:eff_1D_merging}
Let us study the strong pinning problem close to merging, as described by
the two-dimensional Landau-type energy functional \eqref{eq:e_pin_expans_hyp},
\begin{align}\nonumber
  &e_\mathrm{pin}(\Rti; \Ras) = 
  \frac{\Cbar(1-\kappa_s)}{2} \, \uti^2 
  + \frac{\Cbar + \lambda_{+,s}}{2}\, \vti^2  
  +\frac{a_s}{2}\, \uti \vti^2 
  \\
  &\quad+\frac{\alpha_s}{4}\, \uti^2\vti^2
  +\frac{\beta_s}{6}\, \uti^3\vti
  +\frac{\gamma_s}{24}\, \uti^4
  -\Cbar\uas\uti - \Cbar\vas \vti. \label{eq:e_pin_expans_hyp_bis}
\end{align}

As found before for strong pinning close to onset, the energy functional
\eqref{eq:e_pin_expans_hyp_bis} is anisotropic with respect to vortex
displacements in the stable and unstable direction.  Following the strategy of
Sec.\ \ref{sec:eff_1D_onset}, we can use the force-balance equation
\eqref{eq:asymptotic_positions_hyp} to relate the tip position along the
$v$-axis to $\vas$ and $\uti$,
\begin{align}\label{eq:vti_from_vas_hyp}
   \vti \approx\frac{\vas}{1+\lambda_{+,s}/\Cbar}
   \left(1 - \frac{a_s/\Cbar}{1 + \lambda_{+,s}/\Cbar}\>\uti\right).
\end{align}
Inserting \eqref{eq:vti_from_vas_hyp} into the force-balance equation for the
unstable component $\uti$ and integrating, we find that the resulting
effective 1D Landau theory is identical in form to the one close to onset,
\begin{equation}\label{eq:eff_landau_1D_hyp}
   e_\mathrm{pin}^\mathrm{eff}(\uti; \uas,\vas) = \frac{r_s}{2}\uti^2 +
   \frac{w_s}{6}\uti^3 + \frac{\gamma_s}{24}\uti^4 - h_s \uti,
\end{equation}
with a proper replacement of all coefficients involving the parameters
appropriate at merging,
\begin{align}\begin{split}\label{eq:eff_1D_coeffs_hyp}
   r_s&=\left[\Cbar(1-\kappa_s) - \frac{|\delta_s|}{2}\frac{\vas^2}
   {(1+\lambda_{+,s}/\Cbar)^2}\right],\\
   w_s&= \beta_s\frac{\vas}{(1+\lambda_{+,s}/\Cbar)},\\
   h_s&= \Cbar\uas - \frac{a_s}{2}\frac{\vas^2}{(1+\lambda_{+,s}/\Cbar)^2}.
\end{split}\end{align}
The difference to \eqref{eq:eff_1D_coeffs} is the sign change in the term
$\propto |\delta_s| \vas^2$.  This implies a modification of the main equation
determining the shape of $\Uti$ (from which $\Bas$ follows via the force
balance equation \eqref{eq:gen_force_balance}), with the elliptic equation
\eqref{eq:ellipse_A} transforming to the hyperbolic expression
\begin{equation}\label{eq:hyp_A}
   \gamma_s\ujp^2 + 2\beta_s\ujp\vjp - |\delta_s|\vjp^2
   \approx 2\Cbar(\kappa_s-1).
\end{equation}
The results for the jumping and landing hyperbolas in $\Rti$-space and for the
edges of the bistable domain in $\Ras$-space before and after merging can be
derived by following the strategy of Sec.\ \ref{sec:eff_1D_onset} above and
agree with the corresponding results from Sec.\ \ref{sec:hyp_expansion}.

We close with a final remark on the disappearance of critical points after
merging. The critical points are found in the standard manner by setting the
first three derivatives of $e_\mathrm{pin}^\mathrm{eff}(\uti; \uas,\vas)$ to
zero.  This works fine before merging when $1 - \kappa_s > 0$ and we find that
criticality is realized for tip and asymptotic positions as given by Eqs.\
\eqref{eq:hyp_vertices} and \eqref{eq:cusps_hyp} in Sec.\
\ref{sec:hyp_expansion}. However, after merging, the cubic derivative
$\partial^3_{\uti} e_\mathrm{pin}^\mathrm{eff}$ never vanishes, signalling the
absence of a critical point, in agreement with the discussion in Secs.\
\ref{sec:hyp_Bas} and \ref{sec:hyp_Lti}. The merger thus leads to the
disappearance of the two critical (end-)points in asymptotic space, with the
attached first-order lines (the branch crossing line) joining up into a single
line that is framed by two separated spinodals.  We are not aware of such a
disappearance of critical points in a merging process within the standard
discussion of thermodynamic phase transitions.

%

\bibliography{refs_vortices}

\end{document}